\documentclass[uspaper, 12pt]{article}
\usepackage{mathptmx}
\usepackage{type1cm}
\usepackage{mathrsfs}
\usepackage{amsmath}
\usepackage{natbib}
\usepackage{geometry}
\usepackage{type1cm}
\usepackage{mathrsfs}
\usepackage{amssymb}
\usepackage{authblk}
\usepackage{amsfonts}
\usepackage[T2A,T1]{fontenc}
\usepackage{hhline}
\usepackage{amsthm}
\usepackage{multicol}
\usepackage{color}
\usepackage{sgame}
\usepackage{mathtools}
\usepackage{caption}
\usepackage{graphicx}
\usepackage{wrapfig}
\usepackage{changepage}
\usepackage{lipsum}
\usepackage[english]{babel}
\usepackage[utf8]{inputenc}
\usepackage{comment}
\usepackage{subcaption}
\usepackage{url}
\usepackage[pagewise]{lineno}
\usepackage[export]{adjustbox}
\usepackage{tikz}
\usepackage{pgf}
\usepackage{xltabular}
\usepackage{float}
\usepackage{booktabs}
\definecolor{mypurple1}{RGB}{204,51,255}

\usepackage{xcolor}           % For color
\usepackage[colorlinks]{hyperref}  % Make links colored instead of boxed
\hypersetup{
    linkcolor = {blue!50!black},         % Color of internal links (sections, figures, etc.)
    citecolor = {blue!50!black},
    urlcolor  = magenta
}

\newtheorem{proposition}{Proposition}
\newtheorem{corollary}{Corollary}
\newtheorem{lemma}{Lemma}

\newtheorem{example}{Example}
\newtheorem{definition}{Definition}

\newtheorem{hypothesis}{Hypothesis}
\newtheorem{remark}{Remark}

\addto\extrasenglish{%
}

\def\cB{\mathcal{B}}

\usepackage{setspace}
\newcommand{\hide}[1]{} 

\begin{document}

\title{Bundled School Choice\thanks{We received useful feedback from seminars and conferences at Zhejiang University, Korea Advanced Institute of Science and Technology, 2024 Asian Meeting of Econometric Society (Hangzhou), MATCH-UP 2024 (Oxford), Greater Bay Area Market Design Workshop 2024 (Macau), Hitotsubashi Summer Institute Microeconomic Theory Workshop 2025, and ESA Asian Pacific Meeting 2025 (Osaka). All errors are ours.}}
\author{Lingbo Huang\thanks{Center for Economic Research, Shandong University, China. Email: lingbo.huang@outlook.com.} \quad Jun Zhang\thanks{Institute for Social and Economic Research, Nanjing Audit University, China. Email: zhangjun404@gmail.com.}}
\maketitle

\begin{abstract}
    This paper proposes a novel school choice system where schools are grouped into hierarchical bundles and offered to students as options for preference reports. By listing a bundle, a student seeks admission to any school within it without ranking the schools. This approach helps students who struggle to rank schools precisely and expands options on limited preference lists, potentially improving match outcomes. We design a modified deferred acceptance mechanism to handle bundle reports while preserving stability. Two laboratory experiments support our theory, showing that well-constructed bundles aligned with student preferences enhance welfare and match rates without compromising fairness. Practical applications are discussed.
\end{abstract}

\bigskip

\noindent \textbf{Keywords}: School choice; coarse preferences; constrained preference lists; school bundles

\noindent \textbf{JEL Classification}: C78, C91, D71, I20

\thispagestyle{empty}
\setcounter{page}{0}

\newpage

\section{Introduction}

This paper introduces the idea of designing school bundles within the standard school choice system as a means to address two practical challenges. The \textbf{first} challenge arises from the difficulty students face in ranking schools in a strict and precise order. This difficulty often stems from the prohibitively high cost associated with acquiring and processing the extensive information required to evaluate a large number of schools. Conflicting criteria between students and their parents in selecting schools can also complicate the process of assigning strict rankings. 
Consequently, students may prefer to express their preferences in a coarse rather than finely ordered manner. 
The \textbf{second} challenge pertains to the widespread use of constrained preference lists, which can result in a significant number of students remaning unmatched. For instance, in New York City's (NYC) public school admissions, students can rank up to twelve schools. Although this list length may appear sufficient, 7\% (about 5,000) of students were unmatched in the main round of the admissions process during the years 2021 and 2022. In certain districts, such as Manhattan, the unmatch rate was as high as 18\%. Furthermore, these two challenges can interact and reinforce one another. If students were allowed to submit complete preference lists, those with coarse preferences could still construct strict rankings by arbitrarily breaking ties. However, when preference lists are constrained, the pressure to refine preferences becomes more acute, making it harder for students to express their true intentions. For instance, if a student who wishes to attend a school near her parents’ workplace is limited to ranking only one school, she may wish to express a coarse preference, rather than choosing one school to list, even if her underlying preferences are strict and complete.

In our proposed system, some schools are exogenously grouped into bundles and presented to students as options for preference submission on rank-order lists (ROL). When a student ranks a bundle, she indicates a willingness to attend any school within that bundle without specifying a preference among these schools. The system allows multiple bundles, which may overlap; however, to ensure a well-designed admission procedure, the structure of bundles must be hierarchical: among any two overlapping bundles, one must fully encompass the other. Importantly, all individual schools remain available to all students. This ensures that our design not only expands students' choice set compared to the standard system, but also accommodates students' potentially diverse preferences. For example, some students may prefer to express coarse preferences by ranking a bundle, whereas others may opt for expressing precise preferences by ranking individual schools within the bundle. The admission process is organized into two stages. In the first stage, students submit their ROLs and are assigned to their ranked schools or bundles. In the second stage, students who are assigned to bundles are further admitted to specific schools within those bundles. Our paper focuses on the design of the first stage, leaving the details of the second stage to the discretion of policymakers. This flexibility allows policymakers to address context-specific concerns in the second stage. For instance, in an application discussed in \autoref{section:application}, policymakers incorporate randomization into the allocation process in the second stage to promote fairness for both students and schools. That application also illustrates why an explicit tie-breaking rule that converts a bundle report into a strict ranking might be undesirable in practice.

Our proposed system offers two primary advantages over the standard system. First, it enables students to express coarse preferences by ranking bundles, which can potentially enhance their overall welfare if forming precise preferences is costly and bundles are properly designed. Second, by allowing students to rank bundles, the system effectively expands the number of schools that students can rank within the limitations of a constrained ROL, thus potentially increasing the likelihood of securing a match. The following example provides an illustration.

%A bundle system can offer different advantages over the standard system depending on the context. In general, it provides two primary benefits. First, it enables students to express coarse preferences, which can potentially enhance their welfare, especially when forming precise preferences is costly. Second, by enabling students to report bundles, it effectively increases the number of schools students can list on a constrained ROL, thus potentially boosting their match rates. We believe that this latter benefit could be a compelling reason for students to favor reporting bundles when the ROL length is significantly constrained. The following example is an illustration.

\begin{example}\label{example:introduction}
 Consider two schools $ s $ and $s'$ and two students $ i $ and $i'$. Each student prefers both schools over being unmatched. Each school can admit one student and uses the same priority order $i\succ i'$. So, an efficient matching would admit both students, and stability (fairness) would further require that $ i' $ be assigned to his preferred school and $ i $ to the other school. 
 
 However, consider a scenario where students can rank only one school in their ROL. Then, a coordination problem arises: if both students rank the same school, $ i $ will be unmatched. 

Now, suppose we introduce a bundle $\{s,s'\}$. By ranking the bundle instead of a school, a student seeks admission to either school without indicating a preference. We consider three cases:
\begin{itemize}
	\item[(1)] If $ i' $ ranks a school and $ i $ ranks the bundle, then no matter which school $ i' $ ranks, $ i $ is admitted to the bundle. After $ i' $ is assigned to his ranked school, $ i $ is assigned to the other school. This illustrates one potential benefit of bundles: by ranking bundles, a student can increase her match likelihood.

	\item[(2)] If $ i' $ ranks the bundle and $ i $ ranks a school, then no matter which school $i$ ranks, $ i $ can be admitted to his ranked school, because $ i' $ is admitted to the bundle and can then be assigned to the other school. This illustrates another potential benefit of bundles: by reporting bundles, a student provides policymakers flexibility in determining her assignment, which may benefit other students.
	
	\item[(3)] If both students rank the bundle, they are both admitted, with one student finally assigned to $ s $ and the other to $ s' $.   
    However, if students have strict preferences over the two schools, the selected matching may not align with their true preferences. This illustrates a potential cost of bundles: ranking bundles reduces a student's control of her final assignment.
\end{itemize} 
\end{example}

Our proposed system is novel to the literature; however, we emphasize that various bundle policies have been employed in practice, though they have received limited attention from the field.  \autoref{section:application} discusses a bundle policy that had been implemented in the high school admission system of a major city in China for nearly two decades. In the city, three schools were widely regarded by students as the top choices and viewed as nearly indifferent, but before the bundle policy, students were allowed to rank at most one of the three schools in their ROL, which created significant challenges for their strategies. To address this, the government bundled the three schools, allowing students to apply to all three by ranking the bundle. Any student whose exam score exceeded the bundle's cutoff was guaranteed admission to one of the three schools. This bundle policy was embraced by students and considered a successful reform.  Another real-world example can be found in the undergraduate application processes of the University of Cambridge and the University of Oxford. Both institutions consist of multiple colleges from which students can choose when applying. However, if a student prefers not to specify a particular college, she may submit an open application, allowing the university to determine her assignment. In this context, an open application functions as a bundle of colleges. A similar policy exists in Italy's centralized teacher assignment process. Teachers often have geographical preferences and are given the option to rank either individual schools or an entire region in their ROL. Ranking a region is effectively equivalent to ranking a bundle of schools within that region.

We present our proposed system as follows. After outlining the standard school choice system in  \autoref{section:standard:system},  \autoref{section:bundle:system} introduces key concepts in our design. We intentionally present them in a general framework to ensure broad applicability across various contexts while preserving the bundle matching process as closely as possible to the standard system.  
A challenge in our design arises from the fact that bundles do not possess exogenous priority orders over students. Rather than imposing arbitrary admission criteria for bundles, we require that each bundle adhere to the priority rankings jointly agreed upon by the bundled schools. This is achieved by requiring that only schools that use identical priority rankings for a certain group of students can be bundled, and the bundle is only available to those students. This ensures that, for any outcome of our designed mechanism in the first stage, any implementation chosen by policymakers in the second stage must be stable, where the stability concept is adapted to accommodate bundles. Consequently, our design is well-suited for environments where school priorities have homogeneous components. Such settings are common in practice. For instance, in merit-based admission systems, students are often ranked based on standardized test scores; in priority-based systems where priorities are based on coarse criteria (e.g., geographic proximity or demographic attributes), a single tie-breaking lottery can induce homogeneous rankings within each priority class.

\autoref{section:general:DA} presents our modification of the deferred acceptance (DA) mechanism to find a stable outcome for any ROLs submitted by students. This modified DA differs from the standard DA in several critical aspects. First, in each round, the mechanism must address the complexity arising from students applying to schools or bundles that overlap but remain distinct. In particular, the selection of the highest-priority applicant for each school or bundle must be handled with care. When a student applies to a bundle, she is treated as an applicant for every school within that bundle. A student is then defined as the highest-priority applicant for a bundle if she applies to that bundle, and, for every school in that bundle, she holds the highest priority among all students regarded as applicants for the school. The hierarchical structure of the bundle system, together with the identical priority ranking within each bundle, ensures the existence of such highest-priority applicants. In each round, we recursively identify these students and tentatively admit them. 

The second issue concerns updating bundle quotas during the algorithm. The initial quota of a bundle equals the number of seats within it. In each round, if some students are admitted to the schools or bundles encompassed by a bundle $b$, we update not only the quotas of those schools or bundles but also the quota of $b$, as these students occupy seats within $b$. However, if some students are admitted directly to $b$, we update the quota of $b$ but not the quotas of the schools or bundles encompassed by $b$, as the exact assignment of these students within $b$ has not yet been determined. This, however, introduces a potential issue: in subsequent rounds, the sum of the remaining quotas of the schools or bundles within $b$ may exceed the remaining quota of $b$. To address this, if the quota of $b$ reaches zero in any subsequent round, we immediately set the quotas of all schools and bundles within $b$ to zero, even though they do not admit students in that round. We prove that the outcome of our mechanism must be stable with respect to students' ROLs. Moreover, the mechanism generates a maximal set of matched students in a sense formalized in the paper.

Our design can accommodate various environments and potentially allow for a complex bundle system associated with a complicated mechanism. However, in practical applications, policymakers may prefer bundle systems with a simple structure and an easily comprehensible mechanism. Thus, we introduce a class of simplified bundle systems in  \autoref{section:bundle:definition}, in which only schools with identical priority rankings for \textit{all} students are bundled. This restriction significantly simplifies both the structure of the bundle system and the modified DA mechanism. Moreover, the modified DA possesses desirable properties. We prove that a student's match likelihood must weakly increase if she ranks larger bundles in her ROL. Students are also incentivized to be ``truth-telling'': if there is a ROL that represents a student's preference over bundles, she must report that ROL. In general, students face a tradeoff: ranking bundles can increase their chances of being matched but may reduce their control over final assignments. 
When restrictions on the ROL length make the risk of being unmatched a significant concern, the benefits of ranking bundles can outweigh the associated costs. Furthermore, when a student is nearly indifferent over the schools in a bundle such that any assignment within the bundle is almost welfare-equivalent, the cost of ranking the bundle becomes negligible, making the bundle more attractive.

Our focus in the paper is on outlining the foundational principles for designing bundle systems. In general, determining when to adopt a bundle system and how to optimize its structure depends on the specific context, which lies beyond the scope of this paper.  However, regardless of the setting, successful implementation requires policymakers to thoroughly understand both participants' preferences and the relevant environmental constraints.  \autoref{section:practicalguide} offers practical guidelines for implementation, while the application discussed in \autoref{section:application} serves as a valuable case study, illustrating how contextual factors can shape the design process.

\autoref{section:two-stage} discusses an extension of our bundle system, addressing scenarios where preference formation requires costly information acquisition. By allowing students to form preferences in two stages, this extension aims to reduce the associated costs and enhance their welfare. In the first stage, students only need to gather sufficient information to form preferences over bundles. After submitting these preferences and being matched to bundles, they can then acquire more detailed information to form precise preferences over a limited number of schools within their assigned bundles. During the second stage, students report their precise preferences and are matched to specific schools using some algorithm. This extended matching process allows students to allocate their resources more efficiently across the two stages of preference formation. We believe that the potential advantages of this design merit investigation in future research.

%The potential advantages of the bundle system are outlined in this paper, though they are not formally proven within a specific model. This is because our main interest in the paper is to present our design and the benefits of bundles have been exemplified in applications. Another reason is that, in general, 

%Although this paper outlines the potential advantages of the bundle systems in improving matching outcomes, fully characterizing students' ROL strategies in a constrained school choice game, which is a well-known hard problem, remains intractable.\footnote{When students can only rank a limited number of schools, predicting the specific schools students may choose to rank is generally intractable. \cite{ali2024college} characterize the pattern of optimal strategies: students generally apply to a portfolio of schools consisting of ``reaches,'' ``matches,'' and ``safeties.''} The strategic complexity is further exacerbated in the context of our bundle system due to its complex strategy space and the additional uncertainties introduced in the second stage of our admission process. Therefore, we conduct a laboratory experiment as a proof-of-concept to evaluate the impacts of bundle systems on students' strategic behavior and their matching outcomes.  

Prior literature has shown that student strategies under DA with constrained preference lists are generally difficult to characterize, even without information uncertainties \citep{haeringer2009constrained,decerf2021manipulability}. Only recently have \cite{ali2024college} provided a framework for understanding the structure of school portfolios that students should submit when school admission decisions are correlated. In the context of bundle systems, the difficulty is further exacerbated by the complex strategy space and uncertainties in the second stage of the admission process. Therefore,  \autoref{section:experiment} presents laboratory experiments to evaluate the impacts of bundle systems on students' strategic behavior and their matching outcomes. 

%In a designed school choice environment, we compare participants' ROL strategies and matching outcomes under three treatments: a baseline treatment with no bundled schools and two bundle treatments, each representing distinct real-life scenarios. The first bundle treatment simulates a well-designed system where the bundled schools are nearly indifferent to students. In this scenario, participants choose to report the bundle around 62\% of the time, leading to welfare improvements for most students without compromising fairness compared to the baseline. The second bundle treatment represents a poorly designed system where the bundled schools offer disparate payoffs to students. Here, students with higher priorities generally avoid reporting the bundle due to the significant opportunity cost of being assigned to a less preferred school within the bundle. Consequently, the overall bundle reporting rate drops to 45\%, benefiting only lower-priority students who have incentives to report the bundle, while fairness is adversely affected. Furthermore, the observed patterns in participants' ROL strategies align closely with the incentives provided by each bundle treatment. Overall, the experimental results validate our theoretical insights and demonstrate the potential of bundle systems to enhance student welfare, mainly through increasing matching rates.

%To test the benefits of bundle systems, 
We conduct two laboratory experiments. The main experiment adopts a simple environment in which students rank either a single school or a bundle in their preference lists (i.e., ROL length is one). The simple environment allows us to fully characterize students' equilibrium strategies across all treatments. Importantly, our design enables a clean test of the source of efficiency gains associated with the bundle system: whether improvements stem merely from the expansion of reported schools or from the substantive benefits of well-designed bundles that group close rather than dissimilar schools. Our experimental results provide strong support for the theoretical predictions. Offering a well-designed bundle increases average payoffs by 17.7\%, lifts the match rate from 62.7\% to 71.2\%, and cuts the mismatch rate from 20.2\% to 4.4\%. In contrast, offering a poorly-designed bundle yields no payoff gains and slightly increases the mismatch rate (22.4\% vs. 20.2\%) relative to the baseline treatment without bundles. These results indicate that welfare improvements arise primarily from the bundle design itself rather than from the mechanical advantage of allowing students to report more schools through bundles. Further, allowing students to report two individual schools achieves a higher match rate and the highest average payoff, but largely through hedging behavior that lowers average payoffs conditional on matching relative to the well-designed bundle. The second experiment implements a more complex environment with a larger market and allows students to report two options, including both individual schools and bundles. We find qualitatively consistent patterns across both experiments.

\autoref{section:conclusion} concludes the paper by discussing potential directions for future research. The appendix includes proofs and additional results. The online appendix includes the details of the second experiment and instructions for both experiments.

\paragraph{Related literature.}

The concept of bundles has been explored in the market design literature but differs from the role of bundles in our paper. A couple of studies focus on the allocation of bundles in contexts such as course allocation, where agents demand multiple objects (e.g., \citealp{budish2011combinatorial,nguyen2023dynamic}). Unlike these studies, our paper addresses a two-sided matching problem where students demand individual schools, not bundles. In our design, bundles serve as a means in the matching process to an end of being matched with individual schools. 

Another related strand of research has explored how bundling influences the expressiveness of preference reporting languages in matching models, but these studies pursue different objectives from ours. Moreover, our study differs from previous work as it requires significant modifications to the DA mechanism to accommodate bundles. Specifically, in an unpublished paper, \cite{bodoh2013risk} argues that the common practice of constrained preference lists exposes participants to significant risks in large matching markets. He discusses an idea of conflating the description of alternatives to allow participants to express approximate preferences, which represents a broader idea than bundling. However, he only demonstrates the idea through examples, without providing a concrete design as in our paper

In a two-sided matching with contracts model, \cite{hatfield2017contract} examine the impact of bundling primitive contracts on the existence of stable matching. When multiple primitive contracts are bundled as a single new contract, the contractual language becomes less expressive, yet participants' preferences become more likely to be substitutable, rendering stable matching more likely to exist. We differ from this work in that our introduction of bundles makes the preference reporting language more expressive; bundles do not replace individual schools.

In an object assignment model, \cite{fragiadakis2019designing} study a mechanism that allows participants to communicate cardinal preferences via expressing indifference. A participant receives a higher priority if she reports a larger indifference class and ranks it higher in her preference list. Reporting an indifference class resembles reporting a bundle. However, unlike our paper, participants in their mechanism can freely determine the indifference classes, as their focus is on a one-sided assignment problem where stability is not a concern. Additionally, their mechanism improves participants' welfare only when participants play non-equilibrium strategies.

It is important to highlight that our approach to handling bundles in our modified DA mechanism differs significantly from the conventional treatment of indifferent preferences in the literature, which typically incorporates a tie-breaking rule (e.g., \citealp{erdil2017two}). An independent yet closely related paper by \cite{dougan2024geography} studies the teacher assignment process in Italy, which utilizes an official predetermined order of schools to resolve preference ties. In their context, teachers are allowed to report geographical regions in their preference lists. By reporting a region, a teacher indicates indifference among all schools within that region. The main contribution of the paper lies in identifying the flaws of the existing mechanism in resolving preference ties and proposing an improvement. Interestingly, geographical regions naturally form a hierarchy, the condition we impose on a bundle system. However, the Italian mechanism has several critical differences from our design. First, in the Italian system, regions are exogenously defined and available to all teachers for preference submission, with schools within a region potentially ranking teachers differently. In contrast, our system allows policymakers to design bundles, with the restriction that schools within a bundle must rank their targeted students identically. Second, the Italian mechanism operates as a one-stage matching process: when teachers rank a region, they apply sequentially to schools in that region according to the official order, resulting in assignments to specific schools in each step of the mechanism. \cite{dougan2024geography} leverage this exogenous order to achieve stability and other properties of their mechanism. In contrast, our approach employs a two-stage matching process, where students are admitted to bundles in the first stage and assigned to specific schools within bundles in the second stage. Our mechanism does not impose any tie-breaking rule to convert a bundle report into a strict ranking. 
To see how an exogenous tie-breaking rule might be undesirable in applications, consider the application discussed in \autoref{section:application}, where a city's government bundles three schools that students widely regard as nearly indifferent. If an exogenous order is used in that city, it would mean that all students ranking the bundle apply to one of the three schools before applying to the others. This would give that school an undue advantage over the others by letting it admit students with the highest scores, violating the fairness for schools the government wants to achieve. In practice, the government achieves this fairness through randomization in the second stage.

Finally, our discussion of a two-stage preference formation process in  \autoref{section:two-stage} connects to the growing body of literature highlighting the importance of costly information acquisition in market design. For instance, \cite{chen2021information,chen2022information} examine how DA and Immediate Acceptance (IA) mechanisms differ in their ability to incentivize students to acquire information and how different information acquisition costs affect the welfare outcomes of the two mechanisms. \hide{IA incentivizes more information acquisition due to its manipulable nature, while DA, being strategy-proof, reduces the incentive to acquire costly information. In laboratory experiments, while students' willingness to pay for information is
significantly greater under IA than under DA, most students overinvest in information acquisition under IA.} \cite{artemov2021assignment}  explores how common and idiosyncratic components of preferences shape information acquisition behavior under DA and IA, showing that as students acquire more information, their preferences become more heterogeneous, thereby improving their match quality. \cite{hakimov2023costly} demonstrate the advantage of dynamic mechanisms over static ones in reducing information acquisition costs. In dynamic mechanisms, students receive real-time feedback during the matching process, allowing them to avoid acquiring information about unattainable schools. For further related studies, see \cite{maxey2024school} and \cite{koh2022visit}.

\section{Standard School Choice System}\label{section:standard:system}

The standard school choice system involves a finite set of students $ I $ and a finite set of schools $ S $.  Each school $ s\in S $ has a quota $ q_s\in \mathbb{N} $ of seats and ranks students according to a strict priority order $ \succsim_s $. Each student seeks admission by one school. A matching is a mapping $ \mu: I\cup S\rightarrow I\cup S\cup \{\emptyset\} $ such that, for all $ i\in I $, $ \mu(i)\in S\cup \{\emptyset\} $, for all $ s\in S $, $ \mu(s)\subseteq I $ with $|\mu(s)|\le q_s $, and for all $i\in I$ and $s\in S$, $ \mu(i)=s $ if and only if $ i\in \mu(s) $. If $ \mu(i)=\emptyset $, it means that $ i $ is not admitted to any school.

The matching theory literature typically assumes that students both possess and are permitted to submit complete preference rankings over all schools. In reality, however, students are often constrained by a rank-order list (ROL) of limited length, which prevents them from reporting complete preferences. Moreover, from a policymaker’s perspective, students’ true preferences are not directly observable, and the operation of real-world school choice systems does not rely on the assumption that students have complete preference rankings. To accommodate this institutional feature, we consider a constrained school choice environment in which each student may rank up to $ \ell\in \mathbb{N} $ schools in their ROL, where $\ell <|S|$. We use $L_i$ to denote the ROL submitted by any student $i$, which is an order of no more than $\ell$ schools. With abuse of notation, we use $s\in L_i$ to mean that school $s$ is listed in $L_i$, and we use $s L_i s'$ to mean that $s$ is ranked above $s'$ in $L_i$.

In the standard school admission procedure, the policymaker first asks students to submit their ROLs and then runs an algorithm to generate a matching. In practical and legal contexts---such as when a student challenges a matching outcome---the challenge must be based on students' reported preferences rather than their unobservable preferences. Therefore, we define the stability of a matching with respect to students' submitted ROLs. Given a ROL profile $L_I=(L_i)_{i\in I}$, a matching $\mu$ is \textbf{stable} if it satisfies (1) \textit{individual rationality}: for every $i\in I$, if $\mu(i)\neq \emptyset$, then $\mu(i)\in L_i$; (2) \textit{non-wastefulness}: for every $i\in I$ and $s\in S$, if $s L_i \mu(i)$, then $|\mu(s)|=q_s$; (3) \textit{no justified envy}: there do not exist two distinct students $i$ and $j$ such that $\mu(j) L_i \mu(i)$ and $i \succ_{\mu(j)} j$.

The DA algorithm is used to find a student-optimal stable matching for given ROLs. We present the procedure of DA in the following.

\begin{center}
	\textbf{Standard Deferred Acceptance}
\end{center}

\begin{itemize}
	\item \textit{Round 1}: Each student applies to the first school in her ROL. Each school considers all applicants it receives and tentatively admits them one-by-one according to its priority order, up to its capacity. If a school receives more applicants than its capacity, the lowest-priority applicants are rejected. If every student is either tentatively admitted to some school or has been rejected by all schools in her ROL, stop.

	\item \textit{Round $r \ge 2$}: Each student rejected in the previous round applies to the next school in her ROL, if any. Each school considers its current set of tentatively admitted students together with any new applicants, and tentatively admits the highest-priority students among them up to its capacity; all others are rejected. If every student is either tentatively admitted to some school or has been rejected by all schools in her ROL, stop.
\end{itemize}

In the above analysis, we assume that schools use strict priority orders, which are essential for the operation of the DA algorithm. To conclude this section, we briefly discuss how such strict priorities arise in practice, as this helps clarify the applicability of our bundle design in real-world admission systems. In systems that rely on standardized tests, students are naturally strictly ranked according to their test scores. In other systems, schools often assign coarse priority groups based on criteria such as residential zones, sibling attendance, or affirmative action considerations. Because these criteria typically create ties, policymakers employ tie-breaking procedures to refine coarse priorities into strict priority orders. A commonly used procedure is a single-lottery rule: each student is assigned a single random number at the start of the admissions process, and whenever two students fall into the same priority group for a school, their lottery numbers determine their relative ordering. For example, the New York City public school admissions system uses such a single-lottery tie-breaking rule.

\section{Bundled School Choice System}\label{section:bundle:system}

This section defines bundle systems and relevant concepts.  \autoref{section:bundle:definition} presents a general framework of bundle systems.  \autoref{section:bundle-matching} and  \autoref{section:stability} modify the concepts of matching and stability respectively to accommodate bundles.

\subsection{General Bundle System}\label{section:bundle:definition}

A bundle system exogenously groups certain schools together and offers them as additional options for students' preference submission. In a ROL slot, a student can list either an individual school or a school bundle. Unlike the standard system in which every school is available to all students, we require that each bundle be available only to students who are ranked in the same way by all schools in that bundle. This priority uniformity condition ensures that the bundle can apply a common admission criterion shared by all schools in the bundle.

\begin{definition}\label{Defn:bundle}
    A \textbf{bundle} $ b=(S_b,I_b) $ specifies a nonempty subset of schools $ S_b $ and a nonempty subset of students $ I_b$ such that, for every distinct $ s,s'\in S_b $ and every distinct $i,i'\in I_b$, $ i \succ_s i'$ if and only if $i \succ_{s'} i'$. Only students in $ I_b $ are eligible to list $ S_b $ in their ROLs.
\end{definition}

Under this definition, $I_b$ could be any set of students who are ranked in the same order by all schools within $S_b$. Importantly, we do not require that $I_b$ must include \textit{all} such students. This provides a degree of flexibility for the policymaker to select $I_b$ as deemed appropriate. 

For each bundle $ b=(S_b,I_b) $, we say that $ S_b $ is \textbf{available} to $ I_b $ and $ I_b $ is \textbf{targeted} by $ S_b $. We call $ q_{b}\equiv \sum_{s\in S_b} q_s $ the \textbf{quota} of $ b $, which is the total number of seats in $b$. When there is no confusion, we also call $ S_b $ a bundle.

A bundle system consists of finitely many bundles that satisfy the following conditions. These conditions prevent conflicts among bundles and ensure that our algorithm accommodating bundles operates correctly.

\begin{definition}\label{Defn:bundle:system}
    A \textbf{bundle system} $ \mathcal{B} $ consists of finitely many bundles that satisfy:
\begin{itemize}
	\item[(1)]  (Hierarchy) For every distinct $ b, b'\in \cB $, if $ S_{b}\cap S_{b'}\neq\emptyset $, then $ S_{b}\subsetneq S_{b'} $ or $ S_{b'}\subsetneq S_{b} $;
	
	\item[(2)] (Monotonicity in target) For every distinct $ b, b' \in \cB$, if $ S_{b}\subsetneq S_{b'} $, then $I_b\supseteq I_{b'}$; 
	
	\item[(3)] (Individual schools remain available) For every $ s\in S $, $ (\{s\},I)\in \cB $.
\end{itemize}
\end{definition}

Condition (1) requires the bundle system to be hierarchical: whenever two bundles overlap, one must be a superset of the other. Allowing nontrivial overlaps would create significant complications. In any matching algorithm, when a student is tentatively admitted to a bundle $S_b$, her final assigned school within that bundle remains unresolved until the algorithm ends. If bundles could overlap nontrivially, it would be unclear whether the student should be treated as occupying a seat in another overlapping bundle, and thus how to update the remaining quotas of those bundles during the algorithm becomes an issue. Under a hierarchical structure, however, if a student is admitted to a bundle, she must also occupy a seat in any larger bundle containing it. Therefore, we reduce the quota of that bundle and the quota of any larger bundle by one.\footnote{However, because the student’s specific school assignment within the bundle is not yet determined, we do not reduce the quotas of any smaller bundles at that step. The next section describes our algorithm in detail.} Note that this issue does not arise in the standard school choice setting, as individual schools are disjoint by definition.

Condition (1) implicitly requires that there do not exist two bundles $b$ and $b'$ such that $ S_b=S_{b'} $ but $ I_b\neq I_{b'} $. If such bundles were to exist, they could simply be replaced by a bundle $ (S_b,I_b\cup I_{b'}) $.

When $ S_{b}\subsetneq S_{b'} $,  we call $ b $ a \textbf{sub-bundle} of $ b' $ and $ b' $ a \textbf{sup-bundle} of $ b $. We call a bundle $ b $ \textbf{maximal} in $ \cB $ if $ \cB $ does not contain a sup-bundle of $ b $. If a bundle system $ \cB $ contains more than one maximal bundle, then $ \cB $ can be partitioned into multiple distinct sub-hierarchies, each containing a unique maximal bundle, which is a superset of all other bundles in the sub-hierarchy.

Condition (2) mandates that a larger bundle must target a weakly smaller set of students. This condition is natural because the priority uniformity condition for targeted students becomes more stringent for larger bundles. In our algorithm, at any step, when some students apply to a bundle $b$ while others apply to a sup-bundle $b'$, this condition ensures that the schools in $S_b$ rank the union of the applicants in the two sets in the same order. % As a result, ``the highest-priority applicant'' for each bundle is well-defined. 

Condition (3) requires that every individual school, considered as a trivial bundle, remain available to all students. While this condition is not necessary for the operation of our designed system, we impose it to ensure that our design expands each student's choice set compared to the standard system. As discussed in  \autoref{section:application}, this condition may play a crucial role in the success of a bundle system in real-world applications.

\autoref{example:general:bundle} provides an illustration of a bundle system.

\begin{example}\label{example:general:bundle}
    Consider seven schools $ \{s_1,s_2,\ldots,s_7\} $ and eight students $ \{i_1,i_2,\ldots,i_8\} $. Each school has one seat and they use the following priority orders.
	
	\begin{table}[!ht]
		\centering
		\begin{subtable}{.5\linewidth}
			\raggedright
			\small
            \caption{Priority orders}
			\begin{tabular}{cccc|ccc}
				$ \succsim_{s_1} $ & $ \succsim_{s_2} $ & $ \succsim_{s_3} $ & $ \succsim_{s_4} $ & $ \succsim_{s_5} $  & $ \succsim_{s_6} $ & $ \succsim_{s_7} $\\ \hline
				$ i_1 $ & $ i_1 $ & $ i_1 $ & $ i_1 $ & $ i_6 $ & $ i_6 $ & $ i_6 $ \\
				$ i_2 $ & $ i_2 $ & $ i_2 $ & $ i_2 $ & $ i_7 $ & $ i_7 $ & $ i_7 $\\
				$ i_3 $ & $ i_3 $	& $ i_3 $ & $ i_3 $ & $ i_5 $ & $ i_5 $ & $ i_5 $ \\
				$i_8$ & $ i_8 $ & $ i_8 $ & $ i_8 $ & $ i_4 $ & $ i_4 $ & $ i_4 $ \\
				$i_4$ & $ i_4 $ & $ i_4 $ & $ i_4 $ & $ i_8 $ & $ i_8 $ & $ i_8 $ \\
				$i_5$ & $ i_5 $ & $ i_5 $ & $ i_5 $ & $ i_1 $ & $ i_1 $ & $ i_1 $ \\
				$i_6$ & $ i_6 $ & $ i_6 $ & $ i_6 $ & $ i_2 $ & $ i_2 $ & $ i_2 $\\
				$i_7$ & $ i_7 $ & $ i_7 $ & $ i_7 $ & $ i_3 $ & $ i_3 $ & $ i_3 $
			\end{tabular}
		\end{subtable}
		\quad
		\begin{subtable}{.35\linewidth}
			\centering
            \caption{Bundle system}
			\begin{tabular}[c]{ccccccc}%
			\multicolumn{7}{c}{\begin{tabular}{c}
						$\{s_1,s_2,s_3,s_4,s_5, s_6, s_7\} $
			\end{tabular}} \\
				\multicolumn{7}{c}{\begin{tabular}{cc}
							$\{s_1,s_2,s_3,s_4\}$ & $ \{s_5, s_6, s_7\} $
				\end{tabular}} \\
				\multicolumn{7}{c}{ \begin{tabular}{ccc}
							$\{s_1,s_2\}$ & $\{s_3,s_4\}$ & $ \{s_5 , s_6 \}$
				\end{tabular}}  \\
				$s_1$ & $s_2$ & $s_3$ & $s_4$ & $ s_5 $ & $ s_6 $ & $ s_7 $ 
			\end{tabular}
		\end{subtable}
		
		%\caption{Example \ref{example:general:bundle}}\label{table:example:general:bundle}
	\end{table}
	
	Consider a bundle system $\cB$ consisting of individual schools and following bundles: $\{s_1,s_2\}$, $\{s_3,s_4\}$, $\{s_5,s_6\}$, $\{s_1,s_2,s_3,s_4\}$, $\{s_5,s_6,s_7\}$, and $\{s_1,s_2,\ldots,s_7\}$. The bundle $ \{s_1,s_2,\ldots,s_7\} $ is available to $ \{i_1,i_2,i_3\} $. The other bundles are available to all students.
\end{example}

\autoref{remark:application:scope} discusses the environments where our design is applicable.

\begin{remark}\label{remark:application:scope}
    Our design relies on a certain degree of homogeneity in school priorities. In test-based admission systems, where students are strictly ranked according to their test scores, it is natural for schools to adopt identical priority rankings.\footnote{If schools apply different weights to aggregate students' test scores across multiple subjects, schools using the same weighting scheme will employ identical priority orders.} In admission systems where students are ranked coarsely based on their types and a single lottery is used to break priority ties, schools that apply the same criteria for ranking students will ultimately use identical priority orders for students of the same type, even though the specific priority order is not predetermined prior to the lottery. Our design is applicable to both types of admission systems.
\end{remark}

Given a bundle system $ \mathcal{B} $, let $\mathcal{B}_i $ denote the set of bundles (including all individual schools) available to any student $i$. Thus, $\mathcal{B}_i $ is the menu for $i$. Then, $ i $ can rank up to $ \ell $ bundles from $ \mathcal{B}_i$ in her ROL. Let $ P_i $ denote any ROL submitted by $ i $, and let $ \mathcal{B}[P_i] $ denote the set of bundles listed in $ P_i $. For any $ b,b'\in \mathcal{B}[P_i] $, let $b P_i b'$ mean that $i$ ranks $b$ above $b'$. Let $ P_I\equiv \{P_i\}_{i\in I} $ and $ \mathcal{B}[P_I]\equiv \cup_{i\in I} \mathcal{B}[P_i] $. 

After collecting students' ROLs, the admission process we design consists of two stages. In the first stage, students are assigned to reported bundles according to their submitted rankings. This generates a matching between students and bundles. In the second stage, students admitted by bundles are further assigned to the schools within those bundles. This finalizes the matching outcome. We design an algorithm to generate a matching for the first stage, and leave the second stage to the discretion of policymakers to determine the matching outcome. In the remaining subsections, we introduce key concepts to facilitate discussion of this process.

Before proceeding to the next subsection, we remark that our definition of bundles is intentionally general to accommodate a wide range of environments; however, this flexibility comes at the cost of a potentially complex structure of the bundle system. To address this, we introduce a subclass of bundle systems with simpler structures that could be easily comprehensible for participants and well-suited for practical implementation. This subclass naturally arises in admission systems where policymakers only bundle schools that share common features and apply the same admission criteria to all students. For instance, in test-based admission systems where students take standardized tests across multiple subjects, policymakers may choose to bundle schools that emphasize the same subjects and use a common weighting scheme to aggregate scores. In priority-based systems that incorporate geographical considerations, schools within the same region, which categorize students based on equal criteria and use a common lottery to resolve priority ties, may be bundled together for administrative simplicity and policy coherence.

\begin{definition}
	A bundle system $ \cB $ is called \textbf{simple} if, for every bundle $ b\in \cB $ and every two distinct schools $ s,s'\in S_b $, $ \succsim_s=\succsim_{s'} $.
\end{definition}

To illustrate this definition, in \autoref{example:general:bundle}, the bundle system $\cB$ is not simple, but if we remove the bundle $ \{s_1,s_2,\ldots,s_7\} $, then the remaining bundles constitute a simple bundle system.

In a simple bundle system, schools within each bundle rank \textit{all} students in the same order. So, it is technically feasible for each bundle to target all students. However, to maintain generality, we still allow each bundle to target a subset of students.

A convenient way to understand a simple bundle system $\cB$ is to partition bundles into disjoint sub-hierarchies $ \cB_1,\cB_2,\ldots,\cB_K $ such that each $ \cB_k$ contains a unique bundle $S_k$ which is maximal in $\cB$, and all other bundles in $\cB_k$ are sub-bundles of $S_k$. Consequently, we obtain a partition of schools represented by $ S_1,S_2,\ldots,S_K $. In each $ S_k $, all schools use an identical priority order to rank all students. For convenience, we denote this order by $ \succsim_k $.

\subsection{Bundle-matching and implementation}\label{section:bundle-matching}

We first distinguish between matching where students are assigned to bundles and matching where students are assigned to individual schools. We call the former \textbf{bundle-matching} and the latter \textbf{standard matching} (which has been defined in  \autoref{section:standard:system}).

\begin{definition}\label{Defn:bundle:matching}
    Given a bundle system $\cB$, a \textbf{bundle-matching} is a mapping $ \nu: I\cup \cB \rightarrow I\cup \mathcal{B} \cup \{\emptyset\}$ such that, for all $ i\in I $, $ \nu(i)\in \cB_i \cup \{\emptyset\} $,  for all $ b\in \cB $, $ \nu(b)\subseteq I $  and $ \sum_{b'\in\cB:S_{b'}\subseteq S_b} |\nu(b')|\le q_{b} $, and for all $ i\in I $ and all $ b\in \cB $, $ \nu(i)= b $ if and only if $ i\in \nu(b) $. 
\end{definition}

In a bundle-matching, every student is assigned to at most one bundle available to her, and no bundle can admit more students than its quota. Notably, unlike a standard matching where students are directly assigned to individual schools, in a bundle-matching $ \nu $, students who may occupy seats in a bundle $ b $ include not only those explicitly assigned to $b$ but also those assigned to other bundles that overlap with $b$. In particular, the students who are assigned to sub-bundles of $b$ must occupy seats within $b$. Thus, the number of students who must occupy seats within $b$ is $\sum_{b'\in\cB:S_{b'}\subseteq S_b} |\nu(b')|$.  \autoref{Defn:bundle:matching} requires that $\sum_{b'\in\cB:S_{b'}\subseteq S_b} |\nu(b')|\le q_b$. It turns out that this requirement is sufficient for a bundle-matching to be well-defined. For convenience, we define $$ Q^\nu_{b}\equiv \sum_{b'\in\cB:S_{b'}\subseteq S_b} |\nu(b')|.$$

Given a bundle-matching $ \nu $, we say that a standard matching $ \mu $ \textbf{implements} $ \nu $ if every student who is assigned to a bundle is finally assigned to a school in the bundle. 

\begin{definition}
    Given a bundle system $\cB$, a standard matching $\mu$ implements a bundle-matching $\nu$ if, for every $ i\in I $, $ \nu(i)\in \mathcal{B}_i \implies \mu(i)\in S_{\nu(i)} $, and $ \nu(i)=\emptyset \implies \mu(i)=\emptyset $. A bundle-matching $ \nu $ is \textbf{implementable} if it can be implemented by a standard matching. 
\end{definition} 

In general, there exist multiple potential methods for implementing a bundle-matching, but all such implementations must adhere to the following procedure.

\begin{center}
	\textbf{A general procedure to implement a bundle-matching}
\end{center}

\begin{itemize}
	\item[] \noindent Step 0: Unmatched students are not assigned to any school;

	\item[] Step 1: Students admitted by individual schools are directly assigned to those schools;

 \item[] Step 2: Students admitted by every bundle consisting of two schools are assigned to the remaining seats in the bundle;

 $\vdots$

 \item[] Step $ n$: After all students admitted by smaller bundles are assigned, students admitted by every bundle consisting of $n$ schools are assigned to the remaining seats in the bundle.
\end{itemize}

At any step of the above procedure, if a bundle has remaining seats that belong to different schools, there are multiple ways to assign admitted students to these available seats. So, a bundle-matching can be implemented by multiple standard matchings.

\begin{lemma}\label{lemma:implementation}
	Every bundle-matching $\nu$ is implementable. Moreover, a standard matching $\mu$ implements $\nu$ if and only if $\mu$ can be generated by the above procedure.
\end{lemma}

The proof of \autoref{lemma:implementation} is presented in the Appendix \ref{appendix:proofs:lemma}.

\subsection{Stability}\label{section:stability}

\autoref{section:standard:system} has defined stability of a standard matching in the standard system where students rank individual schools in their ROLs. This subsection defines stability for the bundle system. The basic idea is to impose stringent conditions to ensure that once a bundle-matching is determined to be stable, any of its implementations must also be stable.  

Given a bundle system $\cB$ and a ROL profile $ P_I\equiv \{P_i\}_{i\in I} $, we view $ P_i $ as student $i$'s preferences over bundles available to her. Then, a bundle-matching $ \nu $ is \textbf{individually rational} if, for all $i\in I$ such that $\nu(i)\in \cB_i$, $\nu(i)\in \cB[P_i]$; that is, students can be assigned only to bundles they report. 

To define non-wastefulness for a bundle-matching $ \nu $, we must determine when we can say that a bundle $ b $ has vacant seats in $ \nu $. Recall that $ Q^\nu_{b}$ denotes the number of students who must occupy seats within $b$. If $ Q^\nu_{b}=q_{b} $, then all seats within $ b $ must be fully assigned to students. However, even if $ Q^\nu_{b}<q_{b} $, as long as there exists a sup-bundle $ b' $ such that $ Q^\nu_{b'}=q_{b'} $, all seats of $ b $ must also be fully assigned to students. Therefore, the seats of $ b $ must be fully assigned to students if and only if there exists a bundle $ b' $ such that $ S_{b'}\supseteq S_b $ and $ Q^\nu_{b'}=q_{b'} $. So, we call a bundle-matching $ \nu $ \textbf{non-wasteful} if, for any student $ i $ and any bundle $ b\in \cB[P_i] $ such that $ b P_i \nu(i) $, there exists some bundle $ b' $ such that $ S_{b'}\supseteq S_b $ and $ Q^\nu_{b'}=q_{b'} $. This definition is stringent, since once $b$ could potentially have vacant seats in any implementation of $\nu$, we regard $\nu$ as wasteful.

Finally, to define no justified envy for a bundle-matching, we need to carefully handle cases where a student may envy another student. Specifically, for any student $ i $, if there exists a bundle $ b \in \cB[P_i]$ such that $b P_i \nu(i) $, then we say that $ i $ has justified envy towards another student $ j $ if $ j $ could potentially be assigned to a school in $ S_b $ but has lower priority than $ i $ for that school. The students who could potentially be assigned to a school in $ S_b $ are those admitted by any bundle $ b' $ such that $ S_b\cap S_{b'} \neq \emptyset$. Then, we need to consider three cases. As long as one of the three cases holds, we say that $i$ has \textbf{justified envy} towards $j$ in a bundle-matching $\nu$:
\begin{itemize}
	\item[(1)] $ \nu(j)=b $, and $ i $ has higher priority than $ j $ for the schools in $ S_b $;

	\item[(2)] $ S_{\nu(j)}\subsetneq S_b $, and $ i $ has higher priority than $ j $ for the schools in $ S_{\nu(j)} $;
	
	\item[(3)] $ S_{\nu(j)}\supsetneq S_b $, $ Q^\nu_{b'}<q_{b'} $ for every bundle $ b' $ such that $ S_b\subseteq S_{b'} \subsetneq S_{\nu(j)} $, and $ i $ has higher priority than $ j $ for the schools in $ S_b $.
\end{itemize}

The \textit{first} case is a straightforward extension of justified envy in the standard system. In the \textit{second} case, $ j $ is admitted by a sub-bundle of $ b $ and thus must be finally assigned to a school in that bundle. Then, we say that $ i $ has justified envy towards $ j $ if $ i $ has higher priority than $ j $ for every school in that bundle. In the \textit{third} case, $ j $ is admitted by a sup-bundle of $b $ and thus must be finally assigned to a school in that bundle. But if there exists a bundle $ b' $ such that $ S_b\subseteq S_{b'} \subsetneq S_{\nu(j)} $ and $ Q^\nu_{b'}=q_{b'} $, then all seats of $b'$ (including seats of $b$) must be assigned to the students who are admitted by $b'$ or by its sub-bundles. So, $j$ cannot be assigned to any school in $S_b$. Otherwise, $ j $ could potentially be assigned to a school in $ S_b $. Then, we say that $ i $ has justified envy towards $ j $ if $ i $ has higher priority than $ j $ for every school in $S_b $. Note here we also adopt a stringent definition, ensuring that once $ i $ could potentially have justified envy towards $ j $ in any implementation of $\nu$, we say that $i$ has justified envy towards $j$ in $\nu$. A bundle-matching is said to satisfy \textbf{no justified envy} if no student has justified envy towards any other.

\begin{definition}\label{defn:bundle:stability}
 Given a bundle system and a ROL profile, a bundle-matching is \textbf{stable} if it is individually rational, non-wasteful, and satisfies no justified envy.
\end{definition}

We now define stability for a standard matching in the bundle system.
When a student lists a bundle in her ROL, she is expressing a desire for admission to any school in the bundle, without specifying preferences over those schools. From the policymaker's perspective, this indicates that the student is indifferent between the schools in the bundle, thereby granting the policymaker the flexibility in assigning the student. Therefore, to define stability for a standard matching, we interpret every student's submitted rankings of bundles as a preference relation over schools that contains indifference. That is, all schools in every reported bundle are viewed as indifferent to the student, but if some schools are included in different bundles, we compare schools based on their first occurrence in the ROL. 

Formally, for every ROL $ P_i $, we define a preference relation $ \succsim_{P_i} $ that regards only the schools in $ \cup_{b\in \cB[P_i]} S_b $ as acceptable such that, for any two schools $ s,s'\in \cup_{b\in \cB[P_i]} S_b $, if the first occurrence of $s$ is before the first occurrence of $s'$, then $ s\succ_{P_i} s'$; if their first occurrences are in the same bundle, then $ s \sim_{P_i} s' $.
 
\begin{definition}
    Given any ROL profile $ P_I\equiv \{P_i\}_{i\in I} $, a standard matching $ \mu $ is \textbf{stable} if it satisfies (1) individual rationality: for all $ i\in I $, $ \mu(i)\in  \cup_{b\in \cB[P_i]} S_b\cup \{\emptyset\}$; (2) non-wastefulness: there do not exist $ i\in I $ and $ s\in S $ such that $ |\mu(s)|<q_s $ and $ s\succ_{P_i} \mu(i) $; (3) no justified envy: there do not exist distinct $ i,j\in I $ such that $ \mu(j)\succ_{P_i} \mu(i) $ and $ i\succ_{\mu(j)} j $.
  
\end{definition}

\autoref{lemma:implementation:stable} ensures that once a stable bundle-matching is found in the first stage,  any of its implementations in the second stage must be stable. The proof is presented in the Appendix \ref{appendix:proofs:lemma}. %As we will prove in the next section, our modified DA produces a stable bundle-matching for any ROL profile.

\begin{lemma}\label{lemma:implementation:stable}
Given any ROL profile $ P_I\equiv \{P_i\}_{i\in I} $, if a bundle-matching is stable, its every implementation is a stable standard matching. 
\end{lemma}

\begin{comment}
    
\begin{definition}
 Given a ROL profile $ P_I\equiv \{P_i\}_{i\in I} $, a bundle-matching $ \nu $ is \textbf{size-maximally stable} if it is stable and there does not exist another stable bundle-matching $\nu'$ such that
 \[
 \{i\in I: \nu(i)\in \cB[P_i]\}\subsetneq  \{i\in I: \nu'(i)\in \cB[P_i]\}.
 \]
\end{definition}
\end{comment}

A key objective of introducing bundles is to enable students to report more schools in their constrained ROLs, thereby increasing their matching likelihood. Therefore, an important criterion for evaluating a bundle-matching is the number of matched students. So, we define the following concept that identifies a bundle-matching generating a largest possible set of matched students.

\begin{definition}\label{def:SM}
 Given a ROL profile $ P_I\equiv \{P_i\}_{i\in I} $, a bundle-matching $ \nu $ is \textbf{size-maximal} if it is individually rational and there does not exist another bundle-matching $\nu'$ such that
 \[
 \{i\in I: \nu(i)\in \cB[P_i]\}\subsetneq  \{i\in I: \nu'(i)\in \cB[P_i]\}.
 \]
\end{definition}

This concept compares the set of matched students in the two bundle-matchings, without referring to student welfare. The following definition is weaker, as it requires that the other bundle-matching $\nu'$ must generate more matched students and ``Pareto dominate'' the benchmark $\nu$.

\begin{definition}\label{def:PUSM}
 Given a ROL profile $ P_I\equiv \{P_i\}_{i\in I} $, a bundle-matching $ \nu $ is \textbf{Pareto-undominated size-maximal} if it is individually rational and there does not exist another bundle-matching $\nu'$ such that $\{i\in I: \nu(i)\in \cB[P_i]\}\subsetneq  \{i\in I: \nu'(i)\in \cB[P_i]\}$ and for all $i\in I$ with $\nu(i)\in \cB[P_i]$, either $\nu'(i) P_i \nu(i)$ or $\nu'(i)=\nu(i)$.
\end{definition}

In the following section, we will prove that despite the outcome of our algorithm may not be size-maximal, it must be Pareto-undominated size-maximal.

\begin{remark}\label{remark:sizemaximal}
Stable bundle-matchings may never generate the maximum number of matched students, even in the standard setting without any bundled schools. For instance, consider two students $i$ and $i'$ and two schools $s$ and $s'$, each with one seat. Both schools rank $i$ above $i'$. Suppose that $i$ ranks $s$ above $s'$, yet $i'$ lists only $s$ in her ROL. The only stable matching assigns $i$ to $s$ and leaves $i'$ unmatched. In contrast, the matching that assigns $i$ to $s'$ and $i'$ to $s$ matches both students, but it is not stable.
\end{remark}

\section{Deferred Acceptance in Bundle System}\label{section:general:DA}

We modify DA to find a stable bundle-matching for any ROL profile submitted by students in a given bundle system. Our approach seeks to retain as many of the original features of the standard DA as possible. However, the incorporation of bundles introduces additional complexity, necessitating significant adjustments to the original algorithm. To differentiate it from the standard DA, we call our algorithm \textbf{bundle-DA}. 

Similar to the standard DA, in every round of bundle-DA, each unmatched student applies to the highest-ranked bundle in her ROL that has not rejected her. Each bundle then tentatively accepts the highest-priority applicants up to its capacity. However, given that bundles may overlap, two challenges arise in defining the bundle-DA procedure.

The \textbf{first} challenge involves the definition of ``the highest-priority applicants'' for a bundle in each round of bundle-DA. We note that, by applying to a bundle, a student seeks admission to any school in the bundle. Thus, we regard the student as an applicant to every school in the bundle. To illustrate the challenge and explain how we solve it, consider a bundle $b$ with two disjoint sub-bundles, $b'$ and $b''$, each with one seat, resulting in $b$ having a total of two seats. Suppose in a particular round, a student $i$ applies to $b$, while another student $i'$ applies to $b'$, and a third student $i''$ applies to $b''$, as illustrated by \autoref{fig:illustration}. We regard $i$ as an applicant to every school in $S_b$, $i'$ as an applicant to every school in $S_{b'}$, and $i''$ as an applicant to every school in  $S_{b''}$. Then, there are three cases: 

\begin{figure}[!ht]
			\centering
            \caption{An illustration of how to define the highest-priority student}\label{fig:illustration}
			\begin{tikzpicture}[bend angle=20,xscale=0.8,yscale=.8]
				%\draw[help lines](0,0) grid (8,8);
				\node[black] at (0,.3) {\footnotesize $ b $};
				\node[blue] at (-1,0) {\footnotesize $ b' $};
				\node[red] at (1,0) {\footnotesize $ b'' $};
				\node[black] at (0,1.2) {\footnotesize $ i $};
				\node[blue] at (-1,1.2) {\footnotesize $ i' $};
				\node[red] at (1,1.2) {\footnotesize $ i'' $};
				
				\draw[black,thick] (0,0) ellipse (2cm and .7cm);
				\draw[blue,thick] (-1,0) ellipse (.7cm and .4cm);
				\draw[red,thick] (1,0) ellipse (.7cm and .4cm);
				
				\draw[black,thick,->] (0,1) -- (0,.5);
				\draw[blue,thick,->] (-1,1) -- (-1,.2);
				\draw[red,thick,->] (1,1) -- (1,.2);
			\end{tikzpicture}
		\end{figure}

\begin{itemize}
	\item[(1)] If $ i $ has higher priority than $ i' $ for $ S_{b'} $ and also higher priority than $ i'' $ for $ S_{b''} $, then we regard $ i $ as the highest-priority applicant for $ b $ in that round, but we do not regard $i'$ and $i''$ respectively as the highest-priority applicant for $b'$ and $b''$ in that round.
	
	\item[(2)] If $ i $ has higher priority than $ i' $ for $ S_{b'} $ but lower priority than $ i'' $ for $ S_{b''} $, then we regard $ i'' $ as the highest-priority applicant for $ S_{b''} $, but we do not regard $ i $ and $ i' $ respectively as the highest-priority applicant for $ b $ and $ b' $. The symmetric situation, where $ i $ has lower priority than $ i' $ for $ S_{b'} $ but higher priority than $ i'' $ for $ S_{b''} $, is similar.
	
	%Symmetrically, if $ i $ has higher priority than $ i'' $ for $ S_{b''} $ but lower priority than $ i' $ for $ S_{b'} $, then we regard $ i' $ as the highest-priority applicant for $ b' $, but we do not regard $ i $ and $ i'' $ respectively as the highest-priority applicant for $ b $ and $ b'' $. 

	\item[(3)] If $ i $ has lower priority than $ i' $ for $ S_{b'} $ and also lower priority than $ i'' $ for $ S_{b''} $, then we regard $ i' $ and $ i'' $ respectively as the highest-priority applicant for $ b' $ and $ b'' $, but we do not regard $ i $ as the highest-priority applicant for $ b $.
\end{itemize}

When analyzing the three cases, we implicitly use Condition (2) in \autoref{Defn:bundle:system}. Since $i$ is eligible to report $b$, she is also eligible to report $b'$ and $b''$. Then, the schools in $S_{b'}$ must rank $i$ and $i'$ in the same order; the schools in $S_{b''}$ must rank $i$ and $i''$ in the same order.	

In general, in each round of bundle-DA, a student is considered an applicant for a school if she applies to any bundle containing that school. A student is considered the highest-priority applicant for a bundle \textbf{if and only if} she applies to the bundle and, for every school in the bundle, she has the highest priority among all students considered applicants for the school. In each round, we recursively identify the highest-priority applicant for each bundle and tentatively admit them, until all seats in each bundle are filled or all applicants have been tentatively admitted.

The \textbf{second} challenge pertains to updating the quotas of bundles after they have tentatively admitted students. For instance, if a bundle $b$ tentatively admits a student $i$ in a round, it follows that the remaining quota for $b$ should be reduced by one. Moreover, the remaining quotas of all sup-bundles of $ b $ should also be reduced by one, because $ i $ occupies one seat in these bundles. However, we cannot reduce the remaining quota of any sub-bundle of $ b $ at this step, because $ i $ has not yet been assigned to any specific school in $ S_b $. This discrepancy leads to a complication: in subsequent rounds, the sum of the remaining quotas of sub-bundles of $ b $ may exceed the remaining quota of $ b $. To address this, we pay special attention to two scenarios in subsequent rounds.

First, if the remaining quota of $ b $ becomes zero in a round, it means that all seats of $ b $ have been filled. In that case, we set the remaining quotas of all sub-bundles of $ b $ to zero immediately. 

Second, if the sub-bundles of $b$ are supposed to admit applicants based on their remaining quotas in a round, but the remaining quota of $b$ is insufficient to admit all such applicants, we need a tie-breaking rule to allocate the remaining seats of $b$. Consider again case (1) of the illustrative situation depicted in \autoref{fig:illustration}. After $ i $ is admitted by bundle $ b $, $ b $ has only one remaining seat. However, both $b'$ and $b''$ still have a quota of one and are supposed to admit $i'$ and $i''$, respectively. To resolve this conflict, we will need a tie-breaking rule between $ i' $ and $ i'' $. 
In a general bundle system, there is no natural way to break such ties. Therefore, we introduce an exogenous order of students to resolve them (but policymakers may also use other rules).\footnote{Note that these rules are not used to break preference ties within a bundle and hence differ from conventional tie-breaking rules in the literature.}

However, in a simple bundle system, all schools within each sub-hierarchy rank all students in the same order. This common ranking provides a natural basis for breaking ties. As a result, in simple bundle systems, the procedure of bundle-DA can be significantly simplified and possesses desirable properties. In contrast, the mechanism for general bundle systems is intricate and may lack some of these properties. Therefore, we focus on simple bundle systems in the remainder of this section and relegate the definition of general bundle-DA to the Appendix \ref{appendixbundle-DA:general}.

\subsection{Definition of Bundle-DA in Simple Bundle System}

Given a simple bundle system $\cB$, in each round, bundle-DA processes students' applications at the sub-hierarchy level. That is, within each sub-hierarchy $ \cB_k$, applicants are processed one by one according to the priority order $\succsim_k$, which is used by all schools in $ \cB_k$ to rank students. 

\begin{center}
	\textbf{Bundle-DA in Simple Bundle System}
\end{center} 

\begin{itemize}
	\item Round $ r\ge 1 $: Each unmatched student applies to the highest-ranked bundle in her ROL that has not rejected her. If such a bundle does not exist, the student is determined to be unmatched. 
 
 For each sub-hierarchy $\cB_k$, let $A^r_k$ represent the set of students consisting of those who were tentatively admitted to a bundle in $\mathcal{B}_k$ in the previous round and those who are new applicants to a bundle in $\mathcal{B}_k$ in this round. We then deal with the students in $A^r_k$ sequentially according to the priority order $\succsim_k$. In this process, once a student is considered, we check whether the bundle she applies to has a remaining quota. If that bundle has a remaining quota, admit the student to that bundle. After that,  the quota of that bundle and the quota of its every sup-bundle are reduced by one. In this process, once the quota of any bundle becomes zero, the quotas of its sub-bundles are immediately set to zero. At the end of this process, students who are not admitted are rejected in this round. We then proceed to the next round. 
 
 Stop the algorithm when every student is either admitted to some bundle or has been rejected by all bundles reported in her ROL.
\end{itemize}

We present an example to illustrate the above procedure.

\begin{example}\label{example:illustration:simple:bundle-DA}
Consider \autoref{example:general:bundle}. The bundle system $\cB$ is not simple because the schools in the bundle $ \{s_1,s_2,\ldots,s_7\} $ do not use an identical priority order. But if we remove that bundle, the remaining bundles constitute a simple bundle system. It includes two sub-hierarchies, $\cB_1$ and $\cB_2$, where $\cB_1$ includes three bundles, $\{s_1,s_2,s_3,s_4\}$, $\{s_1,s_2\}$, and $\{s_3,s_4\}$, and the individual schools in these bundles, while $\cB_2$ includes two bundles, $\{s_5,s_6,s_7\}$ and $\{s_5,s_6\}$, and the individual schools in these bundles. All these bundles are available to all students.

Suppose that the ROL length is two and students report the following ROLs.
	
	\begin{table}[!ht]
		\centering
		\begin{tabular}{cccccccc}
			$ P_{i_1} $ & $ P_{i_2} $ &  $ P_{i_3} $ & $ P_{i_4} $ & $ P_{i_5} $ & $ P_{i_6} $ & $ P_{i_7} $ & $ P_{i_8} $ \\ \hline
			$ s_1 $ & $ \{s_1,s_2,s_3,s_4\} $ & $ s_3 $  & $ \{s_1,s_2\} $  & $ \{s_3,s_4\} $ & $ \{s_5,s_6,s_7\} $ & $ \{s_5,s_6\} $ & $ s_5 $  \\
			$ s_2 $ & $ \{s_5,s_6\} $ & $ \{s_1,s_2\} $  & $ s_5 $ & $ s_5 $ & $ \{s_3,s_4\} $ & $ s_1 $ & $ \{s_1,s_2,s_3,s_4\} $			
		\end{tabular}
  %\caption{Students' submitted ROL}\label{table:example:ROL}
	\end{table}

 Then, bundle-DA runs as follows.	 
	
	\begin{itemize}
		\item Round 1: All students apply to their first choice. So, $i_1,i_2,i_3,i_4,i_5$ are the applicants for $\cB_1$, and $i_6,i_7,i_8$ are the applicants for $\cB_2$. 
  
Within $\cB_1$, according to its priority order, $i_1$ is first tentatively admitted by $s_1$. Then, the quotas of $s_1$, $\{s_1,s_2\}$, and $\{s_1,s_2,s_3,s_4\}$ are reduced by one. Next, $i_2$ is tentatively admitted by $\{s_1,s_2,s_3,s_4\}$, whose quota is then reduced by one. Next, $i_3$ is tentatively admitted by $s_3$. Then, the quotas of $s_3$, $\{s_3,s_4\}$, and $\{s_1,s_2,s_3,s_4\}$ are reduced by one. Next, $i_4$ is tentatively admitted by $\{s_1,s_2\}$. Then, the quotas of $\{s_1,s_2\}$ and $\{s_1,s_2,s_3,s_4\}$ are reduced by one. At this point, the quota of $\{s_1,s_2,s_3,s_4\}$ has become zero. So, the quotas of the sub-bundles and individual schools in $\cB_1$ are all set to zero. Then, $i_5$ is rejected by $\{s_3,s_4\}$. 

Within $\cB_2$, according to its priority order, $i_6$ is first tentatively admitted by $\{s_5,s_6,s_7\}$, whose quota is then reduced by one. Next, $i_7$ is tentatively admitted by $\{s_5,s_6\}$. Then, the quotas of $\{s_5,s_6\}$ and $\{s_5,s_6,s_7\}$ are reduced by one. Next, $i_8$ is tentatively admitted by $s_5$.

\item Round 2: The rejected student $ i_5 $ in the previous round applies to the next report in her ROL, $s_5$. So we only need to update the admissions for $\cB_2$ in this round. 

Within $\cB_2$, among the tentatively admitted students in the previous round and the new applicants in this round, according to the priority order, $i_6$ and $i_7$ are sequentially tentatively admitted by $\{s_5,s_6,s_7\}$ and $\{s_5,s_6\}$, respectively, as in Round 1, and the quotas of these bundles are updated accordingly. Next, $i_5$ is tentatively admitted by $s_5$. Then, the quotas of $s_5$, $\{s_5,s_6\}$ and $\{s_5,s_6,s_7\}$ are reduced by one. At this point, the quotas of $s_5$, $\{s_5,s_6\}$ and $\{s_5,s_6,s_7\}$ have become zero. So, $i_8$ is rejected by $s_5$. 

\item Round 3: The rejected student $ i_8 $ in the previous round applies to the next option in her ROL, $\{s_1,s_2,s_3,s_4\}$. So we only need to update the admissions for $\cB_1$ in this round. 

Within $\cB_1$, among tentatively admitted students in the previous round and the new applicants in the round, according to the priority order, $i_1$, $i_2$, and $i_3$ are sequentially tentatively admitted by $s_1$, $\{s_1,s_2,s_3,s_4\}$, and $s_3$, respectively, as in Round 1, and the quotas of these schools and relevant bundles are updated accordingly. Next, $i_8$ is tentatively admitted by $\{s_1,s_2,s_3,s_4\}$, whose quota is then reduced by one. At this point, the quota of $\{s_1,s_2,s_3,s_4\}$ has become zero. So, the quotas of the sub-bundles and individual schools in $\cB_1$ are all set to zero. So, $i_4$ is rejected by $\{s_1,s_2\}$. 
		
		\item Round 4: The rejected student $i_4$ in the previous round applies to $s_5$ in this round. But because $i_4$ has lower priority than the tentatively admitted students $i_5,i_6,i_7$ for the schools in $\cB_2$, $i_4$ must be rejected by $s_5$. Then, since $i_4$ has been rejected by all options in her ROL, $i_4$ remains unmatched.
\end{itemize}

So bundle-DA finds the following bundle-matching:

\begin{table}[!htb]
	 	\centering
	 	\begin{tabular}{cccccccc}
   \hline
	 		$ i_1 $ & $ i_2 $ &  $ i_3 $ & $ i_4 $ & $ i_5 $ & $ i_6 $ & $ i_7 $ & $ i_8 $ \\ 
	 		$ s_1 $ & $ \{s_1,s_2,s_3,s_4\} $ & $ s_3 $  & $ \emptyset $  & $ s_5 $ & $ \{s_5,s_6,s_7\}$ & $ \{s_5,s_6\} $ & $ \{s_1,s_2,s_3,s_4\}  $\\
    \hline
	 	\end{tabular}
	 \end{table}
	
	To implement the outcome, students admitted by individual schools must be assigned to those schools. So, $ i_1 $ is assigned to $ s_1 $, $ i_3$ is assigned to $ s_3 $, $ i_4$ is unmatched, and $ i_5 $ is assigned to $ s_5 $. Then, within $\cB_1$, since $i_2,i_8$
are admitted by $\{s_1,s_2,s_3,s_4\}$, they can be assigned to the two schools $ s_2, s_4$ with remaining seats, each assigned to a different school. Within $\cB_2$, since $i_5$ must be assigned to $s_5$, $i_7$, who is admitted by $\{s_5,s_6\}$, must be assigned to $s_6$. Then, $i_6$ must be assigned to $s_7$. So the final matching is: 
	 \begin{table}[!ht]
	 	\centering
	 	\begin{tabular}{cccccccc}
   \hline
	 		$ i_1 $ & $ i_2 $ &  $ i_3 $ & $ i_4 $ & $ i_5 $ & $ i_6 $ & $ i_7 $ & $ i_8 $ \\ 
	 		$ s_1 $ & $ s_2 (s_4) $ & $ s_3 $  & $ \emptyset $  & $ s_5 $ & $ s_7$ & $ s_6 $ & $ s_4 (s_2) $\\
    \hline
	 	\end{tabular}
	 \end{table}
    
\end{example}

\subsection{Properties of Bundle-DA in Simple Bundle System}

Given a simple bundle system, for any ROL profile, bundle-DA finds a stable bundle-matching that is Pareto-undominated size-maximal.

\begin{proposition}\label{prop:stability:simple}
	 For any ROL profile, the outcome of bundle-DA is a stable and Pareto-undominated size-maximal bundle-matching.
\end{proposition}

The proofs of the results in this subsection can be found in the Appendix \ref{appendix:proofs: proposition}.

Bundle-DA may not find a size-maximal bundle-matching. This is not surprising because, similar to the standard DA, bundle-DA is not forward-looking: it does not anticipate whether rejecting a higher-priority student in favor of a lower-priority student could result in more matched students overall (see \autoref{remark:sizemaximal}). However, \autoref{prop:stability:simple} means that it is impossible to increase the number of matched students on the basis of the bundle-DA outcome without making any existing matched student worse off.

In the standard school choice system, when students have complete preferences and can rank all schools, it is a weakly dominant strategy for students to report their complete preferences. When students can only rank a limited number of schools, while it is generally intractable to predict the specific schools students may choose to rank, it remains a weakly dominant strategy for them to truthfully rank reported schools \citep{haeringer2009constrained}. In other words, students cannot achieve a more favorable assignment by misrepresenting the order of reported schools. However, in a bundle system, the strategy space for students is more complex and poses significant challenges in characterizing students' strategies. Despite this,  \autoref{prop:truthtelling:simple} proves that bunlde-DA retains the strategy-proofness property: if a student's reported ROL represents her preferences over bundles, then she cannot manipulate the outcome of bundle-DA by changing the order of the reported bundles. In the following propositions, we use $ \nu $ to denote the outcome of bundle-DA when 
the ROL profile is $ P_I\equiv \{P_i\}_{i\in I} $, and use $ \nu' $ to denote the new outcome of bundle-DA if a student $ i $ unilaterally changes to report a new ROL $ P'_i$.

\begin{proposition}\label{prop:truthtelling:simple}
	Given any ROL profile $ P_I\equiv \{P_i\}_{i\in I} $, for any student $ i $,  if $ i $ changes to report any ROL $P'_i$ such that $ P'_i \neq P_i$ and $ \cB[P_i]=\cB[P'_i] $, then $ \nu(i) =\nu'(i) $ or $ \nu(i) P_i\nu'(i) $. 
\end{proposition}

Our next result demonstrates that bundle-DA incentivizes students to report bundles in the sense that, other things equal, if a student replaces a bundle $b$ reported in her ROL with a sup-bundle $b'$, she will never be assigned to a bundle ranked below her assignment when reporting $b$.

\begin{proposition}\label{prop:biggerbundle:simple}
	Given any ROL profile $ P_I\equiv \{P_i\}_{i\in I} $, for any student $i$, if $ i $ changes to report any ROL $ P'_i $ by replacing some $ b \in \cB[P_i] $ with a sup-bundle $ b'\notin \cB[P_i] $, then:
	\begin{itemize}
		\item[(1)] If $ \nu(i) P_i b $, then $ \nu(i) =\nu'(i) $;
		
		\item[(2)] If $ \nu(i)=b $, then $ \nu'(i)=b' $;
		
		\item[(3)] If $  b P_i \nu(i)$, then $  \nu'(i)=b' $ or $ \nu'(i)=\nu(i)$.
	\end{itemize}  
\end{proposition}

A corollary of this result is that, if a student $ i $ is matched when reporting $P_i$, she must also be matched when replacing some bundles in her ROL with their sup-bundles.
\begin{corollary}
		Given any ROL profile $ P_I\equiv \{P_i\}_{i\in I} $, for any student $i$, if $ i $ changes to report any ROL $ P'_i $ by replacing some $ b \in \cB[P_i] $ with a sup-bundle $ b'\notin \cB[P_i] $, then, as long as $ \nu(i)\neq \emptyset $, $ \nu'(i)\neq \emptyset $.
\end{corollary}

To prove \autoref{prop:truthtelling:simple} and \autoref{prop:biggerbundle:simple}, we embed a simple bundle system into the matching with contracts model of \cite{hatfield2005matching}. Imagine each maximal bundle $ S_k $ as a ``hospital'' that can sign contracts chosen from $ \cB_k $ with students who are regarded as ``doctors''. That is, if a student $ i $ is admitted to a bundle $ b\in \cB_k $, we say that $ i $ signs the contract $ b $ with $ S_k $ and denote the contract by $ (i,b) $. We define choice functions for these hospitals and show that although these choice functions violate substitutability, they satisfy substitutability on all ``observable'' offer processes in the procedure of bundle-DA, along with other conditions defined by \cite{hatfield2021stability}. Leveraging their results on the strategy-proofness of the cumulative offer algorithm, we prove the two propositions. The details can be found in the Appendix \ref{appendix:proofs: proposition}.

We present two examples to further illustrate the properties of bundle-DA. The first example demonstrates that although a student's match likelihood weakly increases when she reports a larger bundle, the impact on other students is ambiguous. It may benefit some students while harming others, or it might only have positive or negative impacts on others who are affected.

\begin{example}
	\autoref{example:introduction} has shown a situation in which a student's bundle report can benefit others. Now, we add another student $i''$ to  \autoref{example:introduction} and assume the following priority order.
	\begin{table}[!ht]
		\centering
		\begin{tabular}{cc}
			$ \succsim_{s} $ & $ \succsim_{s'} $  \\ \hline
			$ i' $ & $ i'$  \\
			$ i $ & $ i $ \\
			$ i'' $ & $ i'' $		
		\end{tabular}
	\end{table}

Suppose that the ROL length is one. If students can only report individual schools, then $ i $ can only be matched if she reports a different school than the one reported by $ i' $, and similarly, $ i'' $ can only be matched if she reports a different school than those reported by both $ i $ and $ i' $.

Now, suppose that there is a bundle $ S_b=\{s,s'\} $ that targets all students. 

Consider a scenario in which $ i' $ reports the bundle $ S_b $, while the other students still report individual schools. Then, student $ i $ will benefit from the bundle report of $i'$ because no matter which school $i$ reports, $i$ must be matched (since $ i' $ can be flexibly assigned to the other school). However, $i''$ could be harmed because she must be unmatched. 

Consider another scenario in which $ i $ reports the bundle, while the others still report individual schools. Student $i'$ is unaffected because she is assigned to her reported school anyway. Then, $i$ benefits from reporting the bundle because she must be assigned to a different school than the one reported by $i'$. However, $ i'' $ could be harmed because she must be unmatched.
\end{example}

In the standard school choice system with complete preference lists, DA is widely recognized for producing the student-optimal stable matching. However, in a bundle system, a student-optimal stable bundle-matching may not exist. Moreover, as illustrated by  \autoref{example:bundleDA:notefficient}, the outcome of bundle-DA may be Pareto dominated by another stable bundle-matching. In the example, several students can exchange their bundle-DA assignments to improve their outcomes without harming others or violating stability.

\begin{example}\label{example:bundleDA:notefficient}
	There are four schools $ \{s_1,s_2,s_3,s_4\} $ and five students $ \{i_1,\ldots,i_5\} $. School $s_1$ has two seats, and the other schools have one seat. Schools use the priority orders in  \autoref{table:example:priority}. There is a bundle $ S_b=\{s_1,s_2\}$ that targets all students. The ROL length is two. Suppose that students report the ROLs in  \autoref{table:example:ROL}. Then, bundle-DA finds the bundle-matching $\nu$ in  \autoref{table:example:outcome}. If $ i_3$ and $i_5 $ exchange their assignments in $\nu$, both of them are strictly better off and the new bundle-matching $\nu'$ is still stable.

 \begin{table}[!htb]
			\centering
			\caption{Priority and ROL}\label{table:priority:ROL}
			\begin{subtable}{.4\linewidth}
				\centering
    \subcaption{Priority orders}\label{table:example:priority}
				\begin{tabular}{cccc}
			%\hline
			 $ \succsim_{s_1} $ & $ \succsim_{s_2} $ &  $ \succsim_{s_3} $ & $ \succsim_{s_4} $ \\ \hline
			 $ i_5 $ &  $ i_5 $ & $ i_3 $  & $ i_3 $ \\
		$ i_2 $ & $ i_2 $ & $ i_2 $ & $ i_2$ \\
  $ i_1 $ & $ i_1 $ & $ i_1 $ & $ i_1$ \\
  $ i_4 $ & $ i_4 $ & $ i_4 $ & $ i_4$ \\
  $ i_3 $ & $ i_3 $ & $ i_5 $ & $ i_5$ 
		\end{tabular}
			\end{subtable}
			\begin{subtable}{.4\linewidth}
				\centering
    \subcaption{ROL profile}\label{table:example:ROL}
				\begin{tabular}{ccccc}
			$ P_{i_1} $ & $ P_{i_2} $ &  $ P_{i_3} $ & $ P_{i_4} $ & $ P_{i_5} $ \\ \hline
			$ s_1 $ & $ \{s_1,s_2\}$ & $ \{s_1,s_2\} $  & $ s_2 $  & $ s_3 $ \\
			$ s_4 $ & $ s_4 $ & $ s_3 $  & $ s_4 $ & $ s_1 $	\\
             & \\
             & \\
             &
		\end{tabular}
			\end{subtable}
		\end{table}

  \begin{table}[!htb]
			\centering
			\caption{Outcome of bundle-DA and its Pareto improvement}\label{table:example:outcome}
			\begin{subtable}{.4\linewidth}
				\centering
    \subcaption{$ \nu $}
				\begin{tabular}{ccccc}
			 \hline
			$ i_1 $ & $ i_2 $ &  $ i_3 $ & $ i_4 $ & $ i_5 $ \\
			$ s_1 $ & $ \{s_1,s_2\}$ & $ s_3 $  & $ s_4 $  & $ s_1 $\\
			\hline		
		\end{tabular}
			\end{subtable}
			\begin{subtable}{.4\linewidth}
				\centering
    \subcaption{$ \nu' $}
				\begin{tabular}{ccccc}
			\hline
			$ i_1 $ & $ i_2 $ &  $ i_3 $ & $ i_4 $ & $ i_5 $ \\
			$ s_1 $ & $ \{s_1,s_2\}$ & $ s_1 $  & $ s_4 $  & $ s_3 $\\ \hline		
		\end{tabular}
			\end{subtable}
		\end{table}
\end{example}

The efficiency loss of bundle-DA in the above example is due to the presence of a rejection cycle in the procedure. In round 1, students $i_1$, $i_2$, $i_4$, and $i_5$ are tentatively admitted to $s_1$, $S_b$, $s_2$, and $s_3$, respectively, while $i_3$ is rejected by $S_b$. Notably, as $s_1$ and $s_2$ together offer three seats, and $i_4$ is the last student admitted to these two schools, it appears that $i_3$ is crowded out by $i_4$.  In round 2, $i_3$ applies to $s_3$ and is tentatively admitted, leading to $i_5$'s rejection by $s_3$. In round 3, $i_5$ applies to $s_1$ and is tentatively admitted, which results in $i_4$'s rejection. This sequence suggests that $i_4$ indirectly crowds out $i_3$ from $S_b$, creating a rejection cycle that ultimately excludes herself from $s_2$. In this scenario, $i_4$'s role is analogous to ``interrupters'' defined by \cite{kesten2010school} in the standard DA procedure. However, their roles differ in that the existence of interrupters is necessary for the stability of the standard DA outcome, while in our example, $i_3$ and $i_5$ can swap seats without violating $i_4$'s priority because $i_4$ does not list $s_1$ in her ROL. In general, as long as the bundle-DA outcome is not Pareto efficient within the set of stable bundle-matchings, there must exist stable improvement cycles similar to those defined by \cite{erdil2008s}. By trading all such cycles, we can restore the constrained efficiency of bundle-DA. We omit the details, since the algorithm would be similar to that of \cite{erdil2008s}. 
Note that trading those cycles can improve students' welfare but cannot increase the number of matched students. \autoref{prop:stability:general} has shown that it is impossible to generate more matched students in any Pareto improvement over the bundle-DA outcome.

\subsection{Tradeoff in Bundle Reporting Strategies}

In general, students confront tradeoffs when deciding whether to report bundles in a simple bundle system. When a bundle includes schools for which a student $i$ holds varying preferences, reporting such a bundle forces $i$ to weigh the benefit of enhancing her match likelihood against the cost of reducing control over her final assignment. Specifically, if $i$ is assigned to such a bundle, the actual school she attends may depend on the specific implementation adopted by the policymaker; this could result in her being placed in either a more preferred or less preferred school within the bundle. In this context, students can strategically select an optimal combination of reported bundles to balance these benefits and costs. For example, if $i$ ranks a bundle $b$ above a sup-bundle $b'$, this indicates a preference to first seek admission to the schools in $S_b$, and, if unsuccessful, to pursue admission to the schools in $S_{b'}\backslash S_b$. Remarkably, the bundle $S_{b'}\backslash S_b$ may not exist in the exogenous bundle system $\cB$, but students can strategically create such a hypothetical bundle in their ROL.
%This ranking is strategic: if the student is not admitted to $S_b$ but gains admission to $S_{b'}$, she will ultimately be assigned to a school in $S_{b'}\backslash S_b$. 

Although characterizing students' strategies in general cases is intractable, we identify certain strategies that are either dominant or dominated for students:

\begin{itemize}
    \item[(1)] If a student is indifferent among all schools in a bundle $ b' $ yet chooses to report a sub-bundle $b$ in her ROL, then it is a weakly dominant strategy for her to replace $ b $ with $ b' $.

    By \autoref{prop:biggerbundle:simple}, replacing $b$ with $ b' $ can never result in an assignment to a bundle ranked below the assignment when reporting $b$ and can only increase the likelihood of being matched to a school within $ b' $. Because she is indifferent among all schools in $ b' $, she cannot be worse off in the final matching outcome regardless of any implementation in the second stage.

    \item[(2)] It is a weakly dominated strategy for a student to rank a bundle $ b' $ above a sub-bundle $ b $.

    Suppose a student $i$ ranks a bundle $ b' $ above a sub-bundle $ b $. Then, if $i$ is rejected by $b'$, she must also be rejected by $b$. So, the ROL slot used to report $b$ is wasted and could have been used to report a different bundle to increase $i$'s match likelihood (unless there are no other bundles or schools that are acceptable to $i$).    
\end{itemize}

\section{A Practical Guide to Bundle Design}\label{section:practicalguide}

So far, we have outlined the structures of bundle systems and the modified DA mechanism, which facilitates the implementation of our bundle approach. However, since this paper does not focus on a concrete application, we do not address questions such as when a bundle system should be designed or how to choose its optimal structure. We believe that the answers to these questions should depend on the specific context of applications and the objectives of policymakers. The next section presents a real-world example. In general, the success of a design requires policymakers to have a thorough understanding of both the participants' preferences and environmental constraints. This section discusses some ideas that guide practical design. 

First, when policymakers believe that their prior knowledge of the environment is insufficient to design a desirable bundle system, they can conduct \textit{surveys} among participants to gain insights. These surveys may ask participants to explain any indifferences in their preferences, propose bundles, or submit bids on a predefined menu of candidate bundles selected by policymakers.

Second, based on survey results, policymakers may design a bundle system that meets their objectives. Objectives may vary across applications, but two potential goals are commonly pursued. If the goal is to simplify participants' strategies by introducing bundles, policymakers should only bundle schools that participants view as indifferent, or at the very least, avoid bundling schools that are perceived as highly distinct. The application in the next section illustrates this case. Alternatively, if the goal is to enhance the expressiveness of the preference reporting language, policymakers may create as many bundles as possible and make them widely available to participants.

Finally, the following fact should be taken into account when designing a bundle system: although the hierarchy constraint typically prevents the system from accommodating all bundle proposals from the surveys, participants can still express preferences over more bundles than those explicitly included in the bundle system. When a student ranks a bundle $S_b$ above a sup-bundle $S_{b'}$, it is equivalent to ranking $S_b$ above a hypothetical bundle $S_{b'}\backslash S_b$, because if she is not admitted to $S_b$ but gains admission to $S_{b'}$, she will ultimately be assigned to a school in $S_{b'}\backslash S_b$. This strategy is observed in our experiments reported in \autoref{section:experiment}.  

\section{Application and Extension}\label{section:application and extension}

\subsection{Application to High School Admission in Hefei}\label{section:application}

Our bundle system can be applied to various school choice environments. In this subsection, we explore its application to the high school admission process in a major city in China and discuss its advantages compared to the existing bundle policy employed in that city. Our paper is partially inspired by the bundle policy that had been implemented in the city for nearly two decades.

Hefei, the largest city of Anhui province in China, employs a centralized test-based admission process for high schools, where students are uniformly ranked based on their scores.\footnote{To have an idea of the market size, in 2023, about 87,000 middle school graduates participated in the high school admission process in Hefei.} Among the many schools in the city, students commonly regard the No.1, No.6, and No.8 High Schools as the best and are nearly indifferent between them compared with others. Before 2006, the constrained ROL restricted students to listing at most one of the three schools in their ROLs, which complicated students' strategies and often led to undesirable admission outcomes. For instance, due to coordination failure, some schools may be ``over-reported,'' while others may be ``under-reported,'' leading to significantly unbalanced school cutoffs. As a result, some high-scoring students were rejected, while some relatively low-scoring students were admitted. Furthermore, as students adjusted their strategies in response to previous years' cutoffs, it led to the ``good year, bad year'' fluctuations in these unstable cutoffs, which were undesirable for both students and schools. 

To address these issues, since 2006, Hefei had bundled the three schools and required that students who wished to attend any of the three schools must list the bundle as first choice in their ROL. During the admission process, the bundle used a common cutoff to admit students. At the end of the admission process, the students admitted to the bundle were randomly assigned to the three schools to achieve fairness for both students and schools.\footnote{For students, although they are nearly indifferent between the three schools, some may slightly prefer one of them over the others. For schools, they all prefer students with higher scores over those with lower scores.} 

Let $\{s_1,s_2,s_3,\ldots,s_L\}$ denote the set of high schools in Hefei, where $s_1,s_6,s_8$ denote the top three schools. In our terminology, Hefei introduced a bundle $S_b=\{s_1,s_6,s_8\}$ that is available to all students. Notably, the three schools were no longer available as individual options after being bundled. Therefore, this system can be viewed as a standard school choice system where the bundle $S_b$ functions as a single school that possesses the combined quota of the three original schools.

This bundle policy was embraced by students and considered a successful reform. However, it encountered challenges in recent years, as students' preferences had evolved. As the city expanded and the three schools established new campuses in different city areas, an increasing number of students developed location preferences and wanted to express precise preferences. But the policy did not allow students to rank individual schools among the three, which led to its ultimate failure. The government decided to abandon the policy altogether in 2024.

Our proposed bundle system can effectively address the challenges faced by Hefei. In our design, the three schools are grouped into a bundle while still remaining available as individual options. Therefore, students who wish to express precise preferences can apply to individual schools, while those who regard the three schools as nearly indifferent or wish to increase their overall admission likelihood to the three schools can choose to apply to the bundle. This bundle system is \textit{simple}, as students are ranked based on their scores. Consequently, the bundle system and the associated bundle-DA algorithm are potentially easy for students to understand and for the Hefei government to implement.

In the two-stage admission process in our design, we focus on designing the first stage, leaving the second stage to the discretion of policymakers. The case of Hefei high school admissions demonstrates one instance of why flexibility in the second stage is important: the Hefei municipal government aims to involve randomization in the second stage to achieve fairness.

\subsection{Two-stage Preference Formation}\label{section:two-stage}

 In this subsection, we discuss another potential advantage of the two-stage admission process in our design. Market design models typically assume that players have exogenous preferences; however, in reality, players often need to invest significant time and effort to acquire information to form their preferences. Recent research has highlighted the importance of costly information acquisition in school choice. The introduction of bundles enables us to design a two-stage preference formation process that reduces information acquisition costs for students.

Given a bundle system, consider a two-stage admission process in which, in the \textbf{first stage}, students only need to acquire sufficient information to form a preference list of bundles. After students submit their ROLs, we run bundle-DA to generate a bundle-matching. In the \textbf{second stage}, for every student $i$ who is admitted to a bundle, $i$ can choose to acquire more information to form precise preferences over the schools within the bundle. Then, we can ask students to submit their precise preferences and run an algorithm within each bundle to match students with schools. For instance, we can run the following algorithm:
    \begin{enumerate}
		\item[] Step 0:  Unmatched students in the first stage are not assigned to any school;
		
		\item[] Step 1:  Students admitted to individual schools in the first stage are assigned to those schools;
		
		\item[] Step 2: Within each bundle consisting of two schools, run DA among admitted students in the first stage, using their reported preferences in the second stage and the remaining quotas of the schools in the bundle; 
		
		$\vdots$
  
		\item[] Step $ n$: Within each bundle consisting of $n$ schools, run DA among admitted students in the first stage, using their reported preferences in the second stage and the remaining quotas of the schools in the bundle. 
	\end{enumerate}

In this two-stage process, students can choose to form coarser preferences in the first stage, followed by precise preferences over a restricted number of schools in the second stage, thereby enabling them to efficiently allocate their resources invested in preference formation. Depending on the specific environments and the costs faced by students, policymakers can optimally design the bundle structure to minimize students' preference formation costs. This approach offers a novel channel to improve the efficiency of the standard admission system.

\section{Putting the Theory to the Test: Two Proof-of-Concept Experiments}\label{section:experiment}

To evaluate the advantages of the bundle system compared to the standard system, we conduct two laboratory experiments in controlled school choice environments. Sections \ref{sec:design-exp1}, \ref{sec:theory-exp1} and \ref{sec:results-exp1} describe the design, theoretical predictions, and results of Experiment 1, respectively. In Experiment 1, we derive precise theoretical predictions for individual equilibrium strategies, which allow for a rigorous test of the model. \autoref{sec:summary-exp2} briefly discusses Experiment 2, which features a more complex environment and serves as a robustness testing bed for the potential benefits and costs of bundle systems. In Experiment 2, although we cannot derive precise predictions for individual equilibrium strategies, the qualitative insights carry over, namely a well-designed bundle system improves student welfare and match rate.\footnote{Experiment 1 was pre-registered on AsPredicted \#249,226.  Experiment 2 was conducted in May 2023 and not pre-registered.}
	
\subsection{Design of Experiment 1} \label{sec:design-exp1}

\paragraph{Environment} 
We design an admission game involving 3 students (from ID1 to ID3) and 3 schools (from A to C). Each school has one seat. Students are played by experimental participants, while schools are simulated by the computer. The three schools are divided into two tiers: Schools A and B belong to the first tier and are roughly equivalent in payoff to students, whereas School C constitutes the second tier and is substantially less desirable. Each student’s payoff vector is privately and independently drawn from the uniform distribution over the following two types: 
\begin{itemize}
    \item Type A: $ u^i(A)=110$, $u^i(B)=100$, $u^i(C)=20 $, $ u^i(\emptyset)=0 $.
    \item Type B: $ u^i(B)=110$, $u^i(A)=100$, $u^i(C)=20 $, $ u^i(\emptyset)=0 $.
\end{itemize}
Each student observes only their own payoff vector but knows that the other students’ payoffs are drawn from the same distribution.

%After payoff types are realized, each student can observe his own payoff vector, but not those of any other students. However, each student knows that each other's payoff is randomly drawn from the above uniform distribution.

Depending on the treatment, each student can report one or two options in their ROLs. All schools share the same priority order over students, which is randomly drawn from the uniform distribution over the six permutations of students. As a result, all students are ex ante symmetric. The priority order is drawn after students submit their ROLs. Once the ROLs are submitted and the priority order is determined, either the standard DA or the bundle-DA is executed to determine the matching outcome.

\paragraph{Treatments} 
We implement four treatments using a between-subjects design. The baseline treatment, \textit{NoBundle-One}, represents a standard system in which students can report only one individual school. The two bundle treatments, \textit{Indiff-Bundle} and \textit{Strict-Bundle}, introduces bundle systems that allows students to report either an individual school or a bundle consisting of two schools. The final treatment, \textit{NoBundle-Two}, is identical to the baseline treatment except that students can report two individual schools, instead of one.

In Indiff-Bundle, Schools A and B—which are similar in payoff to students—are bundled together as AB. This design captures a setting in which the policymaker correctly bundles schools that students regard as close substitutes. In Strict-Bundle, Schools A and C—two schools that students do not consider similar—are bundled as AC, reflecting a case in which the policymaker either misidentifies appropriate bundles or faces constraints that necessitate grouping dissimilar schools. By comparing these two bundle treatments, we can examine how different bundle designs affect students' strategic behavior and the resulting match outcomes. Moreover, as we discuss in  \autoref{sec:theory-exp1}, the chosen parameterization of the two bundle treatments enables us to disentangle the mechanical advantage of allowing students to report more schools from the substantive benefits that arise when bundling is implemented appropriately.

\textit{NoBundle-Two} serves to establish the benchmark advantage of allowing greater ROL length, against which we evaluate the benefits of bundle designs. It is important to note that while increasing ROL length can, in theory, improve matching efficiency under strict preferences beyond the level achievable through bundling, such gains may be limited in practice if students' preferences are imprecise or weakly ordered. The bundle system addresses this limitation by enabling students to express coarse or uncertain preferences directly through reporting bundles.

The (bundle-)DA procedure is simple in this experimental environment. In the standard system, DA is equivalent to assigning students sequentially (in the priority order) to their most-preferred schools among those with available seats. In the bundle system, the same logic applies, except that bundles are treated as additional options. When a student is admitted through a bundle, they are randomly assigned to one of the available seats within that bundle. This procedure ensures that the comparison across treatments isolates the effects of bundling and reporting flexibility on strategic behavior and matching outcomes.

\paragraph{Experimental procedure}
The experiment was conducted at the Nanjing Audit University Economics Experimental Lab in October 2025 with a total of 288 university students, using the software z-Tree \citep{Fischbacher2007}. We ran six sessions for each of the four treatments. Each session consisted of 12 participants who were randomly re-matched into four groups of three in each of the 20 rounds. Within each group, each participant was assigned a student ID and a lottery number, both of which were randomly determined across rounds.\footnote{Specifically, eight different sequences of the pairs of ID and lottery number at the group-round level were generated by a computerized random number generator; four of them were used for a session. However, we ensured that all treatments use the same assignment of sequences to sessions to mitigate the possibility that any treatment difference is simply due to the different random numbers. }  At the end of each round, participants received feedback about their matching outcome and the round payoff. At the end of the session, one round was randomly chosen for each participant to determine her payment. The experimental instructions are presented in Online Appendix~\ref{appendix:instructions-exp1}.

Upon arrival, participants were randomly seated at a partitioned computer terminal. The experimental instructions were given to participants in printed form and were also read aloud by the experimenter. Participants then completed a comprehension quiz before proceeding. After the experiment, they completed a demographic questionnaire. A typical session lasted about one hour with average earnings of 64.8 RMB, including a show-up fee of 15 RMB.\footnote{The average per-hour earnings in the experiment were substantially higher than the minimum hourly wage, which is about 15-20 RMB in the local region. At the time of the experiment, the conversion rate was approximately 1 US dollar to 7.1 RMB.}

\subsection{Theoretical Predictions}\label{sec:theory-exp1}

\autoref{table:theory-exp1} summarizes the theoretical predictions for each treatment. In Nobundle-One, the unique symmetric equilibrium for any realized student payoffs dictates that each student reports the school with the highest payoff. Consequently, Type A students should report School A, while Type B students should report School B. To verify that this strategy constitutes an equilibrium, consider that from each student’s perspective, every other student essentially plays the mixed strategy $ (1/2A,1/2B) $. Reporting either A or B therefore yields the same admission probability, whereas reporting C guarantees admission but at a much lower payoff. Given these incentives, each student optimally reports the school associated with the highest payoff. Under this equilibrium, the expected probability that a student is admitted to their reported first choice is
\[
\sum_{k=0}^2 \binom{2}{k}(\frac{1}{2})^k(\frac{1}{2})^{2-k}\frac{1}{k+1}=\frac{7}{12}\approx 0.5833.
\]
In addition to the overall match rate, we compute the \textit{mismatch rate}, defined as the frequency with which one of the top two students in the (randomly determined) priority order fails to be admitted to a first-tier school (A or B). The mismatch rate captures fairness and coordination inefficiency, which has important implications for both students and schools, as discussed in \autoref{section:application}. In NoBundle-One, mismatch arises whenever the second-ranked student reports the same school as the first-ranked student, which occurs with probability 0.5. For any given student, the ex-ante mismatch probability is therefore $\frac{0.5}{3}$. It follows that the expected number of mismatched students is $ \frac{0.5}{3}\times 3 = 0.5$ and the corresponding group-level mismatch rate is 0.25.

In Indiff-Bundle, the unique symmetric equilibrium predicts that all students report the bundle AB. Each student is admitted with probability 2/3 and, upon admission, is assigned with equal probability to either A or B. The expected match rate is therefore $ \frac{2}{3}\approx 66.67\% $. Because the two highest-ranked students in the priority order must always be admitted to one of the first-tier schools, the mismatch rate is zero. This treatment thus represents a case of maximal coordination efficiency achieved through well-designed bundles that group similar schools together.

In contrast, in Strict-Bundle, the unique symmetric equilibrium mirrors that of the baseline: each student reports the individual school with the highest payoff. When bundles are poorly designed—here, by grouping dissimilar schools A and C—the bundle system yields no theoretical advantage over the standard system. To see why this strategy is an equilibrium, consider a deviation in which a student reports the bundle AC. Such a student would always be admitted, but if any other student reports A, the deviator would be assigned to C. The probability that at least one other student reports A is $\sum_{k=1}^2 \binom{2}{k}(\frac{1}{2})^k(\frac{1}{2})^{2-k}=\frac{3}{4}$. Consequently, the expected payoff from reporting AC ($\approx 40$) is strictly lower than the expected payoff from reporting A or B according to one's type ($\approx 64$). Hence, deviation is not profitable.

\begin{table}[!h]
    \captionsetup{skip=0pt}
    \caption{Theoretical predictions in Experiment 1}\label{table:theory-exp1}
    \begin{center}
    \begin{tabular}{lcccc}
        \toprule
                 & NoBundle-One & Indiff-Bundle & Strict-Bundle & NoBundle-Two \\ 
        \midrule
        Type A equilibrium strategy & A & AB & A & (A, B) \\
        Type B equilibrium strategy & B & AB & B & (B, A) \\
        %Expected  & 0.5833 match & 1/3 A, 1/3 B&  0.5833 match & Type A: 1/2 A, 1/6 B \\
        %Assignment & 0.4167 unmatch & 1/3 unmatch  & 0.4167 unmatch & Type B: 1/2 B, 1/6 A\\ 
        \multicolumn{5}{l}{\textbf{Expected outcome for expected group composition}} \\
        Expected payoff & 64.17 & 70.00 & 64.17 & 71.67 \\ 
        Expected match rate & 58.33\% & 66.67\% & 58.33\% & 66.67\% \\ 
        Expected mismatch rate & 25\% & 0\% & 25\% & 0\% \\ 
        %Mismatch probability & 0.5 & 0 & 0.5 & 0 \\ 
        \multicolumn{5}{l}{\textbf{Predicted average outcome for realized group compositions in the experiment}} \\
        Predicted payoff & 62.56 & 70.03 & 62.56 & 71.79 \\ 
        Predicted match rate & 56.88\% & 66.67\% & 56.88\% & 66.67\% \\ 
        Predicted mismatch rate & 15.42\% & 0\% & 15.42\% & 0\% \\ 
        \bottomrule
    \end{tabular}
    \end{center}
    {\footnotesize \textit{Notes: The rows correspond to expected outcomes calculate the theoretical predictions for the expected group composition given the primitives. The rows correspond to predicted outcomes calculate the theoretical prediction for the realized group compositions (i.e., lottery and type assignments within each group). The mismatch rate computes the probability of the two highest-ranked students not admitted to a first-tier school A or B.}   }
\end{table}

Finally, in NoBundle-Two, the unique symmetric equilibrium requires each student to report the school with the highest payoff in the first slot and the other first-tier school in the second slot. To verify that this is an equilibrium, consider a Type A student. Given that others follow equilibrium strategies, her expected payoff from the possible reporting strategies are as follows: 71.67 by reporting (A, B); 68.33 from (B, A); 65 from (A, C); 60 from (B, C); 20 from either (C, A) or (C, B). Therefore, the optimal strategy for a Type A student is to report (A, B), and symmetrically, a Type B student should report (B, A). It is important to note that compared with Indiff-Bundle, increasing the ROL length does not improve the theoretical match rate or reduce the mismatch rate, despite the fact that the NoBundle-Two treatment retains more flexibility in reporting preferences. Nevertheless, it does allow students to express strict preferences between the two first-tier schools, yielding a slightly higher expected payoff. The comparison between Indiff-Bundle and NoBundle-Two provides the most demanding test of the substantive benefits of the bundle design because students are endowed with strict preferences in our experimental environment, which precludes the optimality of the bundle system from the outset. Thus, by using this environment, we stack the cards against us when evaluating the bundle system.

Importantly, the chosen parameterization of this experiment allows us to isolate the mechanical advantage of allowing students to report more schools through bundles. A comparison between NoBundle-One and Strict-Bundle shows that although Strict-Bundle provides this mechanical advantage, it does not improve matching efficiency. In contrast, the comparison between Indiff-Bundle and Strict-Bundle highlights that the substantive benefits of bundling stem from proper bundle design rather than from the mere expansion of reporting options. Finally, the comparison between Indiff-Bundle and NoBundle-Two demonstrates that a well-designed bundle system can capture almost all theoretical gains associated with increasing the ROL length.

%Mismatch prob = probability of at least one top-two student not being admitted to the first tier. 
%Mismatch rate = the percentage of top-two students who are not admitted to the first tier.

\subsection{Experimental Results}\label{sec:results-exp1}

\subsubsection{Aggregate-level results}
\autoref{table:aggregate-exp1} reports the efficiency outcomes, measured by the average student payoff, across treatments. Indiff-Bundle yields significantly higher efficiency than the baseline NoBundle-One (70.9 vs. 60.2, $p$ = 0.002, Wilcoxon rank-sum test with each session treated as an independent observation; same below). Mirroring this result, \autoref{table:aggregate-exp1} also shows that relative to NoBundle-One, Indiff-Bundle improves both the match rate (71.2\% vs. 62.7\%, $p$ = 0.002) and the mismatch rate (4.4\% vs. 20.2\%, $p$ = 0.002). In contrast, Strict-Bundle does not improve efficiency (60.2. vs. 60.2, $p$ = 0.818) and even slightly increases the mismatch rate (22.4\% vs. 20.2\%, $p$ = 0.037). These findings indicate that the advantage of a well-designed bundle system stems from its substantive benefits—the ability to bundle similar options effectively—rather than from a mere mechanical advantage of allowing students to report more schools.\footnote{Online Appendix Figures \ref{fig:student-payoff-exp1} to \ref{fig:bundle-report-rate-exp1} show the average student payoff, match rate, mismatch rate and bundle report rate over rounds across treatments, suggesting that the aggregate patterns observed in \autoref{table:aggregate-exp1} are largely stable over time.}

\begin{table}[!h]
    \captionsetup{skip=0pt}
    \caption{Aggregate results}\label{table:aggregate-exp1}
    \begin{center}
    \begin{tabular}{lcccc}
        \toprule
                            & NoBundle-One & Indiff-Bundle & Strict-Bundle & NoBundle-Two \\ 
        \midrule
        Average payoff      & 60.2 & 70.9 & 60.2 & 74.1 \\
        Match rate          & 62.7\% & 71.2\% & 71.1\% & 83.8\% \\ 
        Mismatch rate       & 20.2\% & 4.4\%  & 22.4\% & 5.8\% \\ 
        Bundle report rate  & / & 76.3\% & 24.5\% &  /  \\
        \bottomrule
    \end{tabular}
    \end{center}
\end{table}

Although Strict-Bundle is improperly designed, it nonetheless increases the match rate relative to NoBundle-One (71.1\% vs. 62.7\%, $p$ = 0.002). To explore the source of this outcome, we conduct a counterfactual analysis assuming that participants who reported the bundle AC (24.5\% of observations; see \autoref{table:aggregate-exp1}) followed their equilibrium strategy (see \autoref{table:theory-exp1}). This analysis reveals that the match rate would drop to 57.7\%, while the average student payoff would increase slightly to 61.4 and the mismatch rate would reduce to 16.4\%. These results suggests that the improved match rate observed in Strict-Bundle arises from participants' conservative but suboptimal strategy of reporting the bundle AC.

Turning to the comparison between NoBundle-Two and Indiff-Bundle, we find that both the average student payoff and match rate are significantly higher in NoBundle-Two, contrary to theoretical predictions. The mismatch rate, however, does not differ significantly between the two treatments (5.8\% vs. 4.4\%, $p$ = 0.260). This pattern implies that even a theoretically well-designed bundle system cannot fully capture the theoretical efficiency gains associated with the increased ROL length. 

To investigate the source of this discrepancy, we first conduct a similar counterfactual analysis assuming that participants who reported individual schools in Indiff-Bundle (23.7\% of observations; see \autoref{table:aggregate-exp1}) instead followed the equilibrium strategy of reporting the bundle AB (see \autoref{table:theory-exp1}). This exercise, which is equivalent to the approach used to derive the theoretical outcomes for realized group compositions, reveals that neither the average payoff nor the match rate improves, indicating that the less-than-complete adoption of bundle reporting is not responsible for the observed discrepancy.

We then examine the individual reporting strategy in NoBundle-Two and find that nearly half of observations deviate from the equilibrium behavior. In particular, rather than reporting the two first-tier schools dictated by the equilibrium, students often include School C in the second slot. This deviation reduces the conditional payoff upon being matched: the average student payoff upon successful match is 99.56 in Indiff-Bundle, significantly higher than 88.43 in NoBundle-Two ($p$ = 0.002). This is the opposite of the theoretical prediction. However, this reduction in conditional payoff is more than offset by the improved match rate, leading to significantly higher unconditional average payoff. Indeed, in our experimental data, participants achieve average payoffs exceeding the theoretical equilibrium level. We explore these individual reporting behaviors in greater detail in the next subsection.

\subsubsection{Individual reporting strategies}

Observed individual reporting strategies reported in \autoref{table:individual-exp1} are broadly in line with the theoretical benchmarks in  \autoref{table:theory-exp1}, while revealing systematic hedging behavior that helps explain treatment differences. In NoBundle-One, theory predicts that each type reports their top school. Consistent with this, most Type A report A (62.4\%) and most Type B report B (63.5\%). However, a substantial minority shift to the other first-tier school (Type A: 28.5\% report B; Type B: 29.5\% report A), and a non-trivial minority choose the safe but dominated option C (Type A: 9.2\%; Type B: 7.0\%). These deviations indicate hedging against the risk of remaining unmatched and help account for the relatively high mismatch rate.

\begin{table}[!h]
    \captionsetup{skip=0pt}
    \caption{Individual reporting strategies}\label{table:individual-exp1}
    \begin{center}
    \begin{tabular}{lcccc}
        \toprule
                            & NoBundle-One & Indiff-Bundle & Strict-Bundle & NoBundle-Two \\ 
        \midrule
        \textbf{Type A}      &  &  &  &  \\ 
        A           & 62.4\% & 15.7\% & 37.3\% & / \\ 
        B           & 28.5\% & 3.0\% & 37.5\% & / \\ 
        C           & 9.2\% & 5.7\% & 0.7\% & / \\ 
        AB          & / & 75.7\% & / & / \\ 
        AC          & / & / & 24.6\% & / \\ 
        (A, B)      & / & / & / & 52.6\% \\ 
        (A, C)      & / & / & / & 23.8\% \\ 
        (B, A)      & / & / & / & 7.6\% \\ 
        (B, C)      & / & / & / & 15.5\% \\ 
        (C, A)      & / & / & / & 0.3\% \\ 
        (C, B)      & / & / & / & 0.3\% \\ 
        \addlinespace
        \textbf{Type B}      &  &  &  &  \\ 
        A           & 29.5\% & 3.0\% & 26.3\% & / \\ 
        B           & 63.5\% & 15.2\% & 48.8\% & / \\ 
        C           & 7.0\% & 5.0\% & 0.4\% & / \\ 
        AB          & / & 76.8\% & / & / \\ 
        AC          & / & / & 24.5\% & / \\ 
        (A, B)      & / & / & / & 8.0\% \\ 
        (A, C)      & / & / & / & 12.5\% \\ 
        (B, A)      & / & / & / & 52.7\% \\ 
        (B, C)      & / & / & / & 26.3\% \\ 
        (C, A)      & / & / & / & 0.1\% \\ 
        (C, B)      & / & / & / & 0.4\% \\ 
        \bottomrule
    \end{tabular}
    \end{center}
\end{table}

In Indiff-Bundle, theory predicts universal take-up of the bundle AB. Adoption is indeed high: 75.7\% of Type A and 76.8\% of Type B report AB. The remainder mostly choose an individual first-tier school (Type A: A 15.7\%, B 3.0\%; Type B: B 15.2\%, A 3.0\%), with only small fractions selecting C (about 5–6\%). Thus, the bundle substantially simplifies choices and concentrates reports on the set of close substitutes, consistent with the efficiency gains and low mismatch rates documented at the aggregate level.

In Strict-Bundle, theory predicts a return to the baseline pattern, with participants reporting their single top school. Yet roughly one quarter report the bundle AC (Type A: 24.6\%; Type B: 24.5\%). Aside from this, choices split toward the top-tier school in a type-consistent way (Type A: A 37.3\%, B 37.5\%; Type B: B 48.8\%, A 26.3\%), with almost no reports of C. The take-up of AC is off the equilibrium path but consistent with conservative insurance motives that help to raise matching probabilities while depressing conditional payoffs, as documented earlier.

In NoBundle-Two, theory prescribes that each type lists both first-tier schools in a type-dependent order. The modal strategies conform: Type A most often reports (A, B) at 52.6\%, and Type B reports (B, A) at 52.7\%. A large minority, however, pairs a first-tier with C to hedge (Type A: (A, C) 23.8\%, (B, C) 15.5\%; Type B: (B, C) 26.3\%, (A, C) 12.5\%), with dominated combinations such as (C, A) or (C, B) virtually absent (lower than 1\%). This hedging behavior explains why the match rate exceeds the theoretical benchmark while the payoff conditional on matching is lower, as shown in the aggregate results.

Overall, participants' reporting patterns align with the main theoretical incentives: they strongly adopt a well-designed bundle of close substitutes, leading to improved welfare, but are reluctant to use a poorly-designed bundle.

\subsection{Summary of Findings from Experiment 2}\label{sec:summary-exp2}

The simple structure of Experiment 1 is essential for making precise theoretical predictions, enabling a clean test of the mechanisms through which bundle systems improve matching efficiency. However, this environment is intentionally constrained: the matching market involves only three students and three schools, and students are allowed to report only one option in bundle systems. To show the broader applicability of the bundle system, Experiment 2 introduces a larger, incomplete-information market with six students and six schools (labeled A to F).\footnote{See Online Appendix \ref{appendix:experiment2} for a detailed report on Experiment 2. } Students share common cardinal utilities over schools (D=80, A=50, B=45, C=40, E=30, F=20) and are ranked by exam scores independently drawn from a truncated normal distribution on [1,100]; each student only observes her own score. We compare a standard baseline without bundles to two bundle treatments: one that offers a bundle of close substitutes (ABC), and another that offers a bundle of dissimilar options (DEF). Students can report two options, allowing them in the bundle treatments to include both individual schools and bundles in their ROL.

Observed behavior mirrors theoretical incentives. Bundle take-up is high when the bundle groups similar schools: participants report the bundle ABC in 61.6\% of cases (31.3\% as the first rank and 30.4\% as the second), whereas they report the bundle DEF in 44.5\% of cases, concentrated mainly in the second rank (5.1\% first, 39.5\% second). Low- and medium-score students are the primary users of bundles; high-score students largely avoid DEF but readily place ABC second, treating it as a good insurance option.

Both bundle treatments increase the overall match rate and average payoffs relative to the baseline, primarily through higher matching probabilities. Match rates rise from 82.2\% without bundles to 89.2\%–89.9\% with bundles, and average payoffs increase from 36.4 to about 40.4–40.7. Under the well-designed bundle system, improvements are broad-based: students across all score ranges all match more often and earn higher payoffs. Under the poorly-designed bundle system, gains accrue primarily to medium- and low-score students, while high-score students do not improve because they seldom report the bundle DEF. Fairness effects depend on design: when the bundle aligns with preferences (ABC), the share of justified-envy pairs is statistically indistinguishable from the baseline (13.9\% vs. 13.1\%), whereas a poorly aligned bundle (DEF) increases justified-envy pairs to 19.4\%. 

Overall, Experiment 2 corroborates the findings of Experiment 1. Well-designed bundles that group close substitutes improve coordination, raise match rates, and increase welfare without compromising fairness, and observed reporting strategies align closely with these incentives. Poorly-designed bundles still provide insurance to lower-ranked students and improve overall match rates, but they offer little benefits to top-ranked students and may introduce mismatch concerns. These patterns are consistent with those documented in Experiment 1.

\section{Conclusion} \label{section:conclusion}

This paper proposes a novel school choice system where certain schools are grouped into bundles and offered as options for students' preference submission. By ranking a bundle, a student seeks admission to any school within that bundle without ranking the individual schools. This effectively allows the student to report multiple schools within a single preference slot and avoid reporting precise preferences. A bundle system has the potential to enhance students' match likelihood and overall welfare, particularly in addressing the practical challenges when students face limitations on the preference length and wish to express coarse preferences. To maintain generality, we outline the minimal set of conditions for designing a bundle system and an associated admission mechanism. Additionally, we propose a simplified bundle system and discuss its desirable properties. We further discuss its application to the high school admission process in a major city of China, and extend our framework to an admission process that enables students to customize their preference formation process contingent on their information acquisition costs. To evaluate the effectiveness of the proposed system, we conduct two laboratory experiments to examine student strategies and welfare improvements in the designed bundle systems.

Introducing bundles opens up numerous possibilities for advancing school choice design. Several intriguing avenues for future research remain unexplored, such as analyzing the effects of bundles in highly unbalanced markets, where the risk of remaining unmatched is a significant concern for students. In such contexts, competition may incentivize students to actively report bundles, even if they perceive the schools within the bundles as distinct. Examining how bundles affect admission outcomes for schools is also particularly relevant. In the case study of the Chinese city, bundles benefit both students and schools.  Furthermore, it would be valuable to explore the potential roles of bundles in achieving other policy objectives beyond those discussed in this paper.

\setlength{\bibsep}{0pt plus 0.3ex}
\bibliographystyle{aea}
\bibliography{reference}

@article{decerf2021manipulability,
  title={Manipulability in school choice},
  author={Decerf, Benoit and Van der Linden, Martin},
  journal={Journal of Economic Theory},
  volume={197},
  pages={105313},
  year={2021},
  publisher={Elsevier}
}

@article{ali2024college,
  title={Hedging When Applying: Simultaneous Search with Correlation},
  author={Ali, S Nageeb and Shorrer, Ran I},
  journal={American Economic Review},
  volume={115},
  number={2},
  pages={571--598},
  year={2025},
  publisher={American Economic Association 2014 Broadway, Suite 305, Nashville, TN 37203}
}

@article{dougan2024geography,
  title={When Geography Shapes Preferences: Redesigning Teacher Assignment in Italy},
  author={Do{\u{g}}an, Battal and Cavallo, Mariagrazia},
  year={2024},
  journal={working paper}
}

@article{chen2021information,
  title={Information acquisition and provision in school choice: an experimental study},
  author={Chen, Yan and He, Yinghua},
  journal={Journal of Economic Theory},
  volume={197},
  pages={105345},
  year={2021},
  publisher={Elsevier}
}

@article{chen2022information,
  title={Information acquisition and provision in school choice: a theoretical investigation},
  author={Chen, Yan and He, YingHua},
  journal={Economic Theory},
  volume={74},
  number={1},
  pages={293--327},
  year={2022},
  publisher={Springer}
}

@article{maxey2024school,
  title={School choice with costly information acquisition},
  author={Maxey, Tyler},
  journal={Games and Economic Behavior},
  volume={143},
  pages={248--268},
  year={2024},
  publisher={Elsevier}
}

@article{artemov2021assignment,
  title={Assignment mechanisms: common preferences and information acquisition},
  author={Artemov, Georgy},
  journal={Journal of Economic Theory},
  volume={198},
  pages={105370},
  year={2021},
  publisher={Elsevier}
}

@article{nguyen2023dynamic,
  title={Dynamic Combinatorial Assignment},
  author={Nguyen, Th{\`a}nh and Teytelboym, Alexander and Vardi, Shai},
  journal={arXiv preprint arXiv:2303.13967},
  year={2023}
}

@article{budish2011combinatorial,
  title={The combinatorial assignment problem: Approximate competitive equilibrium from equal incomes},
  author={Budish, Eric},
  journal={Journal of Political Economy},
  volume={119},
  number={6},
  pages={1061--1103},
  year={2011},
  publisher={University of Chicago Press Chicago, IL}
}

@article{fragiadakis2019designing,
  title={Designing mechanisms to focalize welfare-improving strategies},
  author={Fragiadakis, Daniel E and Troyan, Peter},
  journal={Games and Economic Behavior},
  volume={114},
  pages={232--252},
  year={2019},
  publisher={Elsevier}
}

@article{kesten2010school,
  title={School choice with consent},
  author={Kesten, Onur},
  journal={The Quarterly Journal of Economics},
  volume={125},
  number={3},
  pages={1297--1348},
  year={2010},
  publisher={MIT Press}
}

@article{hatfield2021stability,
title={Stability, strategy-proofness, and cumulative offer mechanisms},
	author={Hatfield, John William and Kominers, Scott Duke and Westkamp, Alexander},
	journal={The Review of Economic Studies},
	volume={88},
	number={3},
	pages={1457--1502},
	year={2021},
	publisher={Oxford University Press}
	}

@article{bodoh2013risk,
	title={Risk and conflation in matching markets},
	author={Bodoh-Creed, Aaron L},
	year={2014},
	journal={mimeo}
}

@article{hatfield2017contract,
	title={Contract design and stability in many-to-many matching},
	author={Hatfield, John William and Kominers, Scott Duke},
	journal={Games and Economic Behavior},
	volume={101},
	pages={78--97},
	year={2017},
	publisher={Elsevier}
}

@article{hakimov2023costly,
  title={Costly information acquisition in centralized matching markets},
  author={Hakimov, Rustamdjan and K{\"u}bler, Dorothea and Pan, Siqi},
  journal={Quantitative Economics},
  volume={14},
  number={4},
  pages={1447--1490},
  year={2023},
  publisher={Wiley Online Library}
}

@article{erdil2008s,
	title={What's the matter with tie-breaking? Improving efficiency in school choice},
	author={Erdil, Aytek and Ergin, Haluk},
	journal={American Economic Review},
	volume={98},
	number={3},
	pages={669--89},
	year={2008}
}

@article{erdil2017two,
	title={Two-sided matching with indifferences},
	author={Erdil, Aytek and Ergin, Haluk},
	journal={Journal of Economic Theory},
	volume={171},
	pages={268--292},
	year={2017},
	publisher={Elsevier}
}

@article{koh2022visit,
	title={When to Visit: Information Acquisition in College Admissions},
	author={Koh, Youngwoo and Lim, Wooyoung},
	journal={working paper},
	year={2023}
}

@article{haeringer2009constrained,
	title={Constrained school choice},
	author={Haeringer, Guillaume and Klijn, Flip},
	journal={Journal of Economic Theory},
	volume={144},
	number={5},
	pages={1921--1947},
	year={2009},
	publisher={Elsevier}
}

@article{hatfield2005matching,
	title={Matching with contracts},
	author={Hatfield, John William and Milgrom, Paul R},
	journal={American Economic Review},
	pages={913--935},
	year={2005},
	publisher={JSTOR}
}

@article{Fischbacher2007,
        author = {Fischbacher, Urs},
        journal = {Experimental Economics},
        number = {2},
        pages = {171--178},
        title = {{z-Tree: Zurich toolbox for ready-made economic experiments}},
        volume = {10},
        year = {2007}
}

\clearpage
\appendix

\section*{Appendix}

\section{Proofs of  \autoref{lemma:implementation} and  \autoref{lemma:implementation:stable}} \label{appendix:proofs:lemma}

\begin{proof}[\normalfont \textbf{Proof of \autoref{lemma:implementation}}]
    To prove that every bundle-matching $\nu$ is implementable, we prove that the general implementation procedure must produce a matching. Suppose by contradiction that the procedure cannot generate a matching. Then, there must exist some step $ k $ and some bundle $ b $ consisting of $k$ schools such that, after the students admitted by bundles of smaller sizes are assigned to the seats in these bundles, the remaining seats in $ b $ are not enough to accommodate the students admitted by $ b $. However, it is impossible because $ \sum_{b'\in\cB:S_{b'}\subsetneq S_b} |\nu(b')| + |\nu(b)|\le q_{b}$, where $ \sum_{b'\in\cB:S_{b'}\subsetneq S_b} |\nu(b')| $ is the total number of students admitted by sub-bundles of $ b $ and $ |\nu(b)| $ is the number of students admitted by $ b $. 

    Let $\mu$ be any standard matching that implements $\nu$. Then, it is evident that $\mu$ can be found by the general implementation procedure in which, once a student $i$ is dealt with, $i$ is assigned to $\mu(i)$. This is doable because, given that $\mu$ implements $\nu$, when $i$ is dealt with in a step, the remaining seats in the bundle $\nu(i)$ must be enough to accommodate a matching that coincides with $\mu$ for the assignments of the students considered up to the step. 
\end{proof}

\begin{proof}[\normalfont \textbf{Proof of  \autoref{lemma:implementation:stable}}]

 Given a ROL profile, let $ \nu $ be a bundle-matching and $ \mu $ be any implementation of $ \nu $. We prove that if $ \nu $ is stable, then $ \mu $ is stable. First, $ \mu $ is individually rational because it is an implementation of $\nu$. Second, $ \mu $ must be non-wasteful. If $ \mu $ is wasteful, then there exists a student $ i $ and a school $ s $ such that $ |\mu(s)|<q_s $ and $ i $ ranks some bundle $ b $ that includes $ s $ above the bundle $ \nu(i) $ that includes $ \mu(i) $ in her ROL.\footnote{If $ \mu(i)=\emptyset $, we interpret $ \emptyset $ as a null bundle that is ranked below all real bundles in $ i $'s ROL.} But because $ \nu $ is non-wasteful, there must exist a bundle $ b' $ with $ S_{b'}\supseteq S_b $ such that $ Q^\nu_{b'}=q_{b'} $. This means that in any implementation of $ \nu $, all seats of $ b' $, including the seats of $ s $, must be fully assigned to students. This is a contradiction. Finally, if some student $ i $ has justified envy towards some student $ j $  in $ \mu $, then $ i $ must rank some bundle $ b $ that includes $ \mu(j) $ above the bundle $ \nu(i) $ that includes $ \mu(i) $ in her ROL, and $ i $ has higher priority than $ j $ for $ \mu(j) $. Since $ \mu(j)\in S_{\nu(j)}  $, $ S_b $ and $ S_{\nu(j)} $ are not disjoint. Then there are three cases. 
 
 In the first case, $ \nu(j)=b $, which means that $b $ is available to both $ i $ and $ j $. Then, the fact that $ i $ has higher priority than $ j $ for $ \mu(j) $ means that $ i $ has higher priority than $ j $ for every school in $ S_b $. But this means that $ \nu $ is not stable, which is a contradiction. 
 
 In the second case, $ \nu(j) $ is a sub-bundle of $ b $. Then, all schools in $ S_{\nu(j)} $ must rank $ i $ and $ j $ in the same way in their priority orders. So all of them must rank $ i $ above $ j $, which means that $ \nu $ is not stable, a contradiction. 
 
 In the last case, $ \nu(j) $ is a sup-bundle of $ b $. Because $ \mu(j)\in S_b $, it means that, for every bundle $ b' $ such that $ S_b\subseteq S_{b'} \subsetneq S_{\nu(j)} $, $ Q^\nu_{b'}<q_{b'} $. Also, all schools in $ S_b $ must rank $ i $ and $ j $ in the same way in their priority orders. So all of them must rank $ i $ above $ j $. These mean that $ \nu $ is not stable, a contradiction.  So $ \mu $ must be stable.
\end{proof}

\section{Proofs of Propositions \ref{prop:stability:simple}, \ref{prop:truthtelling:simple}, and \ref{prop:biggerbundle:simple}} \label{appendix:proofs: proposition}

 \autoref{prop:stability:simple} is implied by  \autoref{prop:stability:general} in the Appendix \ref{appendixbundle-DA:general}. 
To prove  \autoref{prop:truthtelling:simple} and  \autoref{prop:biggerbundle:simple}, we embed a simple bundle system into the matching with contracts model of \cite{hatfield2005matching}, and apply the results of \cite{hatfield2021stability}.

Specifically, given a simple bundle system $\cB$, imagine each maximal bundle $ S_k $ as a hospital that can sign contracts chosen from the sub-hierarchy $ \cB_k $ with students who are regarded as doctors. If student $ i $ is admitted by a bundle $ b\in \cB_k $, we say that $ i $ signs the contract $ b $ with $ S_k $ and denote the contract by $ (i,b) $. Let $ X $ denote the set of all possible contracts. In any subset of contracts $ X' $, denote the set of contracts involving any student $ i $ by $ X'_i $ and denote the set of contracts involving any hospital $ S_k $ by $ X'_k $. Because every student can sign at most one contract, we call a subset of contracts $ X'$ \textbf{unitary} if for every student $ i $, $ |X'_i|\le 1 $. For each hospital $ S_k $, we need to define its choice function $ C_k $. In every step of bundle-DA, each $ S_k $ makes choices only from unitary sets of contracts because a student applies to a new bundle only after being rejected by the previous bundle in her ROL. However, to apply the standard results in the matching theory, it is more convenient to define each hospital's choices from all possible sets of contracts. 

Formally, let $ C_k $ denote the choice function of $S_k$. We introduce an arbitrary order $ \rhd^* $ of all bundles and individual schools to break ties in the definition of $ C_k $. The choice of $ \rhd^* $ is irrelevant for the properties of $ C_k $ that we will prove.  For any nonempty set of contracts $ X'\subset X_k $, $ C_k(X') $ is generated by the following procedure:
\begin{itemize}
	\item Consider students in decreasing order of their priority ranking $ \succsim_k $. For each considered student $ i $, consider the bundles that are maximal within $ X'_i $.\footnote{We do not consider bundles of smaller sizes because if a student $ i $ cannot be admitted by any maximal bundle within $ X'_i $, she can neither be admitted by any sub-bundle of these maximal bundles.} If there are multiple maximal bundles within $X'_i$, consider them one by one in the order $ \rhd^* $. Admit $ i $ to the first considered bundle that has vacant seats. Then, the quotas of the considered bundle and its every sup-bundle within $X'$ are reduced by one. Student $ i $ is rejected if she cannot be admitted by any maximal bundle within $X'_i$. When the quota of any bundle becomes zero, the quotas of its sub-bundles within $X'$ are also set to zero.
\end{itemize}

Note that $ C_k(X') $ is unitary because every student is admitted by at most one bundle, and if a student is admitted, she is admitted by a maximal bundle within $X'_i$. Define $ R_k(X')=X'\backslash C_k(X') $ as the rejection function of $ S_k $. Then, a few observations follow.

\begin{lemma}\label{lemma:sup-bundle}
	For any $ X'\subset X_k $, any $ i\in I $, and any two bundles $ b,b'\in \cB_k \cap \cB_i $ such that $ b' $ is a sup-bundle of $ b $, if $ (i,b)\in C_k(X'\cup (i,b)) $, then $ (i,b')\in C_k(X'\cup (i,b')) $ and  $ C_k(X'\cup (i,b))\backslash (i,b)\subseteq C_k(X'\cup (i,b')) $.\footnote{With abuse of notation, we write $C_k(X'\cup (i,b))\backslash (i,b)$ for $ C_k(X'\cup \{(i,b)\})\backslash \{(i,b)\}$.}
\end{lemma}

\begin{proof}
	We compare the procedure of $ C_k(X'\cup (i,b)) $  and the procedure of $ C_k(X'\cup (i,b')) $. The two procedures coincide up to the step in which $ i $ is considered. If $ (i,b)\in C_k(X'\cup (i,b)) $, then $ b $ is a maximal bundle in $ X'_i\cup (i,b) $, and in the procedure of $ C_k(X'\cup (i,b)) $, when $ i $ is considered, $ b $ has vacant seats. Because $ b' $ is a sup-bundle of $ b $, $ b' $ is the unique maximal bundle in $ X'_i\cup (i,b') $ and must have vacant seats when $ i $ is considered in the procedure of $ C_k(X'\cup (i,b')) $. Therefore, $ i $ must be admitted by $ b' $ in $ C_k(X'\cup (i,b')) $; that is, $ (i,b')\in C_k(X'\cup (i,b')) $. The only difference between the two procedures is that, after $ i $ is admitted, in the procedure of $ C_k(X'\cup (i,b)) $, the quotas of $ b $ and its sup-bundles are reduced by one, while in the procedure of $ C_k(X'\cup (i,b')) $, the quotas of $b'$ and its sup-bundles are reduced by one, but the quota of $ b $ and the quotas of the possible bundles between $b$ and $b'$ are not reduced. Therefore, the remaining quotas of bundles in the remaining procedure of $ C_k(X'\cup (i,b')) $ are weakly higher than the corresponding quotas in the remaining procedure of $ C_k(X'\cup (i,b)) $. So, $ C_k(X'\cup (i,b))\backslash (i,b)\subseteq C_k(X'\cup (i,b')) $.
\end{proof}

It is easy to show that $ C_k $ does not satisfy substitutability.\footnote{A choice function $ C_k $ satisfies \textbf{substitutability} if, for any $ X'\subset X_k $ and any $ (i,b)\in X_k\backslash X' $, $ R_k(X')\subset R_k(X'\cup (i,b)) $. Consider an example in which there are three students ranked as $ i\succ_k i' \succ_k j $ and two schools $ s $ and $ s' $, each admitting one student. There is only one bundle $ b=\{s,s'\} $. Consider $ X'=\{(i,s'),(j,s'), (i',s)\} $ and assume that $ s \rhd^* s'$. Then $ C_k(X')=\{(i,s'),(i',s)\} $, while $ C_k(X'\cup (i,s))=\{(i,s),(j,s')\} $. So $C_k$ violates substitutability.} However, it satisfies substitutability on all possible offer processes in the procedure of bundle-DA, which are called observable by \cite{hatfield2021stability}.

Formally, an \textbf{offer process} for $ S_k $ is a finite sequence of distinct contracts $ \mathbf{x}=(x^1,\ldots,x^M) $ in which every $ x^m\in \cB_k $. For all $ m\in\{1,\ldots, M\} $, we define $ \mathbf{x}^m\equiv (x^1,\ldots
,x^m)$ and call it a subprocess of $ \mathbf{x} $. With abuse of notation, we also use $ \mathbf{x} $ to denote the set of contracts in the offer process $ \mathbf{x} $. We say $ \mathbf{x} $ is \textbf{observable} if, for all $ m\in\{2,\ldots, M\} $, $ x^m_i\notin [C_k(\mathbf{x}^{m-1})]_I $, where $x^m_i$ denotes the student involved in $x^m$ and $[C_k(\mathbf{x}^{m-1})]_I$ denotes the set of students involved in the contracts $C_k(\mathbf{x}^{m-1})$.
In words, observability means that when $ S_k $ receives the contract application $x^m$, the student involved in $x^m$ must be rejected from the sequence of contracts $\mathbf{x}^{m-1}$. The offer process in bundle-DA is observable because a student applies to a new bundle only if she has been rejected by all of the previous bundles she has applied to. 

In the rest, we prove that $ C_k $ satisfies several properties defined by \cite{hatfield2021stability}. A choice function $ C_k $ satisfies \textbf{irrelevance of rejected contracts} if, for any $ X'\subseteq X_k $ and any $ (i,b)\in R_k(X') $, $ C_k(X'\backslash (i,b))=C_k(X')$. It satisfies \textbf{observable substitutability}  if, for any observable offer process $ \mathbf{x}= (x^1,\ldots,x^M) $, 
$ R_k(\mathbf{x}^{M-1}) \subseteq R_k(\mathbf{x})$. It satisfies
\textbf{observable size monotonicity} if, for any observable offer process $ \mathbf{x}= (x^1,\ldots,x^M) $, 
$
|C_k(\mathbf{x}^{M-1})| \le |C_k(\mathbf{x})|$. It satisfies \textbf{non-manipulability via contractual terms} if, in every constructed market in which $S_k$ is the only hospital and each student $ i $ has a strict preference relation $ \succsim_i $ over contracts and regards only the contracts in $ X_i\cap X_k $ as acceptable, no student can manipulate DA. That is, there do not exist student $ i $ and $ \succsim'_i $ such that $ DA^k_i(\succsim'_i,\succsim_{-i}) \succ_i DA^k_i(\succsim_I) $, where $DA^k$ denotes DA in the constructed market in which $S_k$ is the only hospital.

\begin{lemma}\label{lemma:choicefunction_property}
	Every choice function $ C_k $ satisfies irrelevance of rejected contracts, observable substitutability, observable size monotonicity, and non-manipulability via contractual terms.
\end{lemma}

\begin{proof}
	(Irrelevance of rejected contracts) Consider any $ X'\subseteq X_k $ and any $ (i,b)\in R_k(X') $. There are two cases. If there exists a bundle $ b' $ such that $ (i,b')\in C_k(X') $, then either $ b $ is not a maximal bundle in $ X'_i $, or it is a maximal bundle in $ X'_i $ but $ b'\rhd^* b $. In both cases, removing $ (i,b) $ will not change the procedure of $ C_k(X') $. So $ C_k(X'\backslash (i,b))=C_k(X')$. If there does not exist $ b' $ such that $ (i,b')\in C_k(X') $, it means that in the procedure of $ C_k(X') $, when $ i $ is considered, all maximal bundles in $ X'_i $ do not have vacant seats. Then, removing $ (i,b) $ will not change the procedure of $ C_k(X') $. So $ C_k(X'\backslash (i,b))=C_k(X')$.

	(Observable substitutability and observable size monotonicity) Consider any observable offer process $ \mathbf{x}= (x^1,\ldots,x^M) $. Let $ x^M=(i,b) $. If $ (i,b)\in R_k(\mathbf{x}) $, by irrelevance of rejected contracts, $ R_k(\mathbf{x}^{M-1}) = R_k(\mathbf{x})$  and therefore $ |C_k(\mathbf{x}^{M-1})| = |C_k(\mathbf{x})|$.

	If $ (i,b)\in C_k(\mathbf{x}) $, then $ b $ is a maximal bundle in $ \mathbf{x}_i $, and there does not exist $ (i,b') $ such that $ (i,b')\in C_k(\mathbf{x}^{M-1}) $. Then we compare the procedure of $ C_k(\mathbf{x}^{M-1}) $ and the procedure of $ C_k(\mathbf{x}) $. The two procedures coincide up to the step in which $ i $ is considered. So, for any $ j $ ranked above $ i $, $ (j,b'') \in C_k(\mathbf{x}^{M-1}) $ if and only if $ (j,b'')\in C_k(\mathbf{x}) $. The fact that $ i $ is admitted by the bundle $ b $ in $ C_k(\mathbf{x}) $ but is rejected in $ C_k(\mathbf{x}^{M-1}) $ means that when $ i $ is considered in both procedures, all maximal bundles in $ \mathbf{x}^{M-1}_i $ do not have vacant seats, yet $ b $ has vacant seats. So, compared with the procedure of $ C_k(\mathbf{x}^{M-1}) $ in which $i$ is rejected, in the procedure of $ C_k(\mathbf{x}) $, the quotas of $ b $ and its sup-bundles are reduced by one after $ i $ is admitted by $b$. For the remaining students ranked below $ i $, there are two cases.
	
	\begin{itemize}
		\item Case 1: Every student ranked below $ i $ that is admitted in $ C_k(\mathbf{x}^{M-1}) $ is also admitted in $ C_k(\mathbf{x}) $. Let $j$ be the highest-ranked among such students, and suppose that $ (j,b'')\in C_k(\mathbf{x}^{M-1}) $.  
        In the steps after $i$ is considered but before $j$ is considered, the quotas of bundles in the procedure of $ C_k(\mathbf{x}) $ are weakly smaller than the corresponding quotas in the procedure of $ C_k(\mathbf{x}^{M-1})  $, and when there is a difference, the difference must be one and the relevant bundles must be $ b $ and its sup-bundles. So every rejected student in $ C_k(\mathbf{x}^{M-1})  $ must also be rejected in $ C_k(\mathbf{x}) $. We now argue that, for the student $j$, given $ (j,b'')\in C_k(\mathbf{x}^{M-1}) $, we must have $ (j,b'')\in C_k(\mathbf{x}) $.  Suppose that $ \mathbf{x}_j= (y^1,y^2,\ldots,y^L) $ in the order of the offer process and let $ \mathbf{x}'= (x^1,x^2,\ldots,x^k) $ where $x^{k+1}=y^L$. Since the offer process is observable, $j$ must be rejected by every contract $ y^\ell $ with $ 1\le \ell< L $ in $C_k(\mathbf{x}')$. That is, the quotas of these bundles must have become zero when $ j $ is considered in the procedure of $ C_k(\mathbf{x}') $. Then, when $ j $ is considered in the procedure of $ C_k(\mathbf{x}) $, the quotas of these bundles must also have become zero. So $ j $ can only be admitted by $ y^L $, which must be $b''$. Given this, the same arguments can be inductively applied to the other students ranked below $ j $, and conclude that those rejected in $ C_k(\mathbf{x}^{M-1})  $ must also be rejected in $ C_k(\mathbf{x}) $ and those admitted in $ C_k(\mathbf{x}^{M-1})  $ must be admitted by the same bundle in $ C_k(\mathbf{x}) $.        
        Therefore, $ R_k(\mathbf{x}^{M-1}) = R_k(\mathbf{x})$ and $ |C_k(\mathbf{x}^{M-1})|+1 = |C_k(\mathbf{x})|$. 
		
		\item Case 2: There are students ranked below $ i $ who are admitted in $ C_k(\mathbf{x}^{M-1}) $ but rejected in $ C_k(\mathbf{x}) $. Let $ j $ be the highest-ranked among such students, and suppose that $ (j,b'')\in C_k(\mathbf{x}^{M-1}) $. In the steps after $i$ is considered but before $j$ is considered, the quotas of bundles in the procedure of $ C_k(\mathbf{x}) $ are weakly smaller than the corresponding quotas in the procedure of $ C_k(\mathbf{x}^{M-1})  $, and when there is a difference, the difference must be one and the relevant bundles must be $ b $ and its sup-bundles. Now, $ (j,b'')\notin C_k(\mathbf{x}) $ means that the quota of $ b'' $ has become zero when $ j $ is considered in $ C_k(\mathbf{x}) $. It must be because either the quota of $ b'' $ has become zero or the quota of a sup-bundle $ b^* $ of $ b'' $ has become zero. In the procedure of $ C_k(\mathbf{x}^{M-1})  $, $ (j,b'')\in C_k(\mathbf{x}^{M-1}) $ means that the quota of $ b'' $ and the quotas of its sup-bundles are positive when $j$ is considered. Then, after $ j $ is admitted in the procedure of $ C_k(\mathbf{x}^{M-1})  $, the quota of $ b'' $ or the quota of $ b^* $ must become zero. For the students ranked below $ j $, the procedure of $ C_k(\mathbf{x}^{M-1}) $ and the procedure of  $ C_k(\mathbf{x}) $ coincide again. Therefore, $ C_k(\mathbf{x})= C_k(\mathbf{x}^{M-1})\cup (i,b)\backslash (j,b'')  $. So $ R_k(\mathbf{x}^{M-1})\subset R_k(\mathbf{x}) $ and $ |C_k(\mathbf{x}^{M-1})|= |C_k(\mathbf{x})| $.
	\end{itemize}

	(Non-manipulability via contractual terms) Consider the constructed matching model where $S_k$ is the only hospital and every student $i$ regards the contracts in $X_k\cap X_i$ as acceptable and has a preference relation $\succsim_i$ over such contracts. Given that $C_k$ processes students' applications according to the priority order, $ DA^k(\succsim_I) $ can be obtained by running the serial dictatorship algorithm: in the priority order, once a student is considered, she is assigned to her most preferred bundle that has vacant seats. Then, it is clear that there is no way for any student to manipulate the algorithm.
\end{proof}

Now we are ready to prove the two propositions. Let $ P_i $ be any student $i$'s submitted ROL in our bundle system. In the corresponding matching with contracts model, we define a corresponding preference relation
$ \succsim^{P_i}_i$ over contracts, which places the bundles in $ \cB[P_i] $ above $ \emptyset $ and ranks them in the same order as in $ P_i $. Then, it is easy to see that running bundle-DA in our bundle system is equivalent to running DA in the corresponding matching with contracts model. By  \autoref{lemma:choicefunction_property} and the results of \cite{hatfield2021stability}, DA in the corresponding matching with contracts model is strategy-proof.

\begin{proof}[\normalfont \textbf{Proof of  \autoref{prop:truthtelling:simple}}]
	Let $ \succsim^{P_i}_i $  and $ \succsim^{P'_i}_i $, respectively, denote the preference relation induced by $ P_i $ and $ P'_i $. If $ \nu'(i) P_i \nu(i) $, it means that in the corresponding matching with contracts model, $ i $ can manipulate DA by reporting $ \succsim^{P'_i}_i $ when her true preference relation is $ \succsim^{P_i}_i $, which is a contradiction.	So $ \nu(i) =\nu'(i) $ or $ \nu(i) P_i\nu'(i) $.
\end{proof}

\begin{proof}[\normalfont \textbf{Proof of  \autoref{prop:biggerbundle:simple}}]	
	If $ \nu(i)P_i b $, the procedure of bundle-DA does not change if $ i $ replaces $ b $ with $ b' $ in her ROL. So $ \nu'(i)=\nu(i) $.

	If $ \nu(i)= b $, let $ r $ be the round of bundle-DA in which $ i $ applies to $ b $ and assume $b\in \cB_k$. In the corresponding matching with contracts model, let $ X' $ denote the set of offers other than $ (i,b) $ that $ S_k  $ receives in round $ r $ of DA. Then, $ (i,b)\in C_k(X'\cup (i,b)) $. Now, if $i$ changes to apply to $b'$ in round $r$, by \autoref{lemma:sup-bundle}, $ (i,b')\in C_k(X'\cup (i,b')) $ and $ C_k(X'\cup (i,b))\backslash (i,b)\subseteq C_k(X'\cup (i,b')) $. That is, $i$ is admitted by $b'$ in round $r$, and every other student who is admitted in round $r$ when $i$ applies to $b$ is still admitted by the same bundle. Therefore, in each subsequent round, there are weakly fewer applications to $ S_k $ than the case when $ i $ reports $ b $. So, $ i $ must be admitted by $ b' $ in each subsequent round. Therefore, $ \nu'(i)=b'$. 
	
	If $ b P_i \nu(i) $, let $ \succsim^{P_i}_i $  and $ \succsim^{P'_i}_i $ respectively denote the preference relations induced by $ P_i $ and $ P'_i $ in the corresponding matching with contracts model. If $ \nu'(i) \notin \{b',\nu(i)\} $, then either $ \nu'(i) \succ^{P_i}_i \nu(i) $ or $ \nu(i) \succ^{P'_i}_i \nu'(i) $, both violating strategy-proofness of DA in the matching with contracts model. 
\end{proof}

\section{Definition of Bundle-DA in General Bundle System}\label{appendixbundle-DA:general}
In each round, every unmatched student applies to the highest-ranked bundle in her ROL that has not rejected her. If such a bundle does not exist, the student remains unmatched. In round $r$, we denote the set of students who send new applications by $ N^r $ and call them \textbf{new applicants}; we denote the set of students who were tentatively admitted in the previous round by $ I^{r-1} $. We define $ A^r\equiv I^{r-1}\cup N^r $ and call it the set of \textbf{active} students at the beginning of round $ r $. For each $ i\in A^r $, we use $ S^r_i $ to denote the bundle that either tentatively admitted $ i $ in the previous round or receives $ i $'s application in this round. 
We call a bundle $ b $ \textbf{active} at the beginning of round $ r $ if its remaining quota is positive and there exists a bundle $ b' $ that receives new applications in round $r$ such that $ S_b\cap S_{b'}\neq \emptyset $. We call a school \textbf{active} if it belongs to an active bundle.

\begin{center}
	\textbf{Bundle-DA in General Bundle System}
\end{center}

\paragraph{Round $ r\ge 1 $.} Let each unmatched student apply to the highest-ranked bundle in her ROL that has not rejected her. If such a bundle does not exist, the student remains unmatched.

\begin{itemize}
	\item[--] \textbf{Step 1}: For every active school $ s $, we define 
	\begin{equation*}
		J^r_s\equiv \{i\in A^r:s\in S^r_{i}\}, \text{ and }i_s\equiv \arg\max_{J^r_s}\succ_s.
	\end{equation*}

	In words, $ J^r_s $ is the set of students considered applicants for $ s $, and $ i_s $ is the highest-priority applicant among $ J^r_s $ for $ s $.
	
	We then define the \textbf{set of highest-priority applicants} as
	\begin{align*}
		H^r\equiv\{i\in A^r:i=i_s \text{ for all active }s\in S^r_{i}\}.
	\end{align*}
	In words, $ H^r $ is the set of active students who have the highest priority among all applicants for every active school in the bundle to which they apply.  \autoref{lemma:nonemptyset} below proves that $ H^r $ must be nonempty, and, for every distinct $ i,i'\in H^r $, $ S^r_i\cap S^r_{i'}= \emptyset $.

	Tentatively admit each $ i\in H^r $ to $ S^r_i $. Once a student is admitted by a bundle $ S_b $, we reduce the remaining quota of $ S_b $ and the remaining quota of its every sup-bundle by one. Once a bundle's remaining quota becomes zero, its sub-bundles' remaining quotas are set to zero immediately. After all students in $ H^r $ are tentatively admitted, they are no longer active in this round. A school is no longer active if it belongs to a bundle whose remaining quota becomes zero.

	\item[--] \textbf{Step $ k\ge 2 $}: Update the set of active students and the set of active schools. We then obtain a new set $ H^r $ as defined in Step 1. Tentatively admit each $ i\in H^r $ to $ S^r_i $. After that, they are no longer active in this round. We then update the quotas of bundles as in Step 1. A school is no longer active if it belongs to a bundle whose remaining quota becomes zero. We stop when reaching a step after which $ J^r_s $ is empty for all active school $ s $ or there is no active school.

	\item[--] \textbf{Overdemanded bundle}: In any step, if the sub-bundles of a bundle $ b $ are supposed to admit applicants, but $ b $ does not have enough seats to admit all applicants, we use an exogenous order $ \rhd $ of students to determine the allocation of the remaining seats of $ b $. 
	
	Formally, in any step, given the set $ H^r $, for any active bundle $ S_b $, we define
	\[
	d(S_b)\equiv \{S_{b'}\in \cB:S_{b'}\subsetneq S_b \text{ and }S_{b'}=S^r_i \text{ for some }i\in H^r\}.
	\]
In words, $ d(S_b) $ is the set of sub-bundles of $ S_b $ that are supposed to admit students in this step. If $ d(S_b) $ is nonempty, by  \autoref{lemma:nonemptyset}, all bundles in $ d(S_b) $ must be disjoint. Then we call $ S_b $ \textit{overdemanded} if $q_b< |d(S_b)|$, where $q_b$ is the remaining quota of $S_b$ in that step. We further call $ S_b $ \textbf{maximally overdemanded} if there does not exist a sup-bundle $ S_{b'} $ of $ S_b $ such that $|d(S_b)|-q_b\le |d(S_{b'})|-q_{b'}$. If there exist overdemanded bundles in any step, for each maximally overdemanded bundle $ S_b $, let the bundles in $ d(S_b) $ admit the relevant students in $ H^r $ one by one according to $ \rhd $, until the remaining quota of $ S_b $ is exhausted. After that, the remaining quota of $ S_b $ and the remaining quotas of its sub-bundles become zero, and the remaining quotas of all sup-bundles of $ S_b $ are reduced accordingly.

\item[--] \textbf{Go to next round}: During round $r$, if a student $ i $ was tentatively admitted in the previous round but $ S^r_i $ is not active in this round, then $ i $ is still tentatively admitted by $ S^r_i $ in this round. When round $ r $ ends, we denote the set of students who are tentatively admitted in round $ r $ by $ I^r $. We then go to the next round. The algorithm stops when every student is either tentatively admitted or has been rejected by every bundle reported in her ROL.
 \end{itemize}

\autoref{lemma:nonemptyset} ensures that bundle-DA is well-defined.

\begin{lemma}\label{lemma:nonemptyset}
	In every step of every round $ r $ of bundle-DA, $ H^r $ is nonempty, and for every distinct $ i,i'\in H^r $, $ S^r_i\cap S^r_{i'}= \emptyset $.
\end{lemma}
\begin{proof}[\normalfont \textbf{Proof of  \autoref{lemma:nonemptyset}}]
	We first prove that $ H^r $ is nonempty. Consider any active school $ s $ for which $ J^r_s $ is nonempty. Recall that $ i_{s}\equiv \arg\max_{J^r_s}\succ_s$. If $ i_s \in H^r $, we are done. Otherwise, there must exist some school $ s'\in S^r_{i_{s}} $ such that $ i_{s}\neq i_{s'} $ and $ i_{s'} \succ_{s'}  i_{s} $. Since the bundle system is a hierarchy, either $ S^r_{i_{s}} \subseteq  S^r_{i_{s'}}$ or $ S^r_{i_{s}} \supsetneq  S^r_{i_{s'}}$. If $ S^r_{i_{s}} \subseteq  S^r_{i_{s'}}$, then $ i_{s'}\in J^r_{s} $. Because $ \{s,s'\} \subseteq S^r_{i_{s}} \subseteq  S^r_{i_{s'}} $, $ s $ and $ s' $ must rank $ i_{s}$ and $i_{s'} $ identically in their priority orders. So, $ i_{s'} \succ_{s'}  i_{s} $ implies $ i_{s'} \succ_{s}  i_{s} $, which contradicts the definition of $ i_{s} $. So, it must be that $ S^r_{i_{s}} \supsetneq  S^r_{i_{s'}}$. If $ i_{s'}\in H^r $, we are done. Otherwise, there must exist $ s''\in S^r_{i_{s'}} $ such that $ i_{s'}\neq i_{s''} $ and $ i_{s''} \succ_{s''}  i_{s'} $. By the above argument, $ S^r_{i_{s'}} \supsetneq  S^r_{i_{s''}} $. If $ i_{s''}\in H^r $, we are done. Otherwise, we can repeat the above argument to find another student $ i_{s'''} $, and so on. Since the number of schools is finite, at the end of this process, we must find a student in $ H^r $.
	
	For every two distinct $ i,i'\in H^r $, if $ S^r_i\cap S^r_{i'}\neq \emptyset $, then, for every $ s\in S^r_i\cap S^r_{i'} $, $ \{i,i'\}\subseteq J^r_s $. But this means that at most one of them can have the highest priority for $ s $, which is a contradiction.
\end{proof}

We provide an example to illustrate the procedure of bundle-DA in general bundle systems. Due to its length, the example is placed in the Appendix \ref{appendix:example:bundle-DA}. 

We prove that  \autoref{prop:stability:simple} still holds for bundle-DA in general bundle systems.

\begin{proposition}\label{prop:stability:general}
	 Given a general bundle system, for any ROL profile, the outcome of bundle-DA is a stable and Pareto-undominated size-maximal bundle-matching.
\end{proposition}

\begin{proof}[\normalfont \textbf{Proof of  \autoref{prop:stability:general}}]
    Given a ROL profile $P_I$ in a general bundle system, let $ \nu $ denote the outcome of bundle-DA. We first prove that $\nu$ is stable. First, $ \nu $ is individually rational because every matched student must be assigned to a bundle reported in her ROL.
    
    Second, $\nu$ is non-wasteful. For any student $ i $ and any bundle $b$ such that $b P_i \nu(i)$, $i$ is rejected by $ b $ in some round $ r $ of bundle-DA. This means that all schools in $ S_b $ must become inactive at the end of round $ r $. This is either because the quota of $ S_b $ becomes zero or because the quota of some sup-bundle of $ S_b $ becomes zero in round $r$. In either case, all seats of $S_b$ must be assigned.
    
    Last, $ \nu $ satisfies no justified envy. Still, suppose that a student $ i $ is rejected by a bundle $ S_b $ in some round $ r $.     
    If there exists any student $ j $ who is tentatively admitted by a bundle $S_{b'}$ in round $r$ such that $ S_{b'}\subseteq S_b $, then, for any such $j$, there exists a step of round $r$ in which $ j\in H^r $. So, in that step, for every active $ s\in S_{b'} $, $ j\in J^r_s $. Because $i$ applies to $ S_b $ in round $r$, we also have $ i\in J^r_s $. Then, $ j\in H^r $ means that $ j $ has higher priority than $ i $ for every active $ s\in S_{b'}$. If there exists any student $ j $ who is tentatively admitted to a bundle $S_{b'}$ in round $r$ such that $ S_{b'}\supsetneq S_b $, and, for every $ S_{b''} $ such that $ S_b\subseteq S_{b''} \subsetneq S_{b'} $, $ Q^\nu_{b''}<q_{b''} $, then, for every such $ j $, there exists a step of round $r$ in which $ j\in H^r $. So, in that step, for every active $ s\in S_{b'} $, $j\in J^r_s$. Because for every $ S_{b''} $ such that $ S_b\subseteq S_{b''} \subsetneq S_{b'} $, $ Q^\nu_{b''}<q_{b''} $, it implies that every $ s\in S_b $ is active in that step of round $r$. So, for every active $ s\in S_b $, $ \{i,j\}\subseteq J^r_s $. Then, $ j\in H^r $ means that $ j $ has higher priority than $ i $ for every $ s\in S_b $. This means that $i$ cannot have justified envy towards $j$. Therefore, there is no justified envy in the outcome after the round $r$. Because the priorities of the students admitted to each bundle only weakly increase in subsequent rounds, the final outcome $ \nu $ must not involve any justified envy.

    We then prove that $\nu$ is Pareto-undominated size maximal. Suppose that there exists another bundle-matching $\nu'$ such that $\{i\in I: \nu(i)\in \cB[P_i]\}\subsetneq  \{i\in I: \nu'(i)\in \cB[P_i]\}$ and for all $i\in I$ with $\nu(i)\in \cB[P_i]$, either $\nu'(i) P_i \nu(i)$ or $\nu'(i)=\nu(i)$. Consider any student $i_0$ matched in $\nu'$ but not matched in $\nu$. Compared to $\nu$, $\nu'$ needs to create at least one more vacant seat to admit $i_0$ while ensuring that every matched student in $\nu$ is no worse off. In particular, every matched student in $\nu$ must still be matched in $\nu'$. However, we prove that it is impossible.
    
    Because $\nu$ is non-wasteful and $i_0$ is unmatched in $\nu$, the seats in $S_{\nu'(i_0)}$ must be fully assigned in $\nu$. That is, there exists a bundle $ b $ with $ S_{b}\supseteq S_{\nu'(i_0)} $ such that $ Q^\nu_{b}=q_{b} $.    
    Therefore, the admission of $i_0$ to $\nu'(i_0)$ must crowd out at least one student who potentially occupies a seat in $S_{\nu'(i_0)}$ in the implementation of $\nu$. Let $i_1$ be such a student. So, $S_{\nu(i_1)}\cap S_{\nu'(i_0)}\neq \emptyset$ and $\nu'(i_1) P_{i_1} \nu(i_1)$. Because $\nu$ is non-wasteful, by similar arguments to the above, the seats in $S_{\nu'(i_1)}$ must be fully assigned in $\nu$. So, the admission of $i_1$ to $\nu'(i_1)$ must crowd out at least one student who potentially occupies a seat in $S_{\nu'(i_1)}$ in the implementation of $\nu$. Let $i_2$ be such a student. So, $S_{\nu(i_2)}\cap S_{\nu'(i_1)}\neq \emptyset$ and $\nu'(i_2) P_{i_2} \nu(i_2)$. Repeating the above arguments, we can find another student $i_3$ who is crowded out by $i_2$, and so on. Because the number of students and the number of bundles are both finite, the above search process must reach a student $i_x$ such that $\nu'(i_x) P_{i_x} \nu(i_x)$ and $\nu'(i_x)$ overlaps with a bundle $\nu(i_y)$ for some $y<x$, which has been found in the search process, such that $i_y$ potentially occupies a seat in $S_{\nu'(i_x)}$ in the implementation of $\nu$. If $i_y\neq i_1$, it means that the students $i_y,i_{y+1},\ldots,i_x$ essentially exchange their seats in $\nu$ to obtain their new seats in $\nu'$, the process of which does not create any vacant seat to admit $i_{y-1}$. Then, we can exclude these students and find another matched student $i'_y$ in $\nu$ such that $S_{\nu(i'_y)}\cap S_{\nu'(i_{y-1})}\neq \emptyset$ and $\nu'(i'_y) P_{i'_y} \nu(i'_y)$ to initiate the above search process. If $i_y=i_1$, it means that the students $i_1,i_2,\ldots,i_x$ essentially exchange their seats in $\nu$ to obtain their new seats in $\nu'$, the process of which does not create any vacant seat to admit $i_0$. Then, we can exclude these students and find another matched student $i'_1$ in $\nu$ such that $S_{\nu(i'_1)}\cap S_{\nu'(i_0)}\neq \emptyset$ and $\nu'(i'_1) P_{i'_1} \nu(i'_1)$ to initiate the above search process. However, in either case, similar arguments to the above would lead to the same conclusion that the obtained sequence of students essentially exchange their seats in $\nu$ to obtain their seats in $\nu'$, without creating any vacant seats to admit other students. This is a contradiction.
\end{proof}

However, the remaining properties of bundle-DA in simple bundle systems may not hold in general bundle systems, due to the potential use of a tie-breaking rule when there exist overdemanded bundles. For instance,  \autoref{example:bigger_bundle:worse_off} illustrates a scenario where a matched student becomes unmatched when she reports a larger bundle. Since we impose a minimal set of conditions on a general bundle system, there does not exist a uniform approach to choosing an appropriate tie-breaking rule to avoid the situation in the example. In practical applications, policymakers may design bundle systems with more structured features to address the specific needs of the environment and enable bundle-DA to possess desirable properties. The class of simple bundle systems discussed in the paper is an example.

\begin{example}\label{example:bigger_bundle:worse_off}
	Consider two schools $ \{s_1, s_2\} $ and three students $ \{i_1,i_2,i_3\} $. Each school has one seat and uses the following priority order. There is a bundle $ S_b=\{s_1,s_2\} $ that targets $ \{i_1,i_3\} $. We use the order $ i_3\rhd  i_1\rhd i_2$ to break ties in bundle-DA when necessary.

 \begin{table}[!htb]
			\centering
			   %\caption{Priority and ROL}\label{table:example2}
			\begin{subtable}{.4\linewidth}
				\centering
             \subcaption{Priority orders}\label{table:example2:priority}
				\begin{tabular}{cc}
			$ \succsim_{s_1} $ & $ \succsim_{s_2} $  \\ \hline
			$ i_3 $ & $ i_3$   \\
			$ i_1 $ & $ i_2 $ \\
			$ i_2 $ & $ i_1 $ 			
		\end{tabular}
			\end{subtable}
			\begin{subtable}{.4\linewidth}
				\centering
    \subcaption{ROL profile}\label{table:example2:ROL}
				\begin{tabular}{ccc}
			$ P_{i_1} $ & $ P_{i_2} $ &  $ P_{i_3} $ \\ \hline
			$ s_1 $ & $ s_2$ & $ \{s_1,s_2\} $   \\
			 \\
          & &
		\end{tabular}
			\end{subtable}
		\end{table}
    Suppose that the ROL length is one and students report the above ROLs. In bundle-DA, $ i_3 $ is first admitted by $ S_b $. Then, $ i_1 $ and $ i_2 $ compete for the remaining seat in $ S_b $. According to the tie-breaking order, $ i_1 $ wins and is admitted by $ s_1 $, while $i_2$ is unmatched. 
    
    If $ i_1 $ changes to report $ S_b $, then $i_1$ will lose the competition with $i_2$ for the remaining seat in $S_b$ after $i_3$ is admitted. This is because now $ i_1 $ is regarded as an applicant for both $ s_1 $ and $ s_2 $, yet $ i_2 $ has higher priority than $ i_1 $ for $ s_2 $. So $i_2$ is admitted by $ s_2 $, and $ i_1 $ is unmatched.
\end{example}

\section{Example of Bundle-DA in General Bundle System}\label{appendix:example:bundle-DA}

We present an example to illustrate the procedure of bundle-DA in a general bundle system.

\begin{example}\label{example:illustration:bundle-DA}
    
Consider five schools $ \{s_1,s_2,s_3,s_4,s_5\} $ and eight students $ \{i_1,\ldots,i_8\} $. School quotas are $q_{s_1}=q_{s_2}=q_{s_3}=2$ and $q_{s_4}=q_{s_5}=1$. Schools use the priority orders in  \autoref{table:example:general:priority}. There are two bundles, $ S_b=\{s_2,s_3\} $ that targets $ \{i_2,i_3,i_4,i_5,i_6,i_7,i_8\} $ and $ S_{b'} =\{s_1,s_2,s_3\} $ that targets $ \{i_5,i_8\} $. The ROL length is two. Suppose that students report the ROLs in  \autoref{table:example:general:ROL}. Then, bundle-DA runs as follows.

\begin{table}[ht!]
		\centering
          \caption{Priority structure}\label{table:example:general:priority}
		\begin{tabular}{ccccc}
			%\hline
			$ \succsim_{s_1} $ & $ \succsim_{s_2} $ &  $ \succsim_{s_3} $ & $ \succsim_{s_4} $ & $ \succsim_{s_5} $ \\ \hline
			 $ i_7 $ & $ i_1 $  & $ i_1 $ & $ i_6 $ & $i_7$ \\
           $i_4$ & $i_5$ & $i_5$ & $i_8$ & $i_8$\\
           $i_5$ & $i_8$ & $i_8$ & $i_5$ & $i_5$\\
           $i_8$ & $i_2$ & $i_2$ & $i_2$ & $i_2$\\
           $i_2$ & $i_3$ & $i_3$ & $i_3$ & $i_3$\\
           $i_3$ & $i_7$ & $i_7$ & $i_7$ & $i_6$\\
           $i_6$ & $i_6$ & $i_6$ & $i_4$ & $i_4$\\
           $i_1$ & $i_4$ & $i_4$ & $i_1$ & $i_1$
		\end{tabular}
	\end{table}
	
	\begin{table}[!ht]
		\centering
  \caption{ROL Profile}\label{table:example:general:ROL}
		\begin{tabular}{cccccccc}
			$ i_1 $ & $ i_2 $ &  $ i_3 $ & $ i_4 $ & $ i_5 $ & $ i_6 $ & $ i_7 $ & $ i_8 $ \\ \hline
			$ s_2 $ & $ \{s_2,s_3\} $ & $ \{s_2,s_3\} $  & $ s_1 $  & $ \{s_1,s_2,s_3\} $ & $ \{s_2,s_3\} $ & $ \{s_2,s_3\} $ & $ s_4 $  \\
			$ s_1 $ & $ s_4 $ & $ s_1 $  & $ \{s_2,s_3\} $ & $ s_5 $ & $ s_4 $ & $ s_1 $ & $ \{s_1,s_2,s_3\} $	
		\end{tabular}
	\end{table}

	\begin{itemize}
		\item Round 1: All students apply to their first choice. Then, in step 1,
  \[
J^1_{s_1}=\{i_4,i_5\},  J^1_{s_2}=\{i_1,i_2,i_3,i_5,i_6,i_7\}, J^1_{s_3}=\{i_2,i_3,i_5,i_6,i_7\}, J^1_{s_4}=\{i_8\}.
  \]
  Hence, $H^1=\{i_1,i_4,i_8\}$. So, these three students are admitted by $ s_2 $, $ s_1 $ and $ s_4 $ respectively, and then become inactive. Then relevant quotas are updated: $ q_{s_1}=1 $, $ q_{s_2}=1 $, $ q_{s_4}=0 $, $ q_{S_b}=3 $, $ q_{S_{b'}}=4 $.
	    
  In step 2, $ H^1=\{i_5\} $. So, $ i_5 $ is tentatively admitted by $ S_{b'} $ and then becomes inactive. The quota of $ S_{b'} $ becomes $ q_{S_{b'}}=3 $.
		
  In step 3, $ H^1=\{i_2\} $. So, $ i_2 $ is tentatively admitted by $ S_b $ and then becomes inactive. Then relevant quotas are updated: $ q_{S_b}=q_{S_{b'}}=2 $. 
		
  In step 4, $ H^1=\{i_3\} $. So, $ i_3 $ is tentatively admitted by $ S_b $ and then becomes inactive. Then, relevant quotas are updated: $ q_{S_b}=q_{S_{b'}}=1 $. 
		
  In step 5, $ H^1=\{i_7\} $. So, $ i_7 $ is tentatively admitted by $ S_b $ and then becomes inactive. Then, relevant quotas are updated: $ q_{S_b}=q_{S_{b'}}=0 $. After this step, the two bundles and their schools become inactive.
  
        Hence, $ i_6 $ is rejected in round 1. The tentative assignment found by this round is:
		\begin{table}[!ht]
			\centering
			\begin{tabular}{cccccccc}
   \hline
				$ i_1 $ & $ i_2 $ &  $ i_3 $ & $ i_4 $ & $ i_5 $ & $ i_6 $ & $ i_7 $ & $ i_8 $ \\ 
				$ s_2 $ & $ S_b $ & $ S_b $  & $ s_1 $  & $ S_{b'} $ & $ \emptyset$ & $ S_b $ & $ s_4 $\\
		\hline	\end{tabular}
		\end{table}
			
		\item Round 2: $ i_6 $ applies to $ s_4 $. Then, in step 1, $J^2_{s_4}=\{i_6,i_8\}$.		
		Because $ i_6 $ has higher priority than $ i_8 $ for $ s_4 $, $ i_6 $ is tentatively accepted and $ i_8 $ is rejected. So, the tentative assignment found by this round is:
		\begin{table}[!ht]
			\centering
			\begin{tabular}{cccccccc}
   \hline
				$ i_1 $ & $ i_2 $ &  $ i_3 $ & $ i_4 $ & $ i_5 $ & $ i_6 $ & $ i_7 $ & $ i_8 $ \\
				$ s_2 $ & $ S_b $ & $ S_b $  & $ s_1 $  & $ S_{b'} $ & $ s_4 $ & $ S_b $ & $ \emptyset$ \\
    \hline			\end{tabular}
		\end{table}
		
		\item Round 3: $ i_8 $ applies to $ S_{b'} $. In step 1, 		\[
	    J^3_{s_1}=\{i_4,i_5,i_8\}, J^3_{s_2}=\{i_1,i_2,i_3,i_5,i_7,i_8\}, J^3_{s_3}=\{i_2,i_3,i_5,i_7,i_8\}.
		\]
	So, $H^3=\{i_1,i_4\}$. The two students are tentatively admitted by $s_2$ and $s_1$ respectively. 
 
 In step 2, $H^3=\{i_5\}$. So $i_5$ is tentatively admitted by $S_{b'}$.
 
 In step 3, $H^3=\{i_8\}$. So $i_8$ is tentatively admitted by $S_{b'}$.
 
 In step 4, $H^3=\{i_2\}$. So $i_2$ is tentatively admitted by $S_{b}$. 

 In step 5, $H^3=\{i_3\}$. So $i_3$ is tentatively admitted by $S_{b}$. 
 
 After step 5, the quota of $S_{b'}$ becomes zero. So the quota of $S_b$ is also set to zero. Then, $ i_7 $ is rejected by $S_b$. The tentative assignment found by this round is: 
		\begin{table}[!ht]
			\centering
			\begin{tabular}{cccccccc}
   \\ \hline
			  $ i_1 $ & $ i_2 $ &  $ i_3 $ & $ i_4 $ & $ i_5 $ & $ i_6 $ & $ i_7 $ & $ i_8 $ \\ 
			  $ s_2 $ & $ S_b $ & $ S_b $  & $ s_1 $  & $ S_{b'} $ & $ s_4$ & $ \emptyset $ & $ S_{b'}  $\\
			\hline
   \end{tabular}
		\end{table}
		
		\item Round 4: $ i_7 $ applies to $ s_1 $. In step 1,
		\[
		J^4_{s_1}=\{i_4,i_5,i_7,i_8\}, J^4_{s_2}=\{i_1,i_2,i_3,i_5,i_8\}, J^4_{s_3}=\{i_2,i_3,i_5,i_8\}.
		\]
  So, $H^4=\{i_1,i_7\}$. The two students are tentatively admitted by $s_2$ and $s_1$ respectively. 
  In step 2, $ i_4 $ is tentatively admitted by $s_1$.  
  In step 3, $ i_5 $ is tentatively admitted by $S_{b'}$.   
  In step 4, $i_8$ is tentatively admitted by $S_{b'}$.   
  In step 5, $ i_2 $ is tentatively admitted by $S_b$.
  
  After step 5, the quota of $S_{b'}$ becomes zero. So $ i_3 $ is rejected. The tentative assignment found by this round is: 
		\begin{table}[!ht]
			\centering
	\begin{tabular}{cccccccc}
 	\hline	
			  $ i_1 $ & $ i_2 $ &  $ i_3 $ & $ i_4 $ & $ i_5 $ & $ i_6 $ & $ i_7 $ & $ i_8 $ \\ 
			  $ s_2 $ & $ S_b $ & $ \emptyset $  & $ s_1 $  & $ S_{b'} $ & $ s_4$ & $ s_1 $ & $ S_{b'}  $\\
			\hline
   \end{tabular}
		\end{table}
		
		\item Round 5: $ i_3 $ applies to $ s_1 $. But because her priority is lower than $i_4$ for $s_1$, $ i_3 $ is rejected. Because there are no remaining options in her ROL, $ i_3 $ is determined to be unmatched. 

        So the outcome of bundle-DA is:
		\begin{table}[!ht]
			\centering
			\begin{tabular}{cccccccc}
   \\ \hline
				$ i_1 $ & $ i_2 $ &  $ i_3 $ & $ i_4 $ & $ i_5 $ & $ i_6 $ & $ i_7 $ & $ i_8 $ \\ 
				$ s_2 $ & $ S_b$ & $ \emptyset $  & $ s_1 $  & $ S_{b'} $ & $ s_4$ & $ s_1 $ & $  S_{b'} $
			\\
   \hline
   \end{tabular}
		\end{table}
  \end{itemize}

	To implement this outcome, students admitted by individual schools are assigned to those schools. Then, since there is only one remaining seat in $ S_b $, which belongs to $ s_2 $, $ i_2 $ is assigned to $ s_2 $. After that, only $ s_3 $ has remaining seats. So $ i_5,i_8 $ are assigned to $ s_3 $. Therefore, the final assignment is unique:
	 \begin{table}[!ht]
	 	\centering
	 	\begin{tabular}{cccccccc}
   \hline
	 		$ i_1 $ & $ i_2 $ &  $ i_3 $ & $ i_4 $ & $ i_5 $ & $ i_6 $ & $ i_7 $ & $ i_8 $ \\ 
	 		$ s_2 $ & $ s_2 $ & $ \emptyset $  & $ s_1 $  & $ s_3 $ & $ s_4$ & $ s_1 $ & $ s_3 $\\
    \hline
	 	\end{tabular}
	 \end{table}

 \end{example}

\clearpage

\setcounter{page}{1}

\section*{Online Appendix}

This online appendix includes the additional figures for Experiment 1 (Section \ref{appendix:additionalresults}), the instructions for Experiment 1 (Section \ref{appendix:instructions-exp1}) and the details of Experiment 2 (Section \ref{appendix:experiment2}).

\section{Additional Figures}\label{appendix:additionalresults}

\setcounter{figure}{0}
\setcounter{table}{0}
\renewcommand\thetable{E\arabic{table}}
\renewcommand\thefigure{E\arabic{figure}}

\begin{figure}[H]
    \begin{center}
		\caption{Student's average payoff over round in Experiment 1}
		\label{fig:student-payoff-exp1}
		\includegraphics[width=0.7\linewidth]{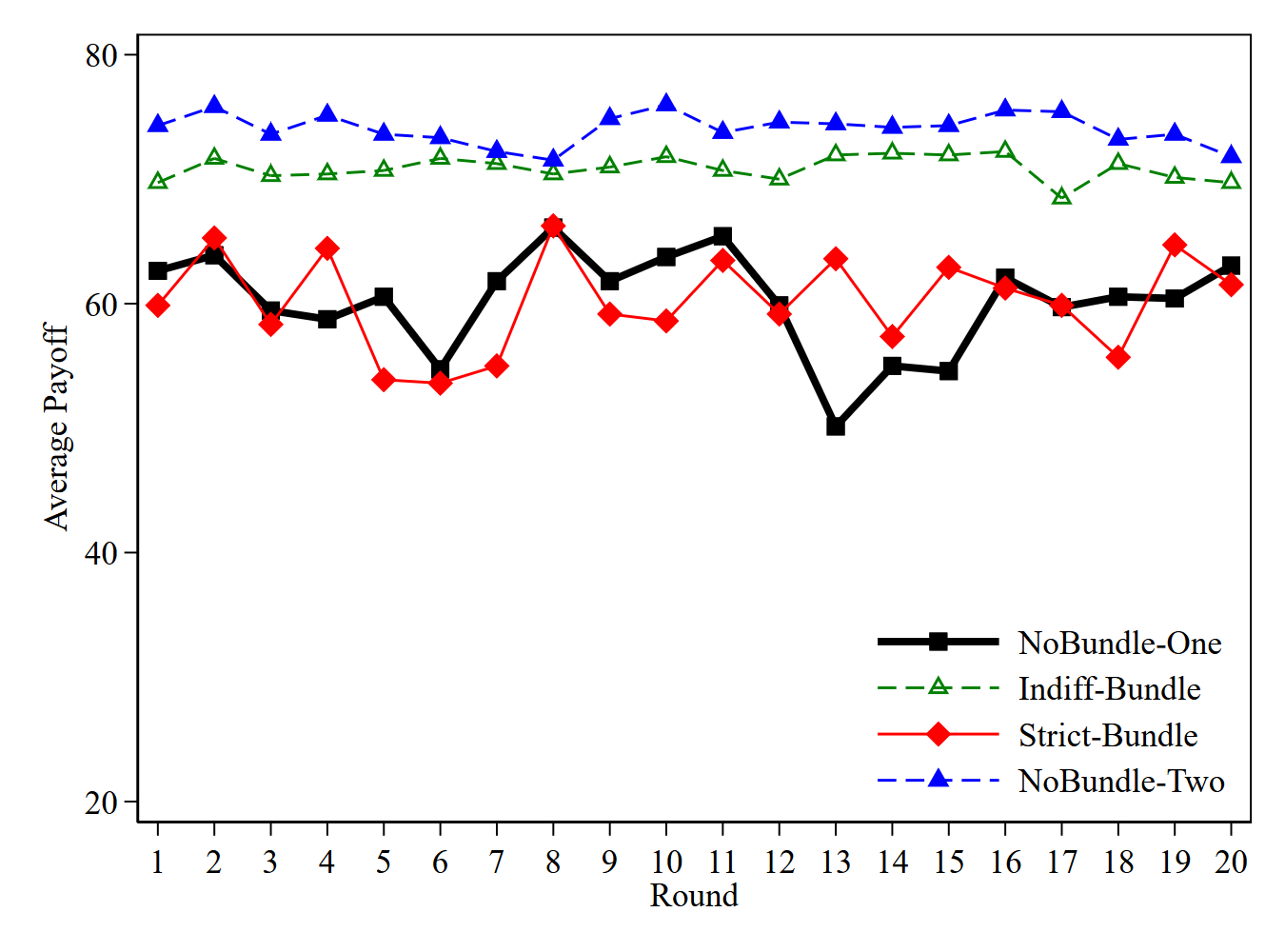}	
    \end{center}
\end{figure}

\begin{figure}[H]
    \begin{center}
		\caption{Match rate over round in Experiment 1}
		\label{fig:match-rate-exp1}
		\includegraphics[width=0.7\linewidth]{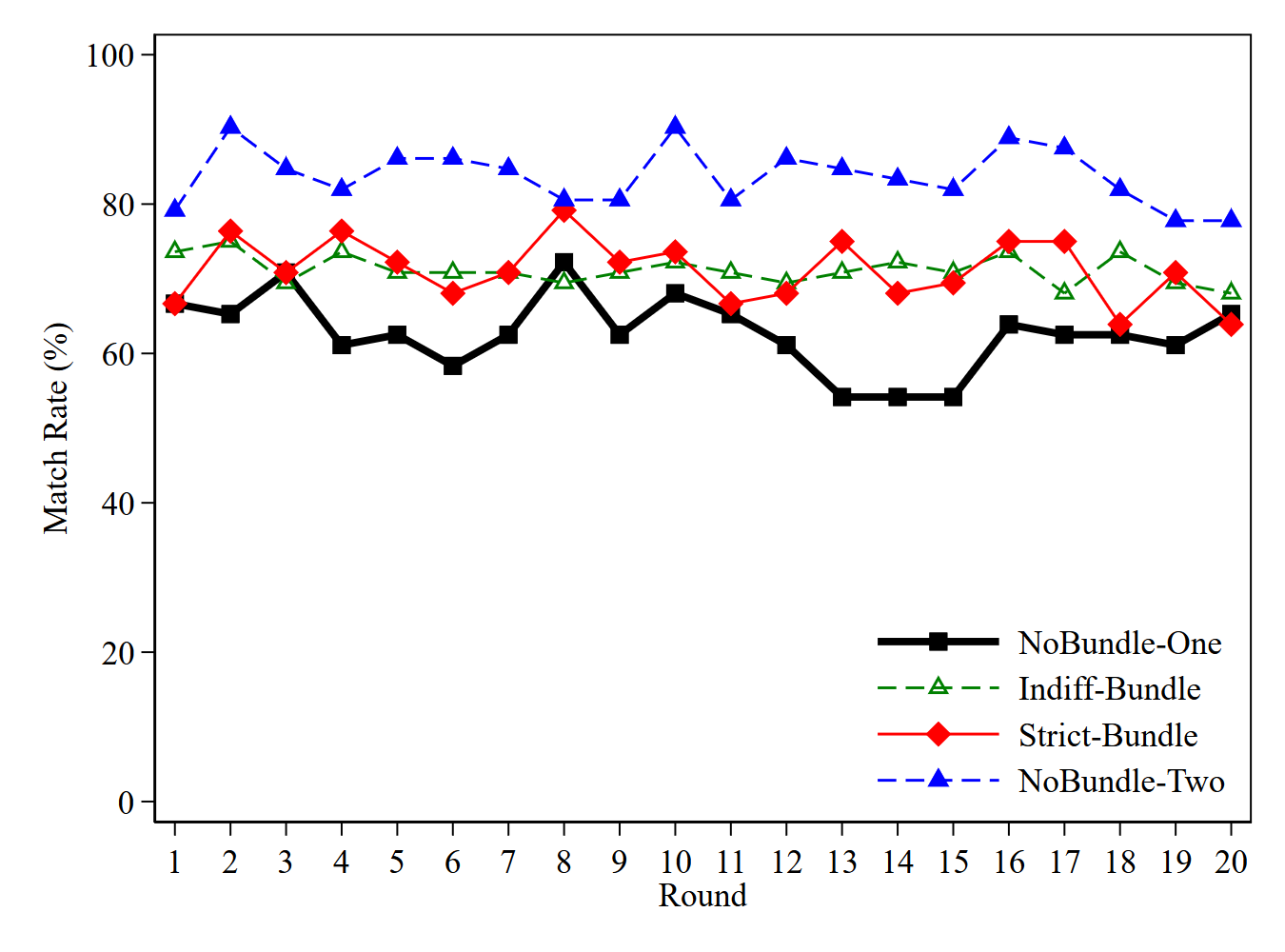}	
    \end{center}
\end{figure}

\begin{figure}[H]
    \begin{center}
		\caption{Mismatch rate over round in Experiment 1}
		\label{fig:mismatch-rate-exp1}
		\includegraphics[width=0.7\linewidth]{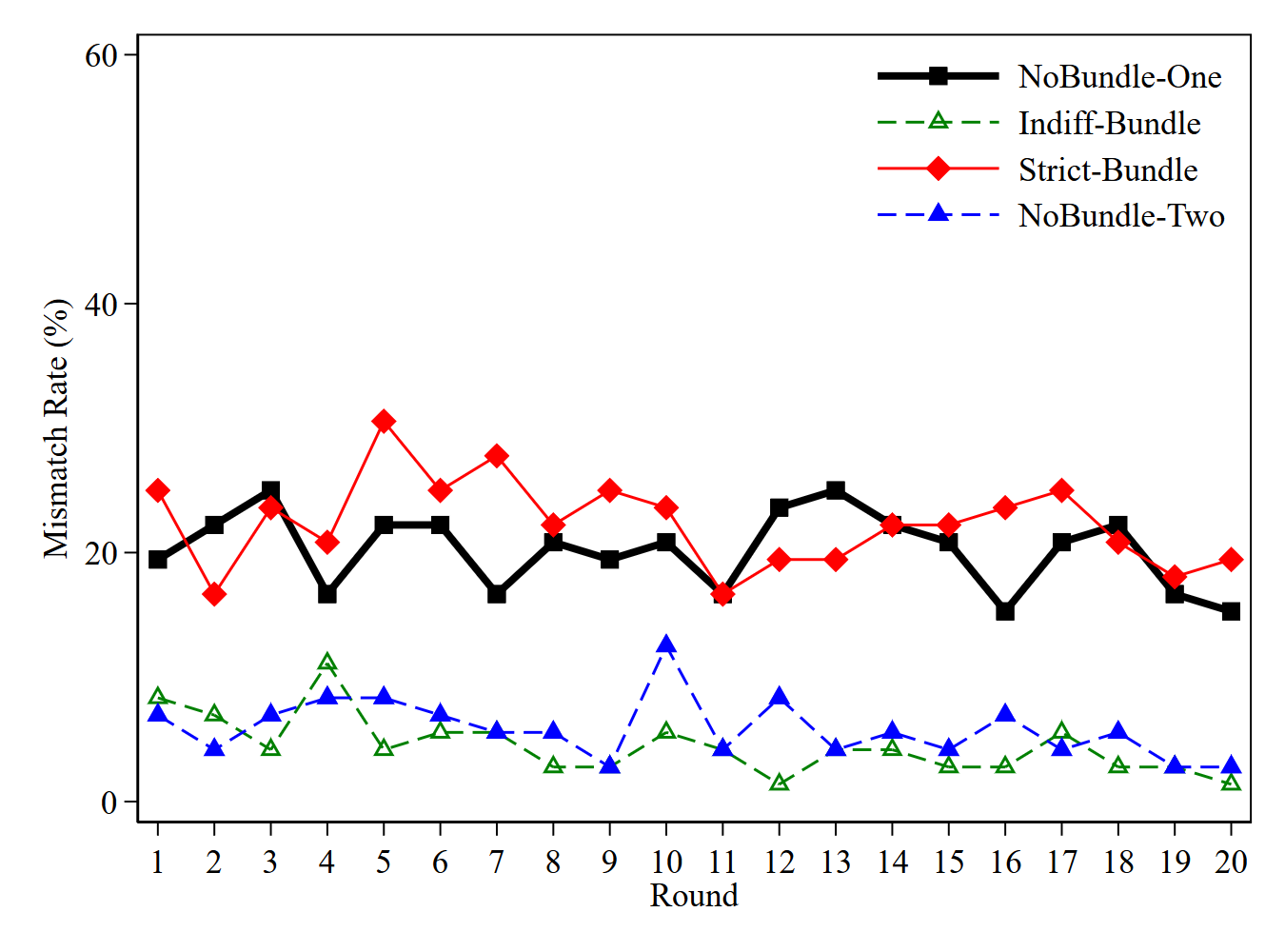}	
    \end{center}
\end{figure}

\begin{figure}[H]
    \begin{center}
		\caption{Bundle report rate over round in Experiment 1}
		\label{fig:bundle-report-rate-exp1}
		\includegraphics[width=0.7\linewidth]{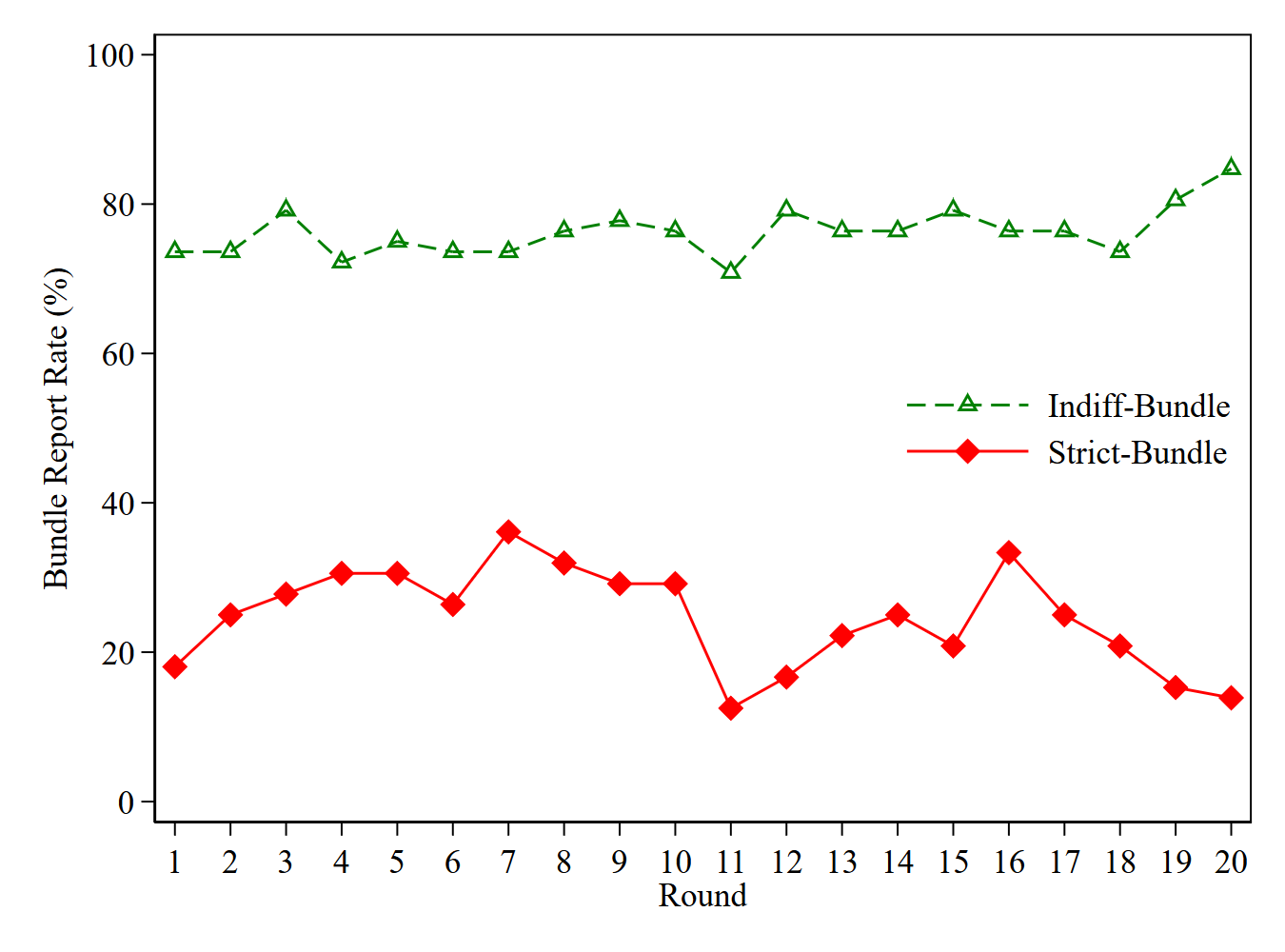}	
    \end{center}
\end{figure}

\clearpage

\section{Instructions for Experiment 1}\label{appendix:instructions-exp1}

In the following, we translate the original instructions in Chinese into English for the Indiff-Bundle, NoBundle\_One and NoBundle\_Two treatments. The instructions for the Strict-Bundle treatment are omitted because they are exactly the same as the instructions for the Indiff-Bundle treatment except for the different schools in the bundle.

\subsection*{Instructions for the Indiff-Bundle treatment}

\textbf{General Information}

You are taking part in a decision-making experiment. Please read the instructions carefully. The instructions are the same for all participants. Please do not communicate with one another during the experiment. Turn off your mobile phone and set it to silent mode. If you have any questions, please raise your hand and an experimenter will come to assist you.

You have earned 15 RMB for showing up on time. In addition, you can earn more money during the experiment. The amount of money you earn will depend on the decisions you make. Your earnings in this experiment are expressed in EXPERIMENTAL CURRENCY UNITS (ECUs), with the conversion rate of 1 ECU = 1 RMB. Your final payment will be transferred to the bank account linked to your university within five working days after the experiment.

All decisions in this experiment are completely anonymous. Your name will remain strictly confidential, and other participants will not know your total payment for today’s experiment.

In today’s experiment, we will simulate the college admission process. You and the other participants will play the roles of students.

\textbf{Experimental Steps}

\begin{itemize}
    \item The experiment consists of 20 rounds of decision-making. Before each round begins, you will be randomly assigned to a group of three participants. Each participant in the group will play the role of a student. Each group faces three different schools, labeled A, B, and C. Each school has one admission slot, and each slot can admit one student. The admission process will be simulated by a computer.
    \item In each round, participants will be randomly reassigned into new groups, so it is unlikely that you will be grouped with the same two participants again. Each student will be randomly assigned an ID number within the group, and IDs will be reassigned in each round.
    \item In each round, you will see the following payoff table, which shows the payoff you would receive if admitted by each school. These payoffs represent your preferences for different schools. Your payoff in each round depends on which school admits you.
        \begin{table}[H]
            \centering
            \begin{tabular}{|l|c|c|c|c|}
                \hline
                Admitting school & A & B & C & Not admitted \\
                \hline
                Payoff (ECU) & X & Y & Z & 0 \\
                \hline
            \end{tabular}
        \end{table}
        \begin{itemize}
            \item If you are admitted by School A, your payoff in that round is X ECUs.
            \item If you are admitted by School B, your payoff in that round is Y ECUs.
            \item If you are admitted by School C, your payoff in that round is Z ECUs.
            \item If you are not admitted by any school, your payoff in that round is 0.
        \end{itemize}
    \item In each round, the payoff table for each student is randomly generated by the computer according to the following rule:
        \begin{itemize}
            \item With probability $1/2$, the payoff table is:
            \begin{table}[H]
                \centering
                \begin{tabular}{|l|c|c|c|c|}
                    \hline
                    Admitting school & A & B & C & Not admitted \\
                    \hline
                    Payoff (ECU) & 110 & 100 & 20 & 0 \\
                    \hline
                \end{tabular}
            \end{table}
            \item With probability $1/2$, the payoff table is:
            \begin{table}[H]
                \centering
                \begin{tabular}{|l|c|c|c|c|}
                    \hline
                    Admitting school & A & B & C & Not admitted \\
                    \hline
                    Payoff (ECU) & 100 & 110 & 20 & 0 \\
                    \hline
                \end{tabular}
            \end{table}
        \end{itemize}
    \item Therefore, in the first type of table, School A yields the highest payoff, while in the second type, School B yields the highest payoff. In each round, the payoff tables for students are independently and randomly generated in this way. That is, regardless of your payoff table, each of the other students in your group has a 1/2 probability of receiving either type of payoff table.
\end{itemize}

\textbf{Admission Rules}

\begin{itemize}
    \item \textbf{Lottery Draw:} In each round, the computer conducts a lottery draw for the three students in each group. Each student is randomly assigned a unique number between 1 and 3.
    
    Note: The lottery number is independent of the student ID. In each round, both the ID and the lottery draw are regenerated.
    \item \textbf{Student Ranking:} In each round, the computer ranks the three students in each group according to the lottery results. The smaller the lottery number, the higher the student’s rank.
    \item \textbf{Preference Submission:} In each round, each student submits one preference. Schools A and B not only admit students individually, but also form a joint admission program labeled AB. When submitting preferences, each student can choose to apply to one of the individual schools (A, B, or C) or to the joint program AB. If a student applies to AB, this means that the student is willing to be admitted by either A or B and accepts whichever school they are assigned to. Note that the total number of admission slots for Schools A and B combined is two, regardless of whether students apply individually or jointly.
    
    Note: Students submit their preferences before the lottery draw, and thus do not know their ranking when filling out their preferences.
    \item \textbf{Admission Mechanism:} After all students have submitted their preferences, the computer processes the admissions in the order of the student ranking determined by the lottery.
        \begin{itemize}
            \item \textbf{Step 1:} Consider the first-ranked student (S1). S1 is always admitted to their chosen preference.
                \begin{itemize}
                    \item If S1 chooses C, then C’s admission slot becomes 0.
                    \item If S1 chooses A or B, that school’s slot becomes 0, and the total combined slots for A and B decrease from 2 to 1.
                    \item If S1 chooses AB, then S1 is jointly admitted by A and B, but the final assignment will depend on the subsequent admission results. The total combined slots for A and B decrease from 2 to 1.
                \end{itemize}
            \item \textbf{Step 2:} Consider the second-ranked student (S2).
                \begin{itemize}
                    \item If S2 chooses C and C still has a slot, S2 is admitted to C and C’s slot becomes 0. Otherwise, S2 is rejected.
                    \item If S2 chooses A or B and that school still has a slot, S2 is admitted to that school, the slot becomes 0, and the total combined slots for A and B decrease by 1. Otherwise, S2 is rejected.
                    \item If S2 chooses AB and the combined slots for A and B are greater than 0, S2 is jointly admitted to A and B, and the total combined slots for A and B decrease by 1. Otherwise, S2 is rejected.
                \end{itemize}
            \item \textbf{Step 3:} Consider the third-ranked student (S3).
                \begin{itemize}
                    \item If S3 chooses C and C still has a slot, S3 is admitted to C. Otherwise, S3 is rejected.
                    \item If S3 chooses A or B and that school has a slot while the total slots for A and B are greater than 0, S3 is admitted to that school. Otherwise, S3 is rejected.
                    \item If S3 chooses AB and the combined slots for A and B are greater than 0, S3 is jointly admitted to A and B. Otherwise, S3 is rejected.
                \end{itemize}
            \item \textbf{Final Assignment:} After all admissions are processed, all students admitted by an individual school remain assigned to that school. For all students admitted through AB, their final assignment depends on which of A or B still has remaining slots:
                \begin{itemize}
                    \item If only one of A or B has a remaining slot, at most one student can be admitted through AB, and that student will be assigned to the school with the remaining slot.
                    \item If both A and B still have slots, and at most two students are admitted through AB, they will be randomly assigned between A and B.
                \end{itemize}
        \end{itemize}
\end{itemize}

\textbf{Examples}

To further explain the admission rules, consider the following three scenarios.

\textbf{Lottery Draw:}
\begin{table}[H]
    \centering
    \begin{tabular}{|l|c|c|c|}
        \hline
        Student ID & 1 & 2 & 3 \\
        \hline
        Lottery number & 3 & 1 & 2 \\
        \hline
    \end{tabular}
\end{table}

\textbf{Ranking of Students:}
\begin{table}[H]
    \centering
    \begin{tabular}{|l|c|c|c|}
        \hline
        Rank & 1st & 2nd & 3rd \\
        \hline
        Student ID & 2 & 3 & 1 \\
        \hline
    \end{tabular}
\end{table}

\textbf{Scenario 1:}
\begin{table}[H]
    \centering
    \begin{tabular}{|l|c|c|c|}
        \hline
        Student ID & 1 & 2 & 3 \\
        \hline
        Preference & AB & A & A \\
        \hline
    \end{tabular}
\end{table}

\textbf{Admission Process:}
\begin{itemize}
    \item Step 1: Student 2 (rank 1) is admitted to School A. A’s slot becomes 0, and the total slots for A and B decrease to 1.
    \item Step 2: Student 3 (rank 2) also applies to A, but A’s slot is 0, so Student 3 is rejected.
    \item Step 3: Student 1 (rank 3) applies to AB. Since the total slots for A and B are still greater than 0, Student 1 is jointly admitted to AB.
\end{itemize}

\textbf{Final Assignment:} Because School A has already admitted Student 2, Student 1 (admitted through AB) is assigned to School B.

\textbf{Final Admission Results:}
\begin{table}[H]
    \centering
    \begin{tabular}{|l|c|c|c|}
        \hline
        Student ID & 1 & 2 & 3 \\
        \hline
        Admitted school & B & A & Not admitted \\
        \hline
    \end{tabular}
\end{table}

\textbf{Scenario 2:}
\begin{table}[H]
    \centering
    \begin{tabular}{|l|c|c|c|}
        \hline
        Student ID & 1 & 2 & 3 \\
        \hline
        Preference & B & A & AB \\
        \hline
    \end{tabular}
\end{table}

\textbf{Admission Process:}
\begin{itemize}
    \item Step 1: Student 2 (rank 1) is admitted to School A. A’s slot becomes 0, and the total slots for A and B decrease to 1.
    \item Step 2: Student 3 (rank 2) applies to AB. Since the total slots for A and B are still greater than 0, Student 3 is jointly admitted to AB, and the total slots for A and B decrease to 0.
    \item Step 3: Student 1 (rank 3) applies to School B, but since the total slots for A and B are already 0, Student 1 is rejected.
\end{itemize}

\textbf{Final Assignment:} Because School A has already admitted Student 2, Student 3 (admitted through AB) is assigned to School B.

\textbf{Final Admission Results:}
\begin{table}[H]
    \centering
    \begin{tabular}{|l|c|c|c|}
        \hline
        Student ID & 1 & 2 & 3 \\
        \hline
        Admitted school & Not admitted & A & B \\
        \hline
    \end{tabular}
\end{table}

\textbf{Scenario 3:}
\begin{table}[H]
    \centering
    \begin{tabular}{|l|c|c|c|}
        \hline
        Student ID & 1 & 2 & 3 \\
        \hline
        Preference & AB & AB & C \\
        \hline
    \end{tabular}
\end{table}

\textbf{Admission Process:}
\begin{itemize}
    \item Step 1: Student 2 (rank 1) is admitted through AB. The total slots for A and B decrease to 1.
    \item Step 2: Student 3 (rank 2) applies to School C and is admitted. C’s slot becomes 0.
    \item Step 3: Student 1 (rank 3) applies to AB. Since the total slots for A and B are greater than 0, Student 1 is admitted through AB.
\end{itemize}

\textbf{Final Assignment:} Since Schools A and B have not admitted any students individually, Students 1 and 2 (both admitted through AB) are randomly assigned to Schools A and B respectively.

\textbf{Final Admission Results:}
\begin{table}[H]
    \centering
    \begin{tabular}{|l|c|c|c|}
        \hline
        Student ID & 1 & 2 & 3 \\
        \hline
        Admitted school & A or B & A or B & C \\
        \hline
    \end{tabular}
\end{table}

\textbf{Payoff Rules}

After each round, you will be informed whether you have been admitted and which school admitted you, as well as your payoff. Your result in each round is independent of the outcomes in other rounds.

After all 20 rounds are completed, one round will be randomly selected as your final payoff. Additionally, you will receive 15 RMB as a show-up fee. Finally, you may earn extra income from a post-experiment questionnaire.

\subsection*{Instructions for the NoBundle\_One treatment}

\textbf{General Information}

You are taking part in a decision-making experiment. Please read the instructions carefully. The instructions are the same for all participants. Please do not communicate with one another during the experiment. Turn off your mobile phone and set it to silent mode. If you have any questions, please raise your hand and an experimenter will come to assist you.

You have earned 15 RMB for showing up on time. In addition, you can earn more money during the experiment. The amount of money you earn will depend on the decisions you make. Your earnings in this experiment are expressed in EXPERIMENTAL CURRENCY UNITS (ECUs), with the conversion rate of 1 ECU = 1 RMB. Your final payment will be transferred to the bank account linked to your university within five working days after the experiment.

All decisions in this experiment are completely anonymous. Your name will remain strictly confidential, and other participants will not know your total payment for today’s experiment.

In today’s experiment, we will simulate the college admission process. You and the other participants will play the roles of students.

\textbf{Experimental Steps}

\begin{itemize}
    \item The experiment consists of 20 rounds of decision-making. Before each round begins, you will be randomly assigned to a group of three participants. Each participant in the group will play the role of a student. Each group faces three different schools, labeled A, B, and C. Each school has one admission slot, and each slot can admit one student. The admission process will be simulated by a computer.
    \item In each round, participants will be randomly reassigned into new groups, so it is unlikely that you will be grouped with the same two participants again. Each student will be randomly assigned an ID number within the group, and IDs will be reassigned in each round.
    \item In each round, you will see the following payoff table, which shows the payoff you would receive if admitted by each school. These payoffs represent your preferences for different schools. Your payoff in each round depends on which school admits you.
        \begin{table}[H]
            \centering
            \begin{tabular}{|l|c|c|c|c|}
                \hline
                Admitting school & A & B & C & Not admitted \\
                \hline
                Payoff (ECU) & X & Y & Z & 0 \\
                \hline
            \end{tabular}
        \end{table}
        \begin{itemize}
            \item If you are admitted by School A, your payoff in that round is X ECUs.
            \item If you are admitted by School B, your payoff in that round is Y ECUs.
            \item If you are admitted by School C, your payoff in that round is Z ECUs.
            \item If you are not admitted by any school, your payoff in that round is 0.
        \end{itemize}
    \item In each round, the payoff table for each student is randomly generated by the computer according to the following rule:
        \begin{itemize}
            \item With probability $1/2$, the payoff table is:
            \begin{table}[H]
                \centering
                \begin{tabular}{|l|c|c|c|c|}
                    \hline
                    Admitting school & A & B & C & Not admitted \\
                    \hline
                    Payoff (ECU) & 110 & 100 & 20 & 0 \\
                    \hline
                \end{tabular}
            \end{table}
            \item With probability $1/2$, the payoff table is:
            \begin{table}[H]
                \centering
                \begin{tabular}{|l|c|c|c|c|}
                    \hline
                    Admitting school & A & B & C & Not admitted \\
                    \hline
                    Payoff (ECU) & 100 & 110 & 20 & 0 \\
                    \hline
                \end{tabular}
            \end{table}
        \end{itemize}
    \item Therefore, in the first type of table, School A yields the highest payoff, while in the second type, School B yields the highest payoff. In each round, the payoff tables for students are independently and randomly generated in this way. That is, regardless of your payoff table, each of the other students in your group has a 1/2 probability of receiving either type of payoff table.
\end{itemize}

\textbf{Admission Rules}

\begin{itemize}
    \item \textbf{Lottery Draw:} In each round, the computer conducts a lottery draw for the three students in each group. Each student is randomly assigned a unique number between 1 and 3.

    Note: The lottery number is independent of the student ID. In each round, both the ID and the lottery draw are regenerated.
    \item \textbf{Student Ranking:} In each round, the computer ranks the three students in each group according to the lottery results. The smaller the lottery number, the higher the student’s rank.
    \item \textbf{Preference Submission:} In each round, each student submits one preference among Schools A, B, and C. After all students have submitted their perferences, the computer conducts the lottery. Thus, students do not know their lottery ranking when filling out their preferences.
    \item \textbf{Admission Mechanism:} After all students have submitted their preferences, the computer processes the admissions in the order of the student ranking determined by the lottery.
    \begin{itemize}
        \item If a school receives more than one student's application, it admits the highest-ranked student and rejects the others.
        \item If a school receives only one student's application, it admits that student directly.
        \item Students who are rejected by their chosen schools are not admitted by any school in that round.
    \end{itemize}
\end{itemize}

\textbf{An Example}

To further explain the admission rules, consider the following scenario.

\textbf{Lottery Draw:}
\begin{table}[H]
    \centering
    \begin{tabular}{|l|c|c|c|}
        \hline
        Student ID & 1 & 2 & 3 \\
        \hline
        Lottery number & 3 & 1 & 2 \\
        \hline
    \end{tabular}
\end{table}

\textbf{Ranking of Students:}
\begin{table}[H]
    \centering
    \begin{tabular}{|l|c|c|c|}
        \hline
        Rank & 1st & 2nd & 3rd \\
        \hline
        Student ID & 2 & 3 & 1 \\
        \hline
    \end{tabular}
\end{table}

\textbf{Submitted Preferences:}
\begin{table}[H]
    \centering
    \begin{tabular}{|l|c|c|c|}
        \hline
        Student ID & 1 & 2 & 3 \\
        \hline
        Preference & B & A & A \\
        \hline
    \end{tabular}
\end{table}

\textbf{Admission Process:}
\begin{itemize}
    \item School A receives applications from Students 2 and 3. According to the ranking, Student 2 ranks higher and is admitted by School A. Student 3 is rejected.
    \item School B receives an application from Student 1 and admits her directly.
\end{itemize}

\textbf{Final Admission Results:}
\begin{table}[H]
    \centering
    \begin{tabular}{|l|c|c|c|}
        \hline
        Student ID & 1 & 2 & 3 \\
        \hline
        Admitted school & B & A & Not admitted \\
        \hline
    \end{tabular}
\end{table}

\textbf{Payoff Rules}

After each round, you will be informed whether you have been admitted and which school admitted you, as well as your payoff. Your result in each round is independent of the outcomes in other rounds.

After all 20 rounds are completed, one round will be randomly selected as your final payoff. Additionally, you will receive 15 RMB as a show-up fee. Finally, you may earn extra income from a post-experiment questionnaire.

\subsection*{Instructions for the NoBundle\_Two treatment}

\textbf{General Information}

You are taking part in a decision-making experiment. Please read the instructions carefully. The instructions are the same for all participants. Please do not communicate with one another during the experiment. Turn off your mobile phone and set it to silent mode. If you have any questions, please raise your hand and an experimenter will come to assist you.

You have earned 15 RMB for showing up on time. In addition, you can earn more money during the experiment. The amount of money you earn will depend on the decisions you make. Your earnings in this experiment are expressed in EXPERIMENTAL CURRENCY UNITS (ECUs), with the conversion rate of 1 ECU = 1 RMB. Your final payment will be transferred to the bank account linked to your university within five working days after the experiment.

All decisions in this experiment are completely anonymous. Your name will remain strictly confidential, and other participants will not know your total payment for today’s experiment.

In today’s experiment, we will simulate the college admission process. You and the other participants will play the roles of students.

\textbf{Experimental Steps}

\begin{itemize}
    \item The experiment consists of 20 rounds of decision-making. Before each round begins, you will be randomly assigned to a group of three participants. Each participant in the group will play the role of a student. Each group faces three different schools, labeled A, B, and C. Each school has one admission slot, and each slot can admit one student. The admission process will be simulated by a computer.
    \item In each round, participants will be randomly reassigned into new groups, so it is unlikely that you will be grouped with the same two participants again. Each student will be randomly assigned an ID number within the group, and IDs will be reassigned in each round.
    \item In each round, you will see the following payoff table, which shows the payoff you would receive if admitted by each school. These payoffs represent your preferences for different schools. Your payoff in each round depends on which school admits you.
        \begin{table}[H]
            \centering
            \begin{tabular}{|l|c|c|c|c|}
                \hline
                Admitting school & A & B & C & Not admitted \\
                \hline
                Payoff (ECU) & X & Y & Z & 0 \\
                \hline
            \end{tabular}
        \end{table}
        \begin{itemize}
            \item If you are admitted by School A, your payoff in that round is X ECUs.
            \item If you are admitted by School B, your payoff in that round is Y ECUs.
            \item If you are admitted by School C, your payoff in that round is Z ECUs.
            \item If you are not admitted by any school, your payoff in that round is 0.
        \end{itemize}
    \item In each round, the payoff table for each student is randomly generated by the computer according to the following rule:
        \begin{itemize}
            \item With probability $1/2$, the payoff table is:
            \begin{table}[H]
                \centering
                \begin{tabular}{|l|c|c|c|c|}
                    \hline
                    Admitting school & A & B & C & Not admitted \\
                    \hline
                    Payoff (ECU) & 110 & 100 & 20 & 0 \\
                    \hline
                \end{tabular}
            \end{table}
            \item With probability $1/2$, the payoff table is:
            \begin{table}[H]
                \centering
                \begin{tabular}{|l|c|c|c|c|}
                    \hline
                    Admitting school & A & B & C & Not admitted \\
                    \hline
                    Payoff (ECU) & 100 & 110 & 20 & 0 \\
                    \hline
                \end{tabular}
            \end{table}
        \end{itemize}
    \item Therefore, in the first type of table, School A yields the highest payoff, while in the second type, School B yields the highest payoff. In each round, the payoff tables for students are independently and randomly generated in this way. That is, regardless of your payoff table, each of the other students in your group has a 1/2 probability of receiving either type of payoff table.
\end{itemize}

\textbf{Admission Rules}

\begin{itemize}
    \item \textbf{Lottery Draw:} In each round, the computer conducts a lottery draw for the three students in each group. Each student is randomly assigned a unique number between 1 and 3.
    
    Note: The lottery number is independent of the student ID. In each round, both the ID and the lottery draw are regenerated.
    \item \textbf{Student Ranking:} In each round, the computer ranks the three students in each group according to the lottery results. The smaller the lottery number, the higher the student’s rank.
    \item \textbf{Preference Submission:} In each round, each student submits two preferences: a first choice and a second choice. Each student can choose among Schools A, B, and C, but may not select the same school for both choices.
        \begin{table}[H]
            \centering
            \begin{tabular}{|wc{40mm}|wc{20mm}|}
            \hline
                First choice  &  \\
            \hline
                Second choice &  \\
            \hline
            \end{tabular}
        \end{table}
    Note: Students submit their preferences before the lottery draw, and thus do not know their ranking when filling out their preferences.
    \item \textbf{Admission Mechanism:} After all students have submitted their preferences, the computer processes the admissions in the order of the student ranking determined by the lottery.
        \begin{itemize}
            \item \textbf{Step 1:} Consider the first-ranked student (S1). S1 is admitted to their first-choice school. That school’s admission slot then becomes 0.
            \item \textbf{Step 2:} Consider the second-ranked student (S2). If their first-choice school still has an available slot, S2 is admitted to that school. Otherwise, S2 is admitted to their second-choice school, which must still have an available slot. That school’s slot then becomes 0.
            \item \textbf{Step 3:} Consider the third-ranked student (S3). If their first-choice school still has an available slot, S3 is admitted to that school. Otherwise, if their second-choice school has an available slot, S3 is admitted there. If both of S3’s choices are full, then S3 is not admitted.
            \item All students who are rejected by both of their choices are not admitted by any school in that round.
        \end{itemize}
\end{itemize}

\textbf{An Example}

To further explain the admission rules, consider the following two scenarios.

\textbf{Lottery Draw:}
\begin{table}[H]
    \centering
    \begin{tabular}{|l|c|c|c|}
        \hline
        Student ID & 1 & 2 & 3 \\
        \hline
        Lottery number & 3 & 1 & 2 \\
        \hline
    \end{tabular}
\end{table}

\textbf{Ranking of Students:}
\begin{table}[H]
    \centering
    \begin{tabular}{|l|c|c|c|}
        \hline
        Rank & 1st & 2nd & 3rd \\
        \hline
        Student ID & 2 & 3 & 1 \\
        \hline
    \end{tabular}
\end{table}

\textbf{Scenario 1:}
\begin{table}[H]
    \centering
    \begin{tabular}{|l|c|c|c|}
        \hline
        Student ID & 1 & 2 & 3 \\
        \hline
        First choice & B & A & A \\
        \hline
        Second choice & C & B & B \\
        \hline
    \end{tabular}
\end{table}

\textbf{Admission Process:}
\begin{itemize}
    \item Step 1: Student 2 (rank 1) is admitted to their first-choice school A. A’s admission slot becomes 0.
    \item Step 2: Student 3 (rank 2) has School A as their first choice, but A has no available slot. Therefore, Student 3 is admitted to their second-choice school B. B’s slot becomes 0.
    \item Step 3: Student 1 (rank 3) has School B as their first choice, but B has no available slot. They are then admitted to their second choice, School C.
\end{itemize}

\textbf{Final Admission Results:}
\begin{table}[H]
    \centering
    \begin{tabular}{|l|c|c|c|}
        \hline
        Student ID & 1 & 2 & 3 \\
        \hline
        Admitted school & C & A & B \\
        \hline
    \end{tabular}
\end{table}

\textbf{Scenario 2:}
\begin{table}[H]
    \centering
    \begin{tabular}{|l|c|c|c|}
        \hline
        Student ID & 1 & 2 & 3 \\
        \hline
        First choice & A & A & B \\
        \hline
        Second choice & B & C & C \\
        \hline
    \end{tabular}
\end{table}

\textbf{Admission Process:}
\begin{itemize}
    \item Step 1: Student 2 (rank 1) is admitted to their first-choice school A. A’s slot becomes 0.
    \item Step 2: Student 3 (rank 2) has School B as their first choice, which still has a slot. Student 3 is admitted to B. B’s slot becomes 0.
    \item Step 3: Student 1 (rank 3) has School A as their first choice, but A has no available slot. Their second-choice school B is also full. Therefore, Student 1 is not admitted.
\end{itemize}

\textbf{Final Admission Results:}
\begin{table}[H]
    \centering
    \begin{tabular}{|l|c|c|c|}
        \hline
        Student ID & 1 & 2 & 3 \\
        \hline
        Admitted school & Not admitted & A & B \\
        \hline
    \end{tabular}
\end{table}

\textbf{Payoff Rules}

After each round, you will be informed whether you have been admitted and which school admitted you, as well as your payoff. Your result in each round is independent of the outcomes in other rounds.

After all 20 rounds are completed, one round will be randomly selected as your final payoff. Additionally, you will receive 15 RMB as a show-up fee. Finally, you may earn extra income from a post-experiment questionnaire.

\clearpage

\section{Details of Experiment 2}\label{appendix:experiment2}

\setcounter{figure}{0}
\setcounter{table}{0}
\renewcommand\thetable{G\arabic{table}}
\renewcommand\thefigure{G\arabic{figure}}

%Intuitively, while reporting a bundle instead of a single school could weakly increase a student's match probability, she may incur a cost from being eventually assigned to a less preferred school in the bundle even though she is eligible for a more preferred school in the same bundle. The cost, however, is expected to be smaller when the bundled schools are nearer to being indifferent to the student and, therefore, should lead to a stronger incentive to report the bundle. Furthermore, reporting a bundle has an ambiguous effect on other students' welfare. For these reasons, it is important to empirically quantify to what extent a bundle system could improve the welfare of students.

This section reports results from Experiment 2 which is more complex than Experiment 1. The designed bundle system is an example of simple bundle systems. We compare participants' ROL strategies and matching outcomes under the bundle system to those under the standard system. The key insights conveyed from our theoretical analysis are confirmed in the experiment. Overall, under the bundle system, a significant proportion of participants choose to report bundles in their ROLs, and they are more likely to do so when the bundled schools are more similar in terms of participants' preferences. Compared with the standard system, the bundle system enhances most students' welfare, mainly through increasing their match rate.

\subsection{Experimental Design}

\subsubsection{Environment} 

We design a test-based admission game involving 6 students (ID1 to ID6) and 6 schools (from A to F). Each school has one seat. Students are played by subjects while schools are simulated by the computer. Students are ranked according to their exam scores, which are integers randomly and independently drawn from the normal distribution $ N(70,10) $ truncated on the interval $ [1,100] $. Since complete information would eliminate any uncertainty for students, we create an incomplete information environment in which each student is informed of her own score but not the scores of others, though the score distribution is commonly known among students. We choose the standard deviation for the score distribution to enable students to infer their ranks from their scores with reasonable precision. Students have identical preferences over schools, which are represented by the following payoffs:

\begin{center}
    $ u^i(D)=80 $, $ u^i(A)=50 $, $ u^i(B)=45 $, $ u^i(C)=40 $, $ u^i(E)=30 $, $ u^i(F)=20 $, $ u^i(\emptyset)=0 $.
\end{center}

\subsubsection{Treatments} 

We implement three treatments using a between-subjects design. The baseline, \textit{NoBundle}, represents a standard system in which participants can only report individual schools. The other two treatments, \textit{Indiff-Bundle} and \textit{Strict-Bundle}, implement a bundle system in which, in addition to individual schools, there is a bundle consisting of three different schools.

The two bundle treatments represent two different cases in which bundles may be designed in practice. In Indiff-Bundle, the three schools A, B, and C, which are close in payoffs for students, are bundled as ABC. This represents the case in which the policymaker properly bundles the schools that students regard as similar. In Strict-Bundle, D, E, and F, which students do not regard as similar, are bundled as DEF. This represents the case in which the policymaker does not properly design bundles or the designer must bundle dissimilar schools for other reasons. We are interested in learning how the different bundles may affect participants' strategies and the final matching outcome. 

Every student can report two options in their ROL. In NoBundle, students can report up to two schools, whereas in the two bundle treatments, students can report up to two schools, or one school and one bundle. 

The procedure of (bundle-)DA is simple in this experimental environment. In the standard system, DA is equivalent to a procedure in which students are sequentially assigned to their favorite schools reported in their ROL that have vacant seats. Similarly, in the bundle system, students are sequentially assigned to their favorite options reported in their ROL that have vacant seats. At the end of the matching process, students who are admitted by the bundle are randomly assigned to available seats in the bundle.

 \subsubsection{Experimental Procedure}

The experiment was conducted at the Nanjing Audit University Economics Experimental Lab with a total of 252 university students, using the software z-Tree \citep{Fischbacher2007}. We ran seven sessions for each of the three treatments. Each session consisted of 12 participants who were randomly re-matched into two groups of six in each of the 20 rounds. Within each group, each participant was assigned a student ID and an exam score, both of which were different across rounds.\footnote{Specifically, four different sequences of the pairs of ID and exam score at the group-round level were generated by a computerized random number generator using the score distribution mentioned earlier; two of them were used for a session. However, we ensured that all treatments use the same assignment of sequences to sessions to mitigate the possibility that any treatment difference is simply due to the different random numbers. We also ensured that students' scores are distinct for each group in each round.}  At the end of each round, participants received feedback about their matching outcome and the round payoff. At the end of the session, one round was randomly chosen for each participant to determine her payment. The experimental instructions are presented in Section~\ref{appendix:instructions-exp2}.

Upon arrival, participants were randomly seated at a partitioned computer terminal. The experimental instructions were given to participants in printed form and were also read aloud by the experimenter. Participants then completed a comprehension quiz before proceeding. After the experiment, they completed a demographic questionnaire. A typical session lasted about one hour with average earnings of 54.9 RMB, including a show-up fee of 15 RMB.\footnote{The average per-hour earnings in the experiment were substantially higher than the minimum hourly wage, which is about 15-20 RMB in the local region. At the time of the experiment, the conversion rate was approximately 1 US dollar to 6.9 RMB.}

\subsubsection{Hypotheses} 
 
Both bundle treatments provide incentives to at least some participants to report the bundle instead of an individual school in their ROL. In Indiff-Bundle, since all schools in the bundle ABC are close in payoffs to all students, we expect that all students should find the bundle to be an attractive option to increase their match probability. In Strict-Bundle, however, since schools in the bundle DEF have disparate payoffs, students would have incurred a much higher payoff loss if they were finally matched to E or F, when they could have been matched to a more preferred school if they had reported better schools. Therefore, we derive the following hypothesis.

\begin{hypothesis}\label{hypo:1}
 Students report bundles more often in the Indiff-Bundle treatment than in the Strict-Bundle treatment.
\end{hypothesis}

In either bundle treatment, students are not guaranteed to be better off by reporting the bundle relative to the baseline treatment. However, given that many students may choose to report the bundle, we still expect that the bundle system should help improve the overall match rate and therefore students' average payoff relative to the baseline.
 
\begin{hypothesis}\label{hypo:2}
The match rate and students' average payoff in both bundle treatments are higher than those in the NoBundle treatment.
\end{hypothesis}

Finally, we discuss students' ROL strategies. In Indiff-Bundle, since the schools in the bundle ABC are close in payoffs and they are between school D and the two remaining schools E and F in students' preferences, we expect students' ROL strategies to be ``truth-telling'' as stated in the following hypothesis. We use D-ABC to represent the strategy of ranking D above ABC. The other abbreviations are interpreted similarly.

\begin{hypothesis}\label{hypo:3}
In the Indiff-Bundle treatment, students either truthfully rank two individual schools, or use one of the following strategies: D-ABC, A-ABC (equivalent to the hypothetical strategy A-BC), ABC-E, ABC-F.
\end{hypothesis}

In contrast, in Strict-Bundle, school D in the bundle DEF has a much higher payoff than E and F, and the other schools not in the bundle are below D but above E and F. Higher-score students would find the tradeoff from reporting such a bundle more difficult, since they have higher chances of being matched to schools better than E and F by reporting such schools individually. Therefore, we expect that students with higher scores are more likely to only report individual schools D, A, B, and C in a ``truth-telling'' order, while students with lower scores are more likely to report one of the above schools in the first rank (to aim high) and then report the bundle DEF in the second rank to increase their overall match probabilities. For students with lower scores, the bundle DEF could be essentially viewed as a hypothetical bundle EF since they are unlikely to be matched to D.

\begin{hypothesis}\label{hypo:4}
In the Strict-Bundle treatment, students with higher scores are more likely to report two individual schools choosing from D, A, B, and C in the ``truth-telling'' manner. Students with lower scores are more likely to report the bundle DEF by using one of the four strategies, D-DEF, A-DEF, B-DEF and C-DEF.
\end{hypothesis}

Finally, in NoBundle, we expect that students report two individual schools truthfully in their ROL, and higher-score students are more likely to report better schools.

\begin{comment}
    
Despite the disparate payoff among the bundled schools in the Strict-Bundle treatment, such a bundle is expected to be more attractive to low-score students than to high-score students. The reason is that the bundle DEF could be viewed essentially as a bundle EF for low-score students: while these students have no realistic chance to be admitted to School D, being able to report the bundle DEF is still marginally better than to report an individual School E or F. Hence the following hypothesis:

\begin{hypothesis}
In the Strict-Bundle treatment, low-score students are more likely to report the bundle DEF than high-score students.
\end{hypothesis}

\end{comment}

\subsection{Experimental Results}

All hypotheses are strongly supported by the experimental results. We first report results on the overall performance in each treatment and then turn to the analysis of the detailed ROL strategies.

\subsubsection{Overall Performance}

\begin{figure}[H]
    \begin{center}
		\caption{Bundle report rate over round in Experiment 2}
		\label{fig:bundle-report-rate-exp2}
		\includegraphics[width=0.7\linewidth]{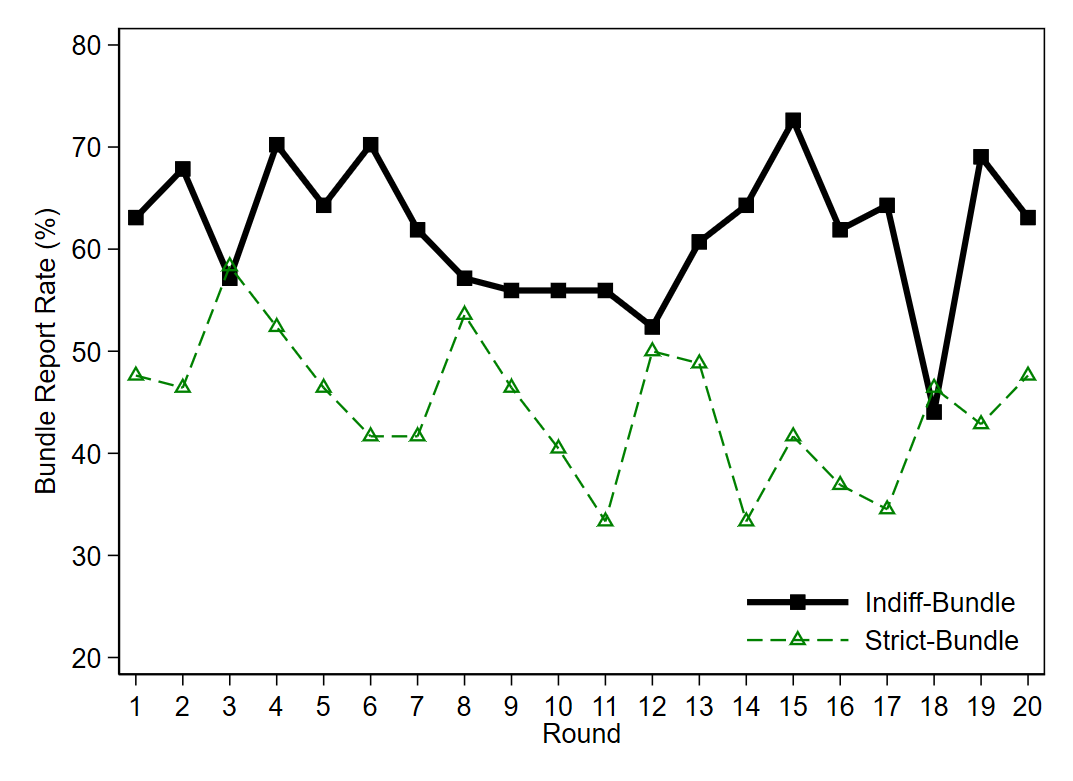}	
    \end{center}
\end{figure}

\autoref{table:summary-stats-exp2} summarizes key statistics across all treatments. We have the following observations. First, many students report the bundle in their ROL in both bundle treatments. The bundle report rate is 61.6\% in Indiff-Bundle and 44.5\% in Strict-Bundle, and the treatment difference is highly significant ($p$ = 0.001, Wilcoxon rank-sum test using each session as an independent observation), consistent with Hypothesis \ref{hypo:1}.\footnote{\autoref{fig:bundle-report-rate-exp2} shows that the difference in bundle report rates between the two bundle treatments is persistent in almost all rounds.} 

\begin{figure}[H]
    \begin{center}
		\caption{Match rate over round in Experiment 2}
		\label{fig:match-rate-exp2}
		\includegraphics[width=0.7\linewidth]{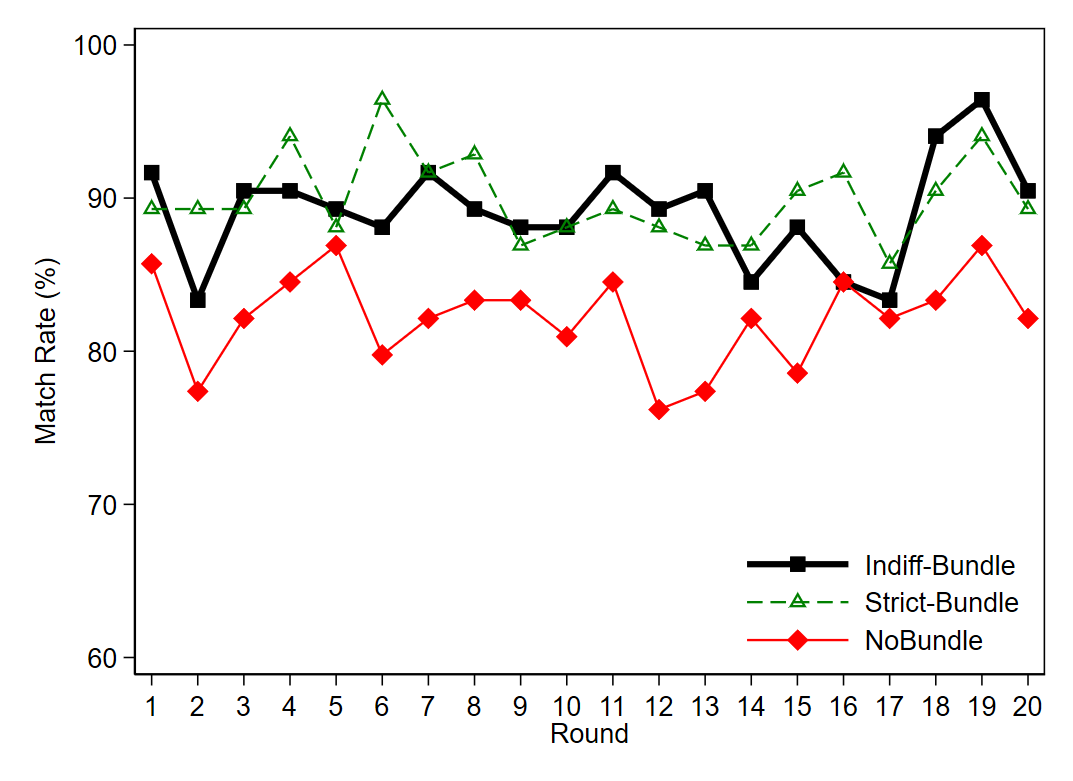}	
    \end{center}
\end{figure}

\begin{figure}[H]
    \begin{center}
		\caption{Student's average payoff over round in Experiment 2}
		\label{fig:student-payoff-exp2}
		\includegraphics[width=0.7\linewidth]{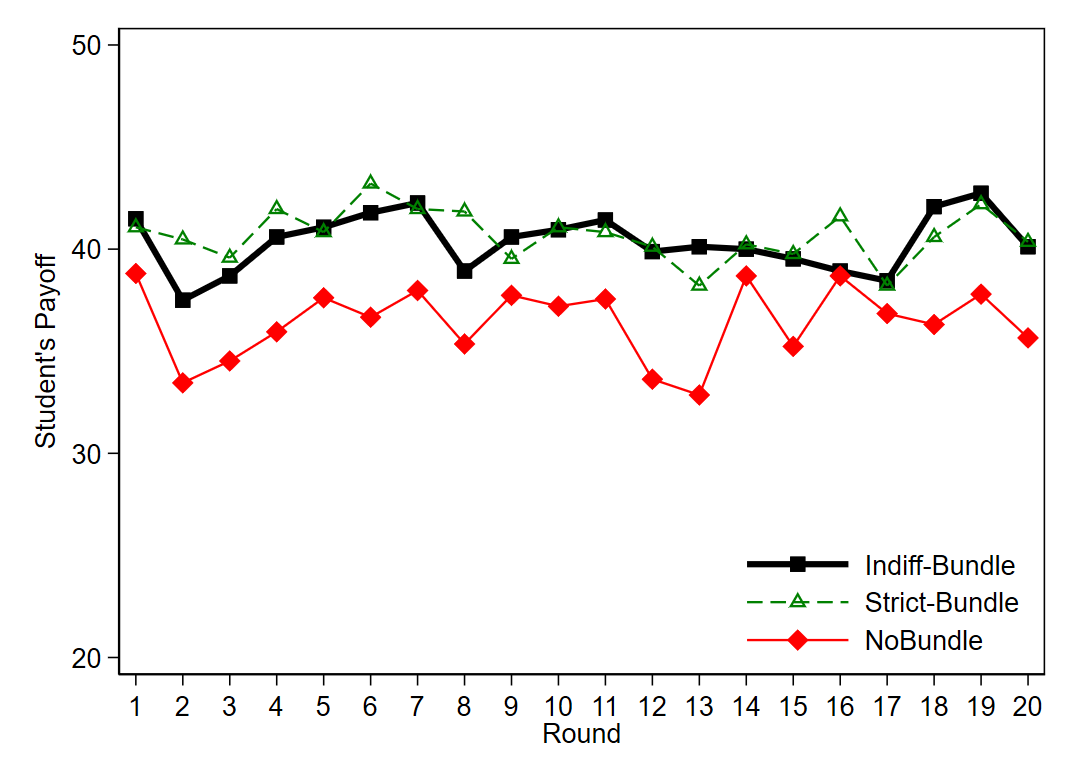}	
    \end{center}
\end{figure}

Second, compared to NoBundle, both bundle treatments significantly improve upon the match rate (89.2\% and 89.9\% vs. 82.2\%, $p$ = 0.001) and students' average payoff (40.7 ECUs and 40.4 ECUs vs. 36.4 ECUs, $p$ = 0.001), supporting Hypothesis \ref{hypo:2}.\footnote{\autoref{fig:match-rate-exp2} and \autoref{fig:student-payoff-exp2} show that the differences in match rates and average payoffs between the bundle treatments and the NoBundle treatment are persistent in almost all rounds.} The payoff increase is primarily driven by the improved match rate; the payoff loss due to unmatch (relative to the payoff under perfect match) is only half of that in NoBundle (8.6\% and 7.9\% vs. 17.5\%, $p$ = 0.001). Between the two bundle treatments, we find no significant differences in either the overall match rate or payoff.

However, the two bundle treatments have heterogeneous impacts on the match rates and payoffs for students in different score ranges. \autoref{table:student-match-rate-exp2} shows that the match rate is higher in Indiff-Bundle than in NoBundle for the students in each of the three score ranges: high (Score$\geq$75), medium (65$<$Score$<$75) and low (Score$\leq$65). The relative improvement appears to be larger for medium-score and low-score students than for high-score students.  In Strict-Bundle, while medium-score and low-score students continue to match with higher probabilities relative to the baseline treatment, high-score students do not benefit or even are slightly worse off under the bundle system, which, as will become clearer later, is attributed to the fact that they are generally not willing to report the bundle DEF. \autoref{table:student-payoff-exp2} shows a similar pattern in terms of average payoff.\footnote{\autoref{table:student-payoff-exp2} additionally shows that when excluding all cases of unmatch, the payoff increase is much diminished, indicating that the overall welfare gain under the bundle system is largely due to the improved match rate.}

Finally, while our theoretical analysis is ambiguous about how bundle systems impact fairness, it is conceivable that the presence of bundles may harm fairness since the final assignment to a school within a bundle is random. To measure fairness, we calculate the proportion of justified-envy pairs in matching outcomes. A pair of students $(i, j)$ is a justified-envy pair if $i$ is ranked higher than $j$ in exam scores but obtains a worse assignment than $j$. We find that, when the bundle is properly designed as in Indiff-Bundle, the bundle system does not compromise fairness compared to the baseline system (13.9\% vs. 13.1\%, $p$ = 0.594). On the contrary, when the bundle is not properly designed as in Strict-Bundle, the fairness measure is worse relative to the baseline system (19.4\% vs. 13.1\%, $p$ = 0.002). Hence, the bundle in Strict-Bundle improves students' average welfare by increasing the match rate but at the expense of fairness.

\begin{table}[!htb]
	\begin{center}
	\caption{Key summary statistics and tests}\label{table:summary-stats-exp2}
 	\begin{adjustbox}{width=1\textwidth}
		\begin{tabular}{lcccccc}
			\hline
			                                & (1) Indiff-Bundle & (2) Strict-Bundle & (3) NoBundle  & \multicolumn{3}{c}{p-value}  \\ 
                                                & \multicolumn{3}{c}{}                       & (1) vs. (3) & (2) vs. (3) & (1) vs. (2) \\
                                            
			\hline
		    Match rate                       & 89.2\% & 89.9\% & 82.2\% & 0.001 & 0.001 & 0.209 \\
		    Payoff loss due to unmatch       & 8.6\%  & 7.9\%  & 17.5\% & 0.001 & 0.001 & 0.403  \\	 
                Student's average payoff         & 40.7   & 40.4   & 36.4   & 0.001 & 0.001 & 0.403   \\
                Perc. of justified-envy pairs    & 13.9\% & 19.4\% &13.1\%  & 0.594 & 0.002 & 0.002   \\
			Bundle report rate               & 61.6\% & 44.5\% & / & / & / & 0.001  \\
			\textit{in rank 1}               & 31.3\% & 5.1\%  & / & / & / & 0.001  \\
			\textit{in rank 2}               & 30.4\% & 39.5\% & / & / & / & 0.013  \\		
			\hline
		\end{tabular}
        \end{adjustbox}
        \end{center}
        {\footnotesize \textit{Notes:} All p-values are produced by the Wilcoxon rank-sum test using each session as an independent observation. }
\end{table}

\subsubsection{ROL Strategies}

\begin{figure}[!htb]
    \begin{center}
		\caption{Prominent ROL strategies in each treatment}
		\label{fig:report-strategy-exp2}
		\begin{subfigure}{0.9\textwidth}
			\caption{Indiff-Bundle}
			\includegraphics[width=\linewidth]{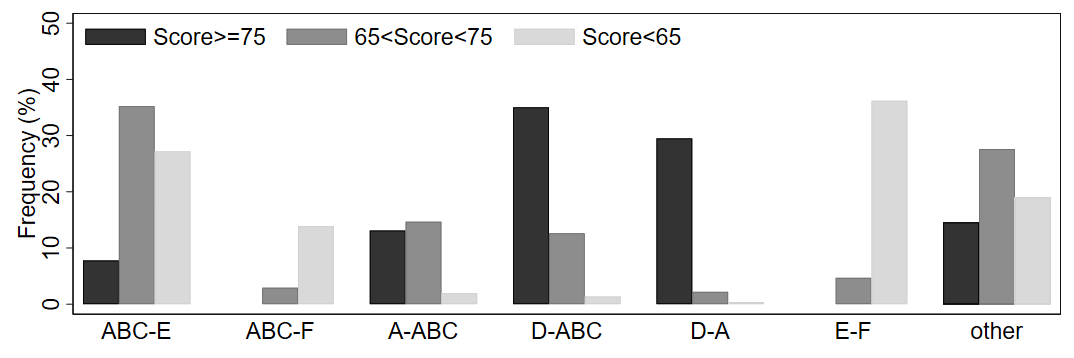}	
		\end{subfigure}
	    \begin{subfigure}{0.9\textwidth}
	    	\caption{Strict-Bundle}
	    	\includegraphics[width=\linewidth]{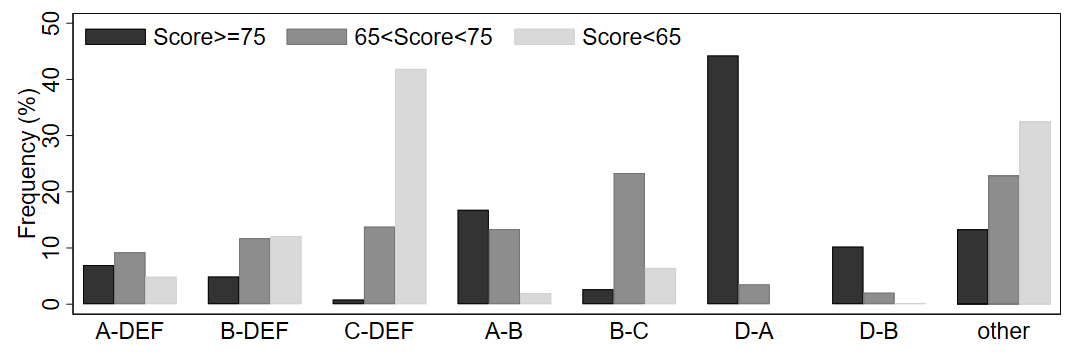}	
	    \end{subfigure}
        \begin{subfigure}{0.9\textwidth}
        	\caption{NoBundle}
        	\includegraphics[width=\linewidth]{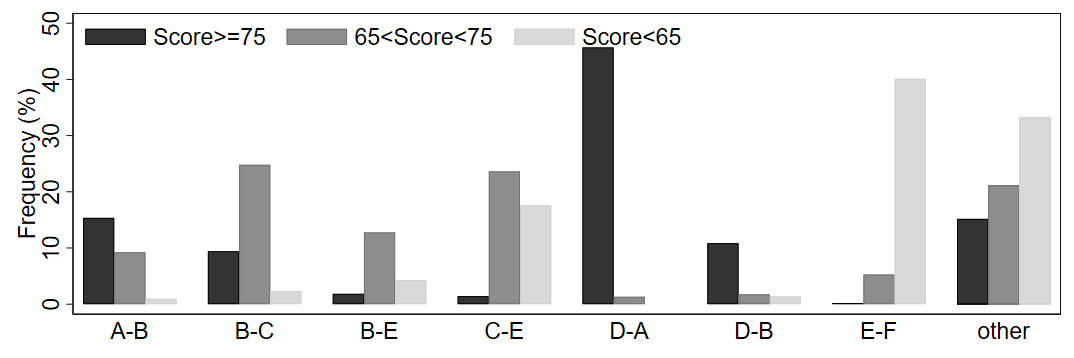}	
        \end{subfigure}
    \end{center}
\end{figure}

Turning to ROL strategies, we first observe that, compared to Strict-Bundle, students in Indiff-Bundle are more likely to report the bundle in the first rank but less likely to do so in the second rank  (first rank: 31.3\% vs. 5.1\%, $p$ = 0.001; second rank: 30.4\% vs. 39.5\%, $p$ = 0.013; see \autoref{table:summary-stats-exp2}). To examine the sources of this difference, \autoref{table:bundle-report-exp2} presents a breakdown of the bundle report rate for students with different score ranges in the two bundle treatments, and  \autoref{fig:report-strategy-exp2} plots the distribution of the most prominent strategies (defined as those occurring with frequency higher than 9\% for at least one score range in a treatment) used by these three types of students in each treatment.\footnote{The choice of the threshold 9\% is somewhat arbitrary. The main concern is to find the number of the prominent strategies, which we settle to be between 7 and 8, that can be comfortably fitted into a figure. For more detailed ROL strategies and their frequencies, \autoref{table:strategy-detail-exp2} reports the frequencies of all possible ROL strategies in all treatments. The table also shows that dominated strategies are only occasionally played by participants, suggesting their good understanding of the underlying incentives.} 

\begin{table}[!htb]
	\centering
	\caption{Bundle report rate by students' exam scores}\label{table:bundle-report-exp2}
	%	\begin{adjustbox}{width=1\textwidth}
		\begin{tabular}{lcc}
			\hline
			                     & Indiff-Bundle & Strict-Bundle  \\ 
			\hline
            Score$\geq$75        & \textbf{59.2\%} & \textbf{16.6\%}  \\
            \textit{in rank 1}   & 8.8\%           & 2.0\%  \\
            \textit{in rank 2}   & 50.4\%          & 14.5\%  \\
            65$<$Score$<$75      & \textbf{74.6\%} & \textbf{41.0\%}  \\
            \textit{in rank 1}   & 39.6\%          & 3.2\%  \\
            \textit{in rank 2}   & 34.9\%          & 37.7\%  \\
            Score$\leq$65         & \textbf{46.6\%} & \textbf{75.9\%}  \\
            \textit{in rank 1}   & 41.5\%          & 10.4\%  \\
            \textit{in rank 2}   & 5.1\%           & 65.6\%  \\
			\hline
		\end{tabular}
		%	\end{adjustbox}
\end{table}

In Indiff-Bundle, all students are generally willing to report the bundle ABC. However, high-score students are much more likely to report the bundle in the second rank (50.4\%) than in the first rank (8.8\%), whereas low-score students are much more likely to do the opposite (41.5\% and 5.1\% in the first and second rank, respectively). Medium-score students are almost equally likely to report the bundle in either rank. \autoref{fig:report-strategy-exp2} further indicates that if any bundle is reported by high-score students, it tends to be D-ABC or A-ABC. In contrast, medium-score and low-score students tend to report ABC-E and ABC-F if they choose to report the bundle. This pattern is thus consistent with Hypothesis \ref{hypo:3} that students are truth-telling in their ROL. In fact, high-score students are most likely to use the strategy D-ABC, followed by D-A. Both strategies are truth-telling and together account for about 65\% of all strategies played by high-score students. In contrast, low-score students are most likely to use the safest strategy E-F, followed by the more risky strategy ABC-E. Both strategies are again truth-telling and together account for 63\% of all strategies played by low-score students. Medium-score students tend to play more diverse strategies, with ABC-E being the most prominent. Choosing ABC-E makes good sense to these students because school D is very likely out of their reach while reporting F is too conservative.

However, in Strict-Bundle, a majority of students, especially high-score students, are not willing to report the bundle DEF, especially in the first rank (2.0\%). In particular, the least preferred schools E and F in the bundle DEF disincentivize high-score students to report such a bundle; the expected payoff of being admitted to the bundle is only 43.33. High-score students have a fairly good chance of doing better by reporting individual schools D, A, or B. As \autoref{fig:report-strategy-exp2} shows, they are most likely to use the strategy D-A, followed by the strategy A-B. Medium-score students play an expanded set of strategies such as the relatively more risky strategies, B-C and A-B, and relatively safer ones, A-DEF, B-DEF and C-DEF. They are more willing to report the bundle than the high-score students, but not as likely as their counterparts in Indiff-Bundle. In contrast, low-score students are much more enthusiastic about reporting the bundle DEF than the others as well as their counterparts in Indiff-Bundle. Reporting individual schools only is risky for them, whereas reporting the bundle would almost guarantee admission. Since these students do not have a realistic chance of being admitted to D, reporting DEF is essentially equivalent to reporting a hypothetical bundle consisting of E and F. Indeed, they most often play the strategy C-DEF, followed by the strategy B-DEF. All of these patterns are consistent with Hypothesis \ref{hypo:4}.

Finally, in NoBundle, high-score students are most likely to use the strategies D-A and A-B, whereas low-score students are most likely to play the strategies E-F and C-E. Medium-score students' strategies are more diverse, including B-C, C-E, B-E, and A-B. All of these strategies are truth-telling and in line with our expectations for each score range.

In summary, the patterns of students' ROL strategies in each treatment align with our hypotheses derived from an intuitive understanding of incentives in bundle systems.

\begin{table}[H]
	\centering
	\caption{Students' match rate by exam scores}\label{table:student-match-rate-exp2}
		\begin{tabular}{lccc}
			\hline
			                & Indiff-Bundle & Strict-Bundle & NoBundle  \\ 
			\hline
			Score$\geq$75   & 98.0\% & 94.7\% & 95.1\%  \\
			65$<$Score$<$75 & 89.3\% & 90.5\% & 84.3\%  \\
			Score$\leq$65   & 80.6\% & 84.7\% & 67.1\%  \\
			\hline
		\end{tabular}
\end{table}

\begin{table}[H]
	\centering
	\caption{Students' average payoff by exam scores}\label{table:student-payoff-exp2}
		\begin{adjustbox}{width=1\textwidth}
		\begin{tabular}{lcccccc}
			\hline
			                & \multicolumn{3}{c}{Overall} & \multicolumn{3}{c}{Excluding unmatch} \\ \cline{2-7}
			                & Indiff-Bundle & Strict-Bundle & NoBundle & Indiff-Bundle & Strict-Bundle & NoBundle \\ 
			\hline
			Score$\geq$75   & 58.9 & 56.2 & 57.1 & 60.2 & 59.4 & 60.1 \\
			65$<$Score$<$75 & 39.0 & 38.5 & 34.1 & 43.6 & 42.5 & 40.4 \\
			Score$\leq$65   & 24.5 & 28.8 & 19.8 & 30.4 & 34.0 & 29.5 \\
			\hline
		\end{tabular}
			\end{adjustbox}
\end{table}

\clearpage

\begin{xltabular}[!htb]{\textwidth}{|r|r|c|c|c|c|c|c|c|c|c|}
	\caption{Proportion of each possible ROL strategy (\%) in each treatment}\label{table:strategy-detail-exp2}  \\
	
	\hline \multicolumn{2}{|c|}{\textbf{Strategy}} & \multicolumn{3}{c|}{\textbf{Indiff-Bundle}} & \multicolumn{3}{c|}{\textbf{Strict-Bundle}} & \multicolumn{3}{c|}{\textbf{NoBundle}}  \\ \hline 
	\textbf{Rank 1} & \textbf{Rank 2} & S$\geq$75 & 65$<$S$<$75 & S$\leq$65 & S$\geq$75 & 65$<$S$<$75 & S$\leq$65 & S$\geq$75 & 65$<$S$<$75 & S$\leq$65 \\ \hline
	\endfirsthead
	
	\multicolumn{11}{c}%
	{\tablename\ \thetable{} -- continued from previous page} \\
	\hline \multicolumn{2}{|c|}{\textbf{Strategy}} & \multicolumn{3}{c|}{\textbf{Indiff-Bundle}} & \multicolumn{3}{c|}{\textbf{Strict-Bundle}} & \multicolumn{3}{c|}{\textbf{NoBundle}}  \\ \hline 
	\textbf{Rank 1} & \textbf{Rank 2} & S$\geq$75 & 65$<$S$<$75 & S$\leq$65 & S$\geq$75 & 65$<$S$<$75 & S$\leq$65 & S$\geq$75 & 65$<$S$<$75 & S$\leq$65 \\ \hline
	\endhead
	
	\hline \multicolumn{11}{|r|}{{Continued on next page}} \\ \hline
	\endfoot
	
	\multicolumn{11}{l}{{\footnotesize \textit{Notes:} The submission strategies marked by * are (weakly) dominated strategies. }} \\
	\endlastfoot
	
	\textbf{*ABC} & \textbf{A} & 0.20 & 0.29 & 0 & & & & & & \\
	\textbf{*ABC} & \textbf{B} & 0.20 & 0.44 & 0 & & & & & & \\
	\textbf{*ABC} & \textbf{C} & 0 & 0.15 & 0 & & & & & & \\
	\textbf{*ABC} & \textbf{D} & 0.61 & 0.59 & 0.39 & & & & & & \\
	\textbf{ABC} & \textbf{E} & 7.79 & 35.24 & 27.20 & & & & & & \\
	\textbf{ABC} & \textbf{F} & 0 & 2.94 & 13.89 & & & & & & \\
	\textbf{ABC} & \textbf{ABC} & 13.11 & 14.68 & 1.96 & & & & & & \\
	\textbf{B} & \textbf{ABC} & 2.05 & 6.75 & 0.59 & & & & & & \\
	\textbf{*C} & \textbf{ABC} & 0.20 & 0.44 & 0.39 & & & & & & \\
	\textbf{D} & \textbf{ABC} & 35.04 & 12.63 & 1.37 & & & & & & \\
	\textbf{*E} & \textbf{ABC} & 0 & 0.44 & 0.20 & & & & & & \\
	\textbf{*F} & \textbf{ABC} & 0 & 0 & 0.59 & & & & & & \\
	\textbf{DEF} & \textbf{A} & & & & 0.82 & 0.29 & 0.39 & & & \\
	\textbf{DEF} & \textbf{B} & & & & 0.82 & 0.88 & 1.76 & & & \\
	\textbf{DEF} & \textbf{C} & & & & 0.41 & 1.91 & 5.68 & & & \\
	\textbf{*DEF} & \textbf{D} & & & & 0 & 0 & 0 & & & \\
	\textbf{*DEF} & \textbf{E} & & & & 0 & 0.15 & 0.59 & & & \\
	\textbf{*DEF} & \textbf{F} & & & & 0 & 0 & 1.96 & & & \\
	\textbf{A} & \textbf{DEF} & & & & 6.97 & 9.25 & 4.89 & & & \\
	\textbf{B} & \textbf{DEF} & & & & 4.92 & 11.75 & 12.13 & & & \\
	\textbf{C} & \textbf{DEF} & & & & 0.82 & 13.80 & 41.88 & & & \\
	\textbf{D} & \textbf{DEF} & & & & 1.84 & 2.64 & 1.76 & & & \\
	\textbf{E} & \textbf{DEF} & & & & 0 & 0.29 & 2.94 & & & \\
	\textbf{F} & \textbf{DEF} & & & & 0 & 0 & 1.96 & & & \\
	\textbf{A} & \textbf{B} & 4.30 & 3.08 & 1.17 & 16.80 & 13.36 & 1.96 & 15.37 & 9.25 & 0.98 \\
	\textbf{A} & \textbf{C} & 0 & 0.15 & 0.20 & 5.74 & 5.87 & 1.37 & 6.56 & 4.85 & 0.98 \\
	\textbf{*A} & \textbf{D} & 0.20 & 0 & 0 & 0.20 & 0 & 0 & 0.41 & 0 & 0 \\
	\textbf{A} & \textbf{E} & 1.23 & 2.64 & 1.37 & 0 & 0.29 & 0.20 & 1.23 &	1.76 & 0.39 \\
	\textbf{A} & \textbf{F} & 0 & 0 & 0.20 & 0 & 0 & 0.98 & 0 & 0.59 & 1.17 \\
	\textbf{*B} & \textbf{A} & 0 & 0 & 0 & 0 & 0.59 & 0 & 0.20 & 0 & 0.20 \\
	\textbf{B} & \textbf{C} & 0.82 & 1.17 & 0.78 & 2.66 & 23.35 & 6.46 & 9.43 & 24.82 & 2.35 \\
	\textbf{*B} & \textbf{D} & 0 & 0.15 & 0 & 0 & 0.15 & 0 & 0.20 & 0.15 & 0 \\
	\textbf{B} & \textbf{E} & 1.02 & 6.02 & 1.76 & 0.20 & 1.03 & 0.98 & 1.84 & 12.78 & 4.31 \\
	\textbf{B} & \textbf{F} & 0 & 0.15 & 1.17 & 0 & 0 & 0.20 & 0.61 & 1.17 & 2.74 \\
	\textbf{*C} & \textbf{A} & 0 & 0 & 0 & 0 & 0 & 0.20 & 0 & 0 & 0 \\
	\textbf{*C} & \textbf{B} & 0 & 0 & 0 & 0 & 0 & 0.78 & 0 & 0.73 & 0.39 \\
	\textbf{*C} & \textbf{D} & 0 & 0 & 0 & 0 & 0.29 & 0 & 0 & 0.73 & 0.59 \\
	\textbf{C} & \textbf{E} & 0 & 2.94 & 3.33 & 0.20 & 6.02 & 4.70 & 1.43 & 23.64 & 17.61 \\
	\textbf{C} & \textbf{F} & 0 & 0.15 & 1.37 & 0 & 0.59 & 1.37 & 0.82 & 3.67 & 11.35 \\
	\textbf{D} & \textbf{A} & 29.51 & 2.20 & 0.39 & 44.26 & 3.52 & 0 & 45.70 & 1.32 & 0 \\
	\textbf{D} & \textbf{B} & 2.25 & 0.88 & 0 & 10.25 & 2.06 & 0.20 & 10.86 & 1.76 & 1.37 \\
	\textbf{D} & \textbf{C} & 0.20 & 0.44 & 0.39 & 2.46 & 0.88 & 0.20 & 3.48 & 4.11 & 1.37 \\
	\textbf{D} & \textbf{E} & 1.23 & 0.44 & 0.59 & 0.61 & 0.29 & 0.20 & 1.43 & 1.62 & 2.94 \\
	\textbf{D} & \textbf{F} & 0 & 0 & 1.96 & 0 & 0.15 & 0 & 0 & 1.03 & 5.09 \\
	\textbf{*E} & \textbf{A} & 0 & 0 & 0 & 0 & 0 & 0 & 0 & 0.15 & 0.39 \\
	\textbf{*E} & \textbf{B} & 0 & 0 & 0 & 0 & 0 & 0 & 0 & 0.15 & 0 \\
	\textbf{*E} & \textbf{C} & 0 & 0 & 0 & 0 & 0 & 0.39 & 0 & 0.29 & 0.39 \\
	\textbf{*E} & \textbf{D} & 0 & 0 & 0 & 0 & 0 & 0 & 0.20 & 0.15 & 0.20 \\
	\textbf{E} & \textbf{F} & 0 & 4.70 & 36.20 & 0 & 0.59 & 3.33 & 0.20 & 5.29 & 40.12 \\
	\textbf{*F} & \textbf{A} & 0 & 0 & 0 & 0 & 0 & 0 & 0 & 0 & 0 \\
	\textbf{*F} & \textbf{B} & 0 & 0 & 0 & 0 & 0 & 0.20 & 0 & 0 & 0.20 \\
	\textbf{*F} & \textbf{C} & 0 & 0 & 0 & 0 & 0 & 0.20 & 0 & 0 & 0.78 \\
	\textbf{*F} & \textbf{D} & 0 & 0 & 0 & 0 & 0 & 0 & 0 & 0 & 0.59 \\
	\textbf{*F} & \textbf{E} & 0 & 0.29 & 2.54 & 0 & 0 & 0.20 & 0 & 0 & 3.52 \\
	\hline
\end{xltabular}

\subsection{Instructions for Experiment 2}\label{appendix:instructions-exp2}

In the following, we translate the original instructions in Chinese into English for the NoBundle and Indiff-Bundle treatments. The instructions for the Strict-Bundle treatment are omitted because they are exactly the same as the instructions for the Indiff-Bundle treatment except for the different schools in the bundle.

\subsubsection*{Instructions for the NoBundle treatment}

\textbf{General Information}

You are taking part in a decision-making experiment. Please read the instructions carefully. The instructions are the same for every participant. Please do not communicate with each other during the experiment. Turn off your mobile phone and put it into the envelop on your desk. If you have a question, feel free to raise your hand, and an experimenter will come to help you.

You have earned 15 RMB for showing up on time. In addition, you can earn more money in this experiment. The amount of money you earn will depend upon the decisions you and other participants make. Your earnings in this experiment are expressed in EXPERIMENTAL CURRENCY UNITS, which we will refer to as ECUs. At the end of the experiment you will be paid using a conversion rate of 1 RMB for every 1 ECU of earnings from the experiment.

Your final payment will be paid to you via bank transfer within 2-3 days on completion of today’s experiment. All decisions are anonymous. That is, other participants will not know about your identity or your final payment.

In today's experiment, we will simulate the process of college admissions. You and the other participants will play the roles of students and submit your own preferences of schools to the admissions system. The system will assign an admission outcome to each student. The description of the experimental steps, payoff rules, and admission rules are as follows.

\textbf{Experimental Steps}

\begin{itemize}
	\item The experiment consists of 20 rounds of decision-making. Before each round, you will be randomly assigned to a group of six participants, and each person in the group will play the role of a student. Each group will face six different schools, represented by A, B, C, D, E, and F. Each school has one admission slot and can admit one student. The admission process will be simulated by a computer.
	\item In each round, participants will be randomly assigned to new groups, so it is unlikely that you will be grouped with the exact same five participants again. 
	\item In each round, you and the other participants will see the following payoff table, which displays the payoff you would receive when admitted to each school, representing your preference for different schools. Your payoff in each round depends on the school that admits you. Note that every student sees the same payoff table.
    	\begin{table}[H]
    		\centering
    			\begin{tabular}{|l|c|c|c|c|c|c|c|}
    				\hline
    				School          & D  & A  & B  & C  & E  & F  & not admitted \\ 
    				\hline
                        Payoff (ECU)    & 80 & 50 & 45 & 40 & 30 & 20 & 0  \\
    				\hline
    			\end{tabular}
    	\end{table}
            \begin{itemize}
            	\item If you are admitted by school D in a round, then your payoff in that round is 80 ECUs.
            	\item If you are admitted by school A in a round, then your payoff in that round is 50 ECUs.
            	\item ... (If you are admitted by another school, your payoff for that round follows the same logic.)
            	\item If you are admitted by school D in a round, then your payoff in that round is 80 ECUs.
            \end{itemize}
        \item In each , each student will see their own exam score. The score for each student is randomly generated by the computer, following a normal distribution with natural numbers ranging from 1 to 100. The graph below illustrates the probability distribution of the exam scores generated and the overall probabilities for each score range. The highest probability is for a score of 70, with a 68\% probability for scores between 60 and 80, and a 95\% probability for scores between 50 and 90.
            \begin{figure}
                \centering
                \includegraphics{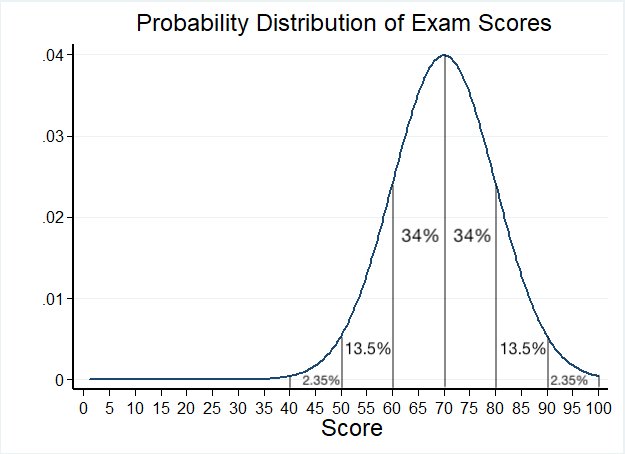}
            \end{figure}
        \item In each round, the scores of the six students in each group are independently and randomly generated. You will be informed of your own score but not the scores of other students in your group. In different rounds, scores for each student will be regenerated. All schools aim to admit students with higher scores.
\end{itemize}

\textbf{Preference Submission Table}

\begin{itemize}
    \item In each round, each student is required to submit a preference table as shown below. Each student can choose to rank two schools, namely the first-ranked school and the second-ranked school. When submitting preferences, each student can select any one of the schools A, B, C, D, E, F, but the two choices cannot be the same school.
        \begin{table}[H]
            \centering
            \begin{tabular}{|wc{40mm}|wc{20mm}|}
            \hline
                First-ranked school  &  \\
            \hline
                Second-ranked school &  \\
            \hline 
            \end{tabular}
        \end{table}
    \item The detailed admission rules will be described below.
\end{itemize}

\textbf{Admission Rules}

\begin{itemize}
    \item In each round, after all students have submitted their preferences, the computer will conduct the admission process for each group according to the following rules. Before the admission process begins, each school has one admission slot available.
        \begin{itemize}
            \item Firstly, all students are sorted in descending order based on their scores.
            \item Step 1: The highest-scoring student is considered first and is directly admitted to her first-ranked school. That school's admission slot will be filled after admitting this student.
            \item Step 2: The second highest-scoring student is considered. If there is an available admission slot in her first-ranked school, she is admitted to that school. Otherwise, if her second-ranked school has an available slot, she is admitted to that school. Similar to Step 1, that school's admission slot will be filled after admitting this student.
            \item Step 3: The process is repeated for the third highest-scoring student. If her first-ranked school has an available slot, she is admitted there. If not, her second-ranked school is checked. If her second-ranked school has an available slot, she is admitted there. If neither school in her submitted preference has an available slot, the student is not admitted. Similar to previous steps, if this student is admitted to a particular school, that school's admission slot will be filled after admitting this student.
            \item This process is repeated for all remaining students in descending order of scores until all students have been assigned admission outcomes.
        \end{itemize}
    \item At the end of each round, each student will be informed whether she has been admitted, along with the school that admitted her and her payoff. It is important to note that each student's admission outcome in each round is independent of the outcomes in previous rounds.
\end{itemize}

\textbf{An Example}

To further explain how the admission rules work, let's consider an example below.

\textbf{Students and Schools:} Six students 1, 2, 3, 4, 5, 6 and six schools A, B, C, D, E, F.

\textbf{Admission slots:} Each school has one admission slot available.

\textbf{Exam scores:} The students' scores, sorted in descending order, are as follows:
    \begin{table}[H]
        \centering
        \begin{tabular}{|l|c|c|c|c|c|c|}
            \hline
            Student ID   & 1  & 2  & 3  & 4  & 5  & 6   \\ 
            \hline
            Score        & 90 & 80 & 70 & 60 & 50 & 40  \\
            \hline
        \end{tabular}
    \end{table}

\textbf{Submitted preferences:} The preferences submitted by the students are as follows:
    \begin{table}[H]
        \centering
        \begin{tabular}{|l|c|c|c|c|c|c|}
            \hline
                                  & Student 1  & Student 2  & Student 3  & Student 4  & Student 5  & Student 6   \\ 
            \hline
            First-ranked school   & D & D & A & D & A & E  \\
            \hline
            Second-ranked school  & A & A & E & E & F & F  \\
            \hline
        \end{tabular}
    \end{table}

\textbf{Admission steps:}

\textbf{Step 1:} Assign Student 1 to her highest-ranked school with an available admission slot. Therefore, Student 1 is admitted to School D, and School D's admission slot becomes 0.
    \begin{table}[H]
        \centering
        \begin{tabular}{|c|c|c|c|c|c|}
            \hline
            Student 1  & Student 2  & Student 3  & Student 4  & Student 5  & Student 6   \\ 
            \hline
            D &  &  &  &  &   \\
            \hline
        \end{tabular}
    \end{table}

\textbf{Step 2:} Assign Student 2 to her highest-ranked school with an available admission slot. Since School D no longer has an admission slot, Student 2 is admitted to their second-ranked school, School A. School A's admission slot becomes 0.
    \begin{table}[H]
        \centering
        \begin{tabular}{|c|c|c|c|c|c|}
            \hline
            Student 1  & Student 2  & Student 3  & Student 4  & Student 5  & Student 6   \\ 
            \hline
            D & A &  &  &  &   \\
            \hline
        \end{tabular}
    \end{table}
    
\textbf{Step 3:} Assign Student 3 to her highest-ranked school with an available admission slot. Since School A no longer has an admission slot, Student 3 is admitted to her second-ranked school, School E. School E's admission slot becomes 0.
    \begin{table}[H]
        \centering
        \begin{tabular}{|c|c|c|c|c|c|}
            \hline
            Student 1  & Student 2  & Student 3  & Student 4  & Student 5  & Student 6   \\ 
            \hline
            D & A & E &  &  &   \\
            \hline
        \end{tabular}
    \end{table}
    
\textbf{Step 4:} Assign Student 4 to her highest-ranked school with an available admission slot. However, both of her choices, School D and School E, have exhausted their admission slots. Therefore, Student 4 is not admitted to any school.
    \begin{table}[H]
        \centering
        \begin{tabular}{|c|c|c|c|c|c|}
            \hline
            Student 1  & Student 2  & Student 3  & Student 4  & Student 5  & Student 6   \\ 
            \hline
            D & A & E & not admitted &  &   \\
            \hline
        \end{tabular}
    \end{table}

\textbf{Step 5:} Assign Student 5 to her highest-ranked school with an available admission slot. Since School A no longer has an admission slot, Student 5 is admitted to her second-ranked school, School F. School F's admission slot becomes 0.
    \begin{table}[H]
        \centering
        \begin{tabular}{|c|c|c|c|c|c|}
            \hline
            Student 1  & Student 2  & Student 3  & Student 4  & Student 5  & Student 6   \\ 
            \hline
            D & A & E & not admitted & F &   \\
            \hline
        \end{tabular}
    \end{table}

\textbf{Step 6:} Assign Student 6 to her highest-ranked school with an available admission slot. However, both of her choices, School E and School F, have exhausted their admission slots. Therefore, Student 6 is not admitted to any school.
    \begin{table}[H]
        \centering
        \begin{tabular}{|c|c|c|c|c|c|}
            \hline
            Student 1  & Student 2  & Student 3  & Student 4  & Student 5  & Student 6   \\ 
            \hline
            D & A & E & not admitted & F & not admitted  \\
            \hline
        \end{tabular}
    \end{table}
    
\textbf{The final admission outcomes and corresponding payoffs are as follows:}
    \begin{table}[H]
        \centering
        \begin{tabular}{|l|c|c|c|c|c|c|}
            \hline
                            & Student 1  & Student 2  & Student 3  & Student 4  & Student 5  & Student 6   \\ 
            \hline
            Admitted school & D & A & E & not admitted & F & not admitted  \\
            \hline
            Payoff (ECU)   & 80 & 50 & 30 & 0 & 20 & 0  \\
            \hline
        \end{tabular}
    \end{table}

\textbf{Payoffs}
\begin{itemize}
    \item After the completion of all 20 rounds of experiments, we will randomly select one round of admission results as your final experimental payoff. Additionally, you will receive a participation fee of 15 RMB. Finally, you may also receive additional income from the post-experiment questionnaire.
\end{itemize}

\subsubsection*{Instructions for the Indiff-Bundle treatment}

\textbf{General Information}

You are taking part in a decision-making experiment. Please read the instructions carefully. The instructions are the same for every participant. Please do not communicate with each other during the experiment. Turn off your mobile phone and put it into the envelop on your desk. If you have a question, feel free to raise your hand, and an experimenter will come to help you.

You have earned 15 RMB for showing up on time. In addition, you can earn more money in this experiment. The amount of money you earn will depend upon the decisions you and other participants make. Your earnings in this experiment are expressed in EXPERIMENTAL CURRENCY UNITS, which we will refer to as ECUs. At the end of the experiment you will be paid using a conversion rate of 1 RMB for every 1 ECU of earnings from the experiment.

Your final payment will be paid to you via bank transfer within 2-3 days on completion of today’s experiment. All decisions are anonymous. That is, other participants will not know about your identity or your final payment.

In today's experiment, we will simulate the process of college admissions. You and the other participants will play the roles of students and submit your own preferences of schools to the admissions system. The system will assign an admission outcome to each student. The description of the experimental steps, payoff rules, and admission rules are as follows.

\textbf{Experimental Steps}

\begin{itemize}
	\item The experiment consists of 20 rounds of decision-making. Before each round, you will be randomly assigned to a group of six participants, and each person in the group will play the role of a student. Each group will face six different schools, represented by A, B, C, D, E, and F. Each school has one admission slot and can admit one student. The admission process will be simulated by a computer.
	\item In each round, participants will be randomly assigned to new groups, so it is unlikely that you will be grouped with the exact same five participants again. 
	\item In each round, you and the other participants will see the following payoff table, which displays the payoff you would receive when admitted to each school, representing your preference for different schools. Your payoff in each round depends on the school that admits you. Note that every student sees the same payoff table.
    	\begin{table}[H]
    		\centering
    			\begin{tabular}{|l|c|c|c|c|c|c|c|}
    				\hline
    				School          & D  & A  & B  & C  & E  & F  & not admitted \\ 
    				\hline
                        Payoff (ECU)    & 80 & 50 & 45 & 40 & 30 & 20 & 0  \\
    				\hline
    			\end{tabular}
    	\end{table}
            \begin{itemize}
            	\item If you are admitted by school D in a round, then your payoff in that round is 80 ECUs.
            	\item If you are admitted by school A in a round, then your payoff in that round is 50 ECUs.
            	\item ... (If you are admitted by another school, your payoff for that round follows the same logic.)
            	\item If you are admitted by school D in a round, then your payoff in that round is 80 ECUs.
            \end{itemize}
        \item In each , each student will see their own exam score. The score for each student is randomly generated by the computer, following a normal distribution with natural numbers ranging from 1 to 100. The graph below illustrates the probability distribution of the exam scores generated and the overall probabilities for each score range. The highest probability is for a score of 70, with a 68\% probability for scores between 60 and 80, and a 95\% probability for scores between 50 and 90.
            \begin{figure}
                \centering
                \includegraphics{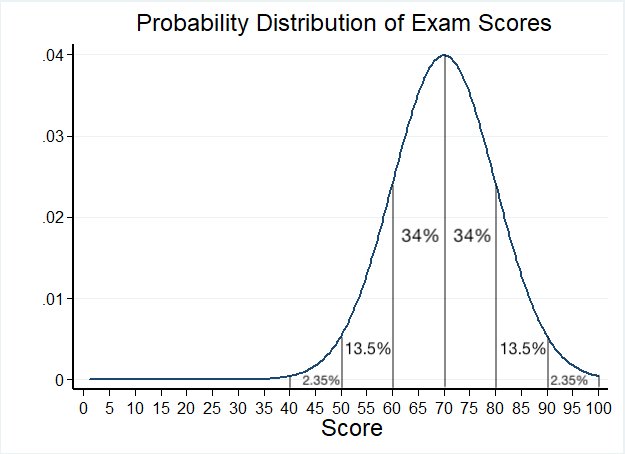}
            \end{figure}
        \item In each round, the scores of the six students in each group are independently and randomly generated. You will be informed of your own score but not the scores of other students in your group. In different rounds, scores for each student will be regenerated. All schools aim to admit students with higher scores.
\end{itemize}

\textbf{Preference Submission Table}

\begin{itemize}
    \item In each round, each student is required to submit a preference table as shown below. Each student can rank two choices, namely the first-ranked choice and the second-ranked choice. In this experiment, schools A, B, and C not only admit students individually but also have a joint admission known as the bundle ABC. When submitting preferences, each student can select individual schools from A, B, C, D, E, or F, as well as choose the bundle ABC. If a student includes ABC in their submission, it indicates their willingness to be jointly admitted by schools A, B, and C, and eventually be assigned to one of these schools.
        \begin{table}[H]
            \centering
            \begin{tabular}{|wc{40mm}|wc{20mm}|}
            \hline
                First-ranked choice  &  \\
            \hline
                Second-ranked choice &  \\
            \hline 
            \end{tabular}
        \end{table}
    \item Specifically, if a student includes ABC in one of her submitted choices, during the admission process, whenever we consider that preference for the student, if any of the three schools (A, B, C) still has an admission slot available, the student will be jointly admitted by the three schools. After the admission process concludes, the student will be assigned to one of the three schools. The specific assignment depends on which of the three schools still has an available admission slot. If more than one school has an admission slot available, the student will be randomly assigned to one of those schools.
    \item The detailed admission rules will be described below.
\end{itemize}

\textbf{Admission Rules}

\begin{itemize}
    \item In each round, after all students have submitted their preferences, the computer will conduct the admission process for each group according to the following rules. Before the admission process begins, each school has one admission slot available, and the bundle ABC, formed by schools A, B, and C, has three admission slots. It should be noted that during the admission process, the total number of students admitted by schools A, B, and C, whether individually or through the bundle, will not exceed three.
        \begin{itemize}
            \item Firstly, all students are sorted in descending order based on their scores.
            \item Step 1: The highest-scoring student is considered first and is directly admitted to her first-ranked choice. If her first-ranked choice is an individual school, that school's admission slot becomes 0. Additionally, if the school is one of the schools A, B, or C, because the student occupies one slot from the total slots of the three schools, the remaining slots for the bundle ABC decrease by 1. If this student is admitted through the bundle ABC, the remaining slots for ABC directly decrease by 1.            
            \item Step 2: The second highest-scoring student is considered. If there is an available admission slot in her first-ranked choice, she is admitted to that choice. Otherwise, if her second-ranked choice has an available slot, she is admitted to that choice. Similar to Step 1, if this student is admitted to an individual school, that school's admission slot becomes 0. Additionally, if the school is one of the schools A, B, or C, because the student occupies one slot from the total slots of the three schools, the remaining slots for the bundle ABC decrease by 1. If this student is admitted through the bundle ABC, the remaining slots for ABC directly decrease by 1.
            \item Step 3: The process is repeated for the third highest-scoring student. If there is an available admission slot in her first-ranked choice, she is admitted to that choice. Otherwise, if her second-ranked choice has an available slot, she is admitted to that choice. If both choices have no available slots, the student is not admitted. Similar to previous steps, if this student is admitted to an individual school, that school's admission slot becomes 0. Additionally, if the school is one of the schools A, B, or C, because the student occupies one slot from the total slots of the three schools, the remaining slots for the bundle ABC decrease by 1. If this student is admitted through the bundle ABC, the remaining slots for ABC directly decrease by 1.

            Note: Starting from this step and in each subsequent step, if the remaining slots for the bundle ABC become 0, it means that the total admission slots for schools A, B, and C have been filled. Therefore, regardless of whether schools A, B, and C admit students individually, their individual admission slots are immediately set to 0.
            
            \item This process is repeated for all remaining students in descending order of scores until all students have been assigned admission outcomes.
            \item Finally, for all students admitted through the bundle ABC, randomly assign them to the schools A, B, and C that still have available admission slots. If only one school has an admission slot left, then only one student must be admitted to ABC, and that student will be assigned to that school. With this, the admission process is completed.
        \end{itemize}
    \item At the end of each round, each student will be informed whether she has been admitted, along with the school that admitted her and her payoff. It is important to note that each student's admission outcome in each round is independent of the outcomes in previous rounds.
\end{itemize}

\textbf{An Example}

To further explain how the admission rules work, let's consider an example below.

\textbf{Students and Schools:} Six students 1, 2, 3, 4, 5, 6 and six schools A, B, C, D, E, F.

\textbf{Admission slots:} Each school has one admission slot available.

\textbf{Exam scores:} The students' scores, sorted in descending order, are as follows:
    \begin{table}[H]
        \centering
        \begin{tabular}{|l|c|c|c|c|c|c|}
            \hline
            Student ID   & 1  & 2  & 3  & 4  & 5  & 6   \\ 
            \hline
            Score        & 90 & 80 & 70 & 60 & 50 & 40  \\
            \hline
        \end{tabular}
    \end{table}

\textbf{Submitted preferences:} The preferences submitted by the students are as follows:
    \begin{table}[H]
        \centering
        \begin{tabular}{|l|c|c|c|c|c|c|}
            \hline
                                  & Student 1  & Student 2  & Student 3  & Student 4  & Student 5  & Student 6   \\ 
            \hline
            First-ranked choice   & D & D & ABC & B & ABC & C  \\
            \hline
            Second-ranked choice  & ABC & ABC & E & E & F & F  \\
            \hline
        \end{tabular}
    \end{table}

\textbf{Admission steps:}

\textbf{Step 1:} Assign Student 1 to her highest-ranked choice with an available admission slot. Therefore, Student 1 is admitted to School D, and School D's admission slot becomes 0.
    \begin{table}[H]
        \centering
        \begin{tabular}{|c|c|c|c|c|c|}
            \hline
            Student 1  & Student 2  & Student 3  & Student 4  & Student 5  & Student 6   \\ 
            \hline
            D &  &  &  &  &   \\
            \hline
        \end{tabular}
    \end{table}

\textbf{Step 2:} Assign Student 2 to her highest-ranked choice with an available admission slot. Since School D no longer has an admission slot, Student 2 is admitted to their second-ranked choice, bundle ABC. ABC's joint admission slots become 2.
    \begin{table}[H]
        \centering
        \begin{tabular}{|c|c|c|c|c|c|}
            \hline
            Student 1  & Student 2  & Student 3  & Student 4  & Student 5  & Student 6   \\ 
            \hline
            D & ABC &  &  &  &   \\
            \hline
        \end{tabular}
    \end{table}
    
\textbf{Step 3:} Assign Student 3 to her highest-ranked choice with an available admission slot. Therefore, Student 1 is admitted to the bundle ABC, and ABC's joint admission slots become 1.
    \begin{table}[H]
        \centering
        \begin{tabular}{|c|c|c|c|c|c|}
            \hline
            Student 1  & Student 2  & Student 3  & Student 4  & Student 5  & Student 6   \\ 
            \hline
            D & ABC & ABC &  &  &   \\
            \hline
        \end{tabular}
    \end{table}
    
\textbf{Step 4:} Assign Student 4 to her highest-ranked choice with an available admission slot. Therefore, Student 4 is admitted to her first-ranked choice, School B. School B's admission slot becomes 0. Additionally, since School B belongs to the bundle ABC, the remaining admission slots for ABC decrease to 0. It is important to note that because the remaining slots for ABC are already 0, the individual admission slots for School A and School C are immediately set to 0 too.
    \begin{table}[H]
        \centering
        \begin{tabular}{|c|c|c|c|c|c|}
            \hline
            Student 1  & Student 2  & Student 3  & Student 4  & Student 5  & Student 6   \\ 
            \hline
            D & ABC & ABC & B &  &   \\
            \hline
        \end{tabular}
    \end{table}

\textbf{Step 5:} Assign Student 5 to her highest-ranked choice with an available admission slot. Since the bundle ABC no longer has any admission slot, Student 5 is admitted to her second-ranked choice, School F. School F's admission slot becomes 0.
    \begin{table}[H]
        \centering
        \begin{tabular}{|c|c|c|c|c|c|}
            \hline
            Student 1  & Student 2  & Student 3  & Student 4  & Student 5  & Student 6   \\ 
            \hline
            D & ABC & ABC & B & F &   \\
            \hline
        \end{tabular}
    \end{table}

\textbf{Step 6:} Assign Student 6 to her highest-ranked choice with an available admission slot. However, both of her choices, School C and School F, have exhausted their admission slots. Therefore, Student 6 is not admitted to any school.
    \begin{table}[H]
        \centering
        \begin{tabular}{|c|c|c|c|c|c|}
            \hline
            Student 1  & Student 2  & Student 3  & Student 4  & Student 5  & Student 6   \\ 
            \hline
            D & ABC & ABC & B & F & not admitted  \\
            \hline
        \end{tabular}
    \end{table}

\textbf{Step 7:} In the above admission process, Students 2 and 3 were admitted through the bundle ABC. However, since School B has already admitted Student 4 individually, Students 2 and 3 can only be randomly assigned to School A and School C. Let's assume that the random assignment results in Student 2 being admitted to School A and Student 3 being admitted to School C. The final admission outcomes and corresponding payoffs are as follows:
    \begin{table}[H]
        \centering
        \begin{tabular}{|l|c|c|c|c|c|c|}
            \hline
                            & Student 1  & Student 2  & Student 3  & Student 4  & Student 5  & Student 6   \\ 
            \hline
            Admitted school & D & A & C & B & F & not admitted  \\
            \hline
            Payoff (ECU)   & 80 & 50 & 40 & 45 & 20 & 0  \\
            \hline
        \end{tabular}
    \end{table}

\textbf{Payoffs}
\begin{itemize}
    \item After the completion of all 20 rounds of experiments, we will randomly select one round of admission results as your final experimental payoff. Additionally, you will receive a participation fee of 15 RMB. Finally, you may also receive additional income from the post-experiment questionnaire.
\end{itemize}

\end{document}